%% file: multi_purchase_msom_all.tex
\documentclass[msom]{informs3} 

\DoubleSpacedXI

\usepackage{natbib}
 \bibpunct[, ]{(}{)}{,}{a}{}{,}%
 \def\bibfont{\small}%
\TheoremsNumberedThrough     

\EquationsNumberedThrough    

\RUNAUTHOR{Tulabandhula et al.} 

\RUNTITLE{Multi-Purchase Behavior: Modeling, Estimation and Optimization}

\input{header_info}

\newcommand{\appModelDerivation}{Appendix~\ref{sec:model-derivation}}
\newcommand{\appEstimation}{Appendix~\ref{sec:estimation}}
\newcommand{\appEstimationAbbriv}{App.~\ref{sec:estimation}}
\newcommand{\appAlgEst}{Alg.~\ref{alg:est}}
\newcommand{\appAlgMIP}{Alg.~\ref{alg:mip}}

\newcommand{\appLemsCombined}{Lem.~\ref{lem:shdbelong}-\ref{lem:monotone}}
\newcommand{\appStructural}{Appendix~\ref{sec:structural}}
\newcommand{\appThmMMC}{Thm.~\ref{thm:mmc_hard}}
\newcommand{\appHardness}{App.~\ref{sec:hardness}}
\newcommand{\appObjAbbriv}{App.~\ref{sec:objective}}
\newcommand{\appObj}{Appendix~\ref{sec:objective}}
\newcommand{\appThmSmallVij}{Thm.~\ref{thm:small_vij}}
\newcommand{\appThmValCons}{Thm.~\ref{thm:valCons}}
\newcommand{\appLemAdxopt}{Lem.~\ref{lem:adxoptSoln}}
\newcommand{\appMainBenchmarks}{Appendix~\ref{sec:main-benchmarks}}
\newcommand{\appMainBenchmarksAbbriv}{App.~\ref{sec:main-benchmarks}}
\newcommand{\appLinearConstr}{Appendix~\ref{sec:linear-constr}}
\newcommand{\appAlgBSE}{Alg.~\ref{alg:bin_srch_outline_eff}}
\newcommand{\appAlgBS}{Alg.~\ref{alg:bin_srch_outline}}
\newcommand{\appAlgEfficient}{App.~\ref{sec:algo-efficient}}
\newcommand{\appAlgFull}{Appendix~\ref{sec:algorithms-full}}
\newcommand{\appAddExp}{Appendix~\ref{sec:add_exp}}
\newcommand{\appTabProdPcts}{Table~\ref{tab:product-percentages-full}}
\newcommand{\appTabEstDataset}{Table ~\ref{tab:estimdatasetresults}}

\begin{document}

\TITLE{Multi-Purchase Behavior: Modeling, Estimation and Optimization}

\ARTICLEAUTHORS{%
\AUTHOR{Theja Tulabandhula}
\AFF{University of Illinois at Chicago, \EMAIL{tt@theja.org}, \URL{}}
\AUTHOR{Deeksha Sinha}
\AFF{Massachusetts Institute of Technology, \EMAIL{deeksha@mit.edu}, \URL{}}
\AUTHOR{Saketh Reddy Karra}
\AFF{University of Illinois at Chicago, \EMAIL{skarra7@uic.edu}, \URL{}}
\AUTHOR{Prasoon Patidar}
\AFF{Carnegie Mellon University, \EMAIL{prasoonpatidar@gmail.com}, \URL{}}
}

\ABSTRACT{%
 \input{0_abstract}
}


\maketitle
          
\input{1_introduction}
\input{2_model}
\input{3_algos}
\input{4_experiments}

\input{5_discussion}
\input{6_conclusion}

 
{
\small
\bibliographystyle{informs2014}
\bibliography{references.bib}
}

\newpage

\begin{APPENDICES}

\input{90_appendix}
\input{91_model_and_estimation}
\input{92_hardness}
\input{93_algorithms}
\input{94_more_experiments}

\end{APPENDICES}

\end{document}

%% file: header_info.tex
\usepackage{ifthen}
\newif\ifisEC
\isECfalse

\newif\ifisDraft
\isDrafttrue

\newif\ifisVerbose
\isVerbosefalse

\usepackage{comment,amsfonts,mathtools,bm,booktabs,lmodern, microtype,xcolor}
\usepackage{appendix}

\usepackage[utf8]{inputenc} 
\usepackage[T1]{fontenc}    
\usepackage{hyperref}       
\usepackage{url}            
\usepackage{booktabs}       
\usepackage{amsfonts}       
\usepackage{nicefrac}       
\usepackage{microtype}      

\usepackage{amsmath}
\usepackage{breqn}
\usepackage{bbm}
\usepackage{dsfont}

\usepackage{pgfplots}
\usepackage[noend]{algorithmic}
\usepackage{algorithm}
\usepackage{subcaption}
\usepackage[font=small,labelfont=bf]{caption}
\usepackage{graphicx}
\usepackage[shortlabels]{enumitem}

\makeatletter
\renewcommand{\ALG@name}{Alg.}
\makeatother

\newcommand{\prodUniverse}{W}
\newcommand{\offeredAsmt}{C}
\newcommand{\numberOfItems}{n}
\newcommand{\prob}{\mathsf{P}}
\newcommand{\feasibleAsmtSet}{\mathcal{C}}
\newcommand{\rcm}{\bmvl-2}
\newcommand{\tcm}{\bmvl-3}
\newcommand{\bmvl}{BundleMVL}

\newcommand{\bmvlk}{\bmvl-K}
\newcommand{\ucm}{MMC}
\newcommand{\binsearch}{\textsc{BinarySearchAO}}
\newcommand{\noisybinsearch}{\textsc{NoisyBinarySearchAO}}
\newcommand{\binsearcheff}{\textsc{BinarySearchAO(Efficient)}}
\newcommand{\noisybinsearcheff}{\textsc{NoisyBinarySearchAO(Efficient)}}
\newcommand{\adxopt}{\textsc{Adxopt}}
\newcommand{\adxopttwo}{\textsc{Adxopt2}}
\newcommand{\adxoptk}{\textsc{AdxoptL}}
\newcommand{\anyasmt}{C}
\newcommand{\anychosenset}{S}

\newcommand{\Vij}{{V_{\{i,j\}}}}
\newcommand{\Vik}{{V_{\{i,k\}}}}
\newcommand{\Vjk}{{V_{\{j, k\}}}}
\newcommand{\Vim}{{V_{\{i,m\}}}}

\newcommand{\Vil}{{V_{\{i,l\}}}}

\newcommand{\Vi}{{V_{\{i\}}}}
\newcommand{\Vj}{{V_{\{j\}}}}
\newcommand{\Vm}{{V_{\{m\}}}}
\newcommand{\Vl}{{V_{\{l\}}}}
\newcommand{\capconst}{d}
\newcommand{\compthresh}{\kappa}
\newcommand{\revk}{R_K}
\newcommand{\revtwo}{R_2}
\def\prob{P}
\newcommand{\ucmao}{\textsc{2-products mmc ao}}
\newcommand{\mmnlao}{\textsc{2-class logit ao}}
\newcommand{\maxcut}{\textsc{Max-Cut}}
\newcommand{\ind}{\mathds{1}}
\newcommand{\specialcell}[2][c]{%
\begin{tabular}[#1]{@{}c@{}}#2\end{tabular}}

%% file: 0_abstract.tex
\emph{Problem definition:} We study the problem of modeling purchase of multiple products and utilizing it to display optimized recommendations for online retailers and e-commerce platforms.
\emph{Relevance:} Rich modeling of users and fast computation of optimal products to display given these models can lead to significantly higher revenues and simultaneously enhance the user experience. 
\emph{Methodology:} We present a parsimonious multi-purchase family of choice models called the \bmvlk \ family, and develop a binary search based iterative strategy that efficiently computes optimized recommendations for this model. We establish the hardness of computing optimal recommendation sets, and derive several structural properties of the optimal solution that aid in speeding up computation. This is one of the first attempts at operationalizing multi-purchase class of choice models. 
\emph{Results:}  We show one of the first quantitative links between modeling multiple purchase behavior and revenue gains. The efficacy of our modeling and optimization techniques compared to competing solutions is shown using several real world datasets on multiple metrics such as model fitness, expected revenue gains and run-time reductions. For example, the expected revenue benefit of taking multiple purchases into account is observed to be $\sim5\%$ in relative terms for the Ta Feng and UCI shopping datasets, when compared to the MNL model for instances with $\sim 1500$ products. Additionally, across $6$ real world datasets, the test log-likelihood fits of our models are on average $17\%$ better in relative terms.
\emph{Managerial implications:} Our work contributes to the study multi-purchase decisions, analyzing consumer demand and the retailers optimization problem. The simplicity of our models and the iterative nature of our optimization technique allows practitioners meet stringent computational constraints while increasing their revenues in practical recommendation applications at scale, especially in e-commerce platforms and other marketplaces. 

%% file: 1_introduction.tex
\section{Introduction}
\label{sec:intro}

While shopping, consumers typically purchase multiple products. Multiple purchases can happen due to complementarity, neighborhood effects (i.e., due to certain products shown or placed next to each other), or due to overlap in the purchase frequencies of various products (sometimes across unrelated categories)~\citep{manchanda1999shopping}. All these effects may not be active simultaneously, and difficult to measure directly. 
Further, given an expressive choice model that captures this behavior, it is a priori unclear how to optimize for the product recommendations that the platform should show to consumers, as these tend to be hard combinatorial optimization problems.
The ability to capture a rich enough choice behavior of each consumer/customer as well as to display real-time optimized recommendations to them, can have a tremendous impact on the user experience and hence the bottom-line of online shopping platforms. Given that online shopping is one of the most popular activities on the Internet world wide, with sales projected to exceed 6.3+ trillion US dollars by 2024~\citep{marketsize}, such an effort can yield significant dividends. 

Addressing the above challenges, we consider a parsimonious family of multi-purchase choice models (i.e. models capturing purchase of more than one products in an interaction), called the \emph{bundle multivariate logit models} (\bmvlk), which is inspired by multiple variations proposed in the marketing literature~\citep{cox1972analysis,russell2000analysis,hruschka1999cross,singh2005modeling}. It models the conditional probability of purchasing a product given the purchase/no-purchase of all other products, and parsimoniously extends the popular multinomial choice model (MNL) to the setting where the customer purchases one or more products (represented by the suffix K $\in \mathbb{Z}_{+}$ in the acronym above) when they are shown a set of recommendations. While modeling multiple purchases has been addressed in the marketing literature, with the end goal of improving customer understanding, the application of \bmvlk{} to identify and optimize the set of products (not just categories) to recommend/display, with a focus on its use in web scale settings, is new to this work. We contrast \bmvlk \ to existing single-choice and multi-choice models~\citep{benson2018discrete}, deliberate on the optimal choice of K based on practical considerations, and validate the model performance (in terms of likelihood) on $6$ real world datasets. For instance, \bmvlk's parameters allow for transparently interpreting certain substitutable and complementary relationships between products. Model fit (assessed using out of sample log likelihood values) when compared against state of the art multi-choice models shows that the \rcm~(i.e., with K$=2$) is a strong candidate in capturing rich multi-choice behavior of customers. In particular, we show that the benefit of taking multiple purchases into account in terms of expected revenue is at least $\sim5\%$ in relative terms when compared to the MNL (this is for $1500$ products; for other revenue improvement comparisons including other models such as the Mixture of MNLs (MMNL), see Section~\ref{sec:experiments}). In addition, across $6$ real world datasets, the test log-likelihood fits of our multi-purchase model are on average $17\%$ better. Our approach addresses the key future direction discussed in the practice paper by ~\cite{alibaba2019}, who establish strong evidence of the utility of single choice models in a product display setting at the firm Alibaba.

This work is one of the first attempts focused on optimizing revenue for multi-purchase choice models, allowing us to make a link between the model and the expected revenue gains that it allows. Most prior work, especially in the marketing literature, has not been able to establish this, understandably due to complexity of the optimization problems involved. In particular, assuming that each customer has a distinct \bmvlk \ model associated with them, we develop an iterative binary search based optimization scheme for computing approximately revenue optimal recommendations, while balancing its computational time and solution quality (extending a similar idea for MNL by~\cite{sinha2017optimizing}).  We complement this by deriving several structural results about the optimal solution that help efficiently explore the search space, as well as by establishing the hardness of the problem (showing that it is indeed NP-complete when $K=2$, compared to linear time when $K=1$). The structural results allow our algorithm to determine if a product is part of the optimal solution or not with certainty in constant time, which allows us to work with smaller problem instances computationally. Our algorithm solves a quadratic unconstrained binary optimization (QUBO)  problem in each comparison step (we show how to solve this practically using state of the art heuristic approximation techniques as well). Our approach is compared against a mixed integer programming benchmark, a natural greedy approach that extends the  \adxopt~algorithm developed for the MNL single-choice model ~\citep{jagabathula2014assortment} and the revenue-ordered heuristic, among others. We also shed light on the properties of the solutions obtained using these benchmarks and when they perform optimally. Our solution to the revenue optimal recommendation problem is one of the first that is also practical and scalable for this class of models. The operational value of our algorithm is that it allows practitioners to capture additional revenue by being able to compute near optimal highly relevant recommendations at scale, hence minimizing the impact on customer disengagement due to computational delays that is typically observed in online platforms~\citep{speed}. Along the way, we also make progress on the theoretical and computational tractability of recommendation set optimization under another natural multi-choice model proposed by \cite{benson2018discrete}, and illustrate how the \bmvlk \ model and recommendations based on it provide superior performance. Table \ref{table:summary} outlines the list of models/algorithmic techniques developed in this work.

\begin{minipage}{.96\textwidth}
\begin{minipage}[b]{0.48\textwidth}
\scriptsize
\centering
\resizebox{\textwidth}{!}{%
\begin{tabular}{| l |c|} 
 \hline
 \textbf{Results} & \textbf{Section/} \\
                  & \textbf{Appendix} \\ [0.5ex] 
 \hline
 \textbf{Model:} \rcm & \\
  - Estimation (Lem.~\ref{lemma:estimate}, \appAlgEst) & Sec.~\ref{sec:model} \& \appEstimationAbbriv\\
  \specialcell{
               - Hardness (Thm.~\ref{thm:bmvl_hard}), and structural \\
                properties (\appLemsCombined)}
                                                     & Sec.~\ref{sec:alg-rec-set} \& \appObjAbbriv\\
  - \binsearch{} (\appAlgBS) 		&  Sec.~\ref{sec:alg-rec-set} \& \appAlgEfficient \\
  - \binsearcheff{} (\appAlgBSE) 	&  Sec.~\ref{sec:alg-rec-set} \& \appAlgEfficient \\
  - \noisybinsearcheff &  Sec.~\ref{sec:alg-rec-set} \& \appAlgEfficient \\
  - Integer Program (\appAlgMIP)  &  Sec.~\ref{sec:alg-rec-set} \& \appMainBenchmarksAbbriv\\
  - \adxoptk~(\appLemAdxopt) &  Sec.~\ref{sec:alg-rec-set} \& \appMainBenchmarksAbbriv\\ 
  - Revenue-ordered (\appThmValCons{} \&\appThmSmallVij) &  Sec.~\ref{sec:alg-rec-set} \& \appMainBenchmarksAbbriv\\
 \textbf{Model:} \ucm & 		\\
  - Hardness (\appThmMMC)   &  Sec.~\ref{sec:alg-rec-set} \& \appHardness \\ [1ex] 
 \hline
\end{tabular}%
}
  \captionof{table}{Summary of contributions.\label{table:summary}}
\end{minipage}
\hfill
\begin{minipage}[b]{0.48\textwidth}
\centering
	\includegraphics[width=0.75\linewidth]{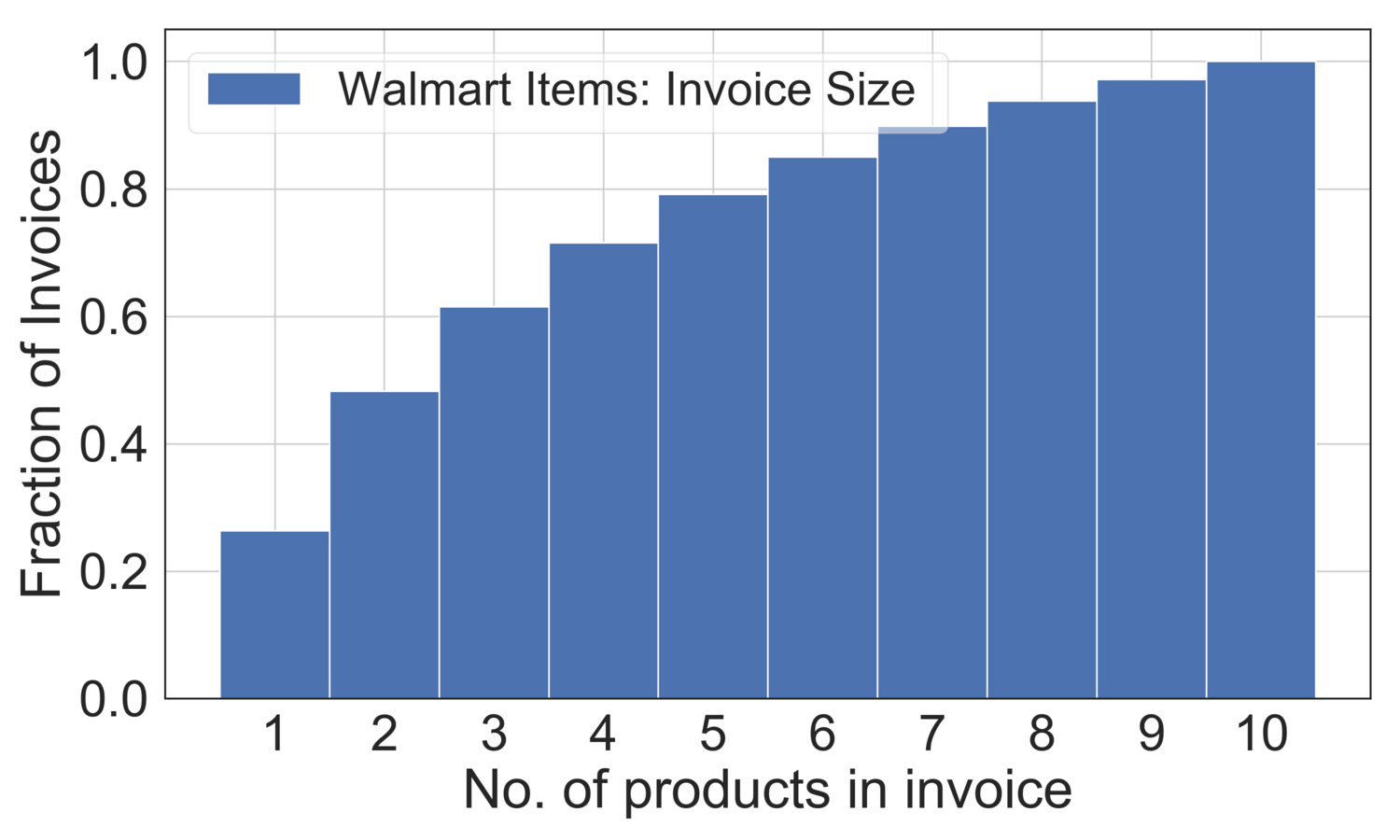}
	\captionof{figure}{Cumulative distribution function (cdf) of the number of products purchased per transaction (truncated to $\leq 10$) in the Walmart Items dataset.
	\label{fig:walmart_tran_hist}}
\end{minipage}
\end{minipage}


\subsection{Relevant Literature} \label{sec:related}

Though purchasing multiple products by customers is extremely common in both online platforms and brick and mortar stores, the majority of research in choice modeling in the operations management literature has focused on single product purchases. \cite{train2009discrete} gives a good overview of the commonly used choice models, which include the MNL model~\citep{plackett1975analysis}, the nested logit model 
and others. More recently, a variety of other choice models such as the Markov chain choice model,
the distribution over ranking model 
have been proposed and studied. Most work here concerns both the estimation of the model from data, as well as the design of algorithms for revenue maximizing recommendation/choice sets (assortments) and other related objectives. Some works have pursued robust algorithms, such as the \adxopt~\citep{jagabathula2014assortment}, which was designed to be choice model agnostic. While others have tried to capture various business-driven preferences, such as: a constraint on the number of products that can be recommended, scalable optimization, precedence constraints among products etc.~\citep{rusmevichientong2010dynamic,sinha2017optimizing}. 
We refer the readers to \cite{kok2008assortment} for an overview of these optimization methods. In many datasets (see Section~\ref{sec:stats}), only a minority percentage of transactions are single purchases, supporting the need for multi-purchase modeling that better reflects reality.

In the marketing literature, multi-choice models have appeared in the context of modeling purchase of products in multiple categories~\citep{seetharaman2005models} as well as in bundle choice modeling. Two types of choice models have been predominant here - the multivariate probit (MVP) and variants of the multivariate logit (MVL). Both of these models are based on the \emph{random utility theory}. The earliest variation of MVL was proposed in \cite{cox1972analysis}  and has been used and improved upon in various subsequent works such as \cite{russell2000analysis} and \cite{singh2005modeling}. 
In bundle choice modeling~\citep{mccardle2007bundling}, consumer choice is modeled at the level of product bundles instead of the category of products. 
Both MVL variants~\citep{kopalle1999role}
and the MVP model have also been considered for this task. 
Recently, neural network based multi-choice models have also been proposed~\citep{yang2019examining}
\ to capture multi-purchase behavior. While they are able to fit observed data much better than the parametric models considered here, the resulting recommendation set optimization problems become quickly intractable due to lack of structure. Even with parametric models (such as ours), the optimization problems tend to be NP-hard, and in this paper, we devote significant effort to tame this complexity to make recommendation set computations scalable. 

A key prior work is by~\cite{benson2018discrete} who propose a MVL variant, which we refer to as the  Mixture Multi-Choice (\ucm) model. This model assumes that the mean utility of any subset of products is the sum of the utilities of each product in the subset and an optional correction term. By limiting the number of sets which receive a correction, this model can have a sparse parameterization. A crucial drawback of this model is that the random variables that affect the probability of choosing bundles with overlapping products are assumed independent. 



\subsection{Evidence of Multi-Purchase Behavior} \label{sec:stats}

Before introducing our model, we make some preliminary observations regarding multiple purchases as observed in various real world datasets (see Table~\ref{tab:datasummary} for a summary of their key characteristics).
Five datasets (Bakery, Walmart Items, Kosarak, Instacart, LastFM Genres) have fixed recommendation sets across all interactions, and one dataset (Yoochoose Items) has variable recommendation sets (see ~\cite{benson2018discrete} for their descriptions). The Ta Feng Grocery (\url{https://www.kaggle.com/chiranjivdas09/ta-feng-grocery-dataset}) and the UCI Online Retail (\url{https://archive.ics.uci.edu/ml/datasets/online+retail}) datasets additionally contain information about the revenues/prices, which will be useful for comparing prescriptive gains (Section~\ref{sec:experiments}).

\begin{table}
\footnotesize
\centering
\resizebox{.99\textwidth}{!}{
 \begin{tabular}{|c|c|c|c|c|c|c|c|c|} 
 \hline
 \textbf{Dataset} &  \textbf{Bakery} & \textbf{Walmart Items} & \textbf{Kosarak} & \textbf{Instacart} & \textbf{LastFM Genres} & \textbf{Yoochoose Items} &  \textbf{Ta Feng} & \textbf{UCI Online Retail} \\\hline
 
  \textbf{No. of products} & 50 & 1075 & 2621 & 5981 & 443 & 22915 & 3357 & 3350 \\ \hline
  \textbf{No. of unique purchased bundles} & 1267 & 3895  & 24921 & 100238 & 9866 & 41127 & 68597 & 7739\\ \hline
   \textbf{No. of observations} & 17171 & 25135 & 286399 & 298332 & 471638 & 208049 & 95001 & 11056\\ \hline
   \textbf{No. of unique recommended sets} & 1 &1 & 1 & 1 &1 &160711 & 1 & 1\\ \hline
 
  \end{tabular}
 }
\caption{List of datasets used for comparing predictive and prescriptive gains across different choice models.}
\label{tab:datasummary}
\end{table}

Our first observation is that a vast majority of purchases involve multiple products. Figure~\ref{fig:walmart_tran_hist} shows that the fraction of invoices that contain exactly one purchase is less than $22\%$ for the Walmart Items dataset. In other words, more than three out of every four transactions involve the purchase of multiple items. Similar trends are observed in all the other datasets, as shown in Table~\ref{tab:product-percentages}. Note that we have removed extremely infrequent products ($\sim 10\%$ on average) and have displayed transactions of size less than or equal to $8$ (for the full list of proportions, see \appTabProdPcts{} in \appAddExp).

\begin{table}[]
\centering
\resizebox{\textwidth}{!}{%
\begin{tabular}{|c|c|c|c|c|c|c|c|c|}
\hline
\textbf{Purchase bundle size} &
  \textbf{Bakery} &
  \textbf{Instacart} &
  \textbf{Kosarak} &
  \textbf{LastFM Genres} &
  \textbf{Walmart Items} &
  \textbf{Yoochoose Items} &
  \textbf{Ta Feng} &
  \textbf{UCI Online Retail} \\ \hline
1                & 4.8\%  & 4.9\% & 15.4\% & 37.8\% & 21.3\%  & 0.0\%  & 14.9\% & 22.6\% \\ \hline
2                & 18.1\% & 5.8\% & 20.0\% & 15.6\% & 17.8\%   & 2.0\%  & 12.7\% & 6.2\%  \\ \hline
3                & 32.9\% & 6.4\% & 17.6\% & 9.0\%  & 10.8\%  & 7.8\%  & 11.2\% & 4.2\%  \\ \hline
4                & 22.7\% & 6.9\% & 12.1\% & 6.1\%  & 8.1\%    & 13.7\% & 9.3\%  & 3.1\%  \\ \hline
5                & 11.5\% & 7.1\% & 7.4\%  & 4.5\%  & 6.1\%    & 22.4\% & 8.0\%  & 3.0\%  \\ \hline
6                & 5.1\%  & 7.1\% & 4.6\%  & 3.5\%  & 4.8\%    & 0.4\%  & 6.8\%  & 2.6\%  \\ \hline
7                & 2.9\%  & 6.8\% & 3.0\%  & 2.8\%  & 3.9\%    & 1.3\%  & 5.6\%  & 2.5\%  \\ \hline
8                & 2.0\%  & 6.3\% & 2.2\%  & 2.4\%  & 3.2\%    & 2.0\%  & 4.7\%  & 2.4\%  \\ \hline
\end{tabular}%
}
\caption{Summary of the fractions of product bundles purchased across datasets.}
\label{tab:product-percentages}
\end{table}

\begin{figure}
    \begin{subfigure}[b]{.25\linewidth}
        \centering
        \includegraphics[width=\linewidth]{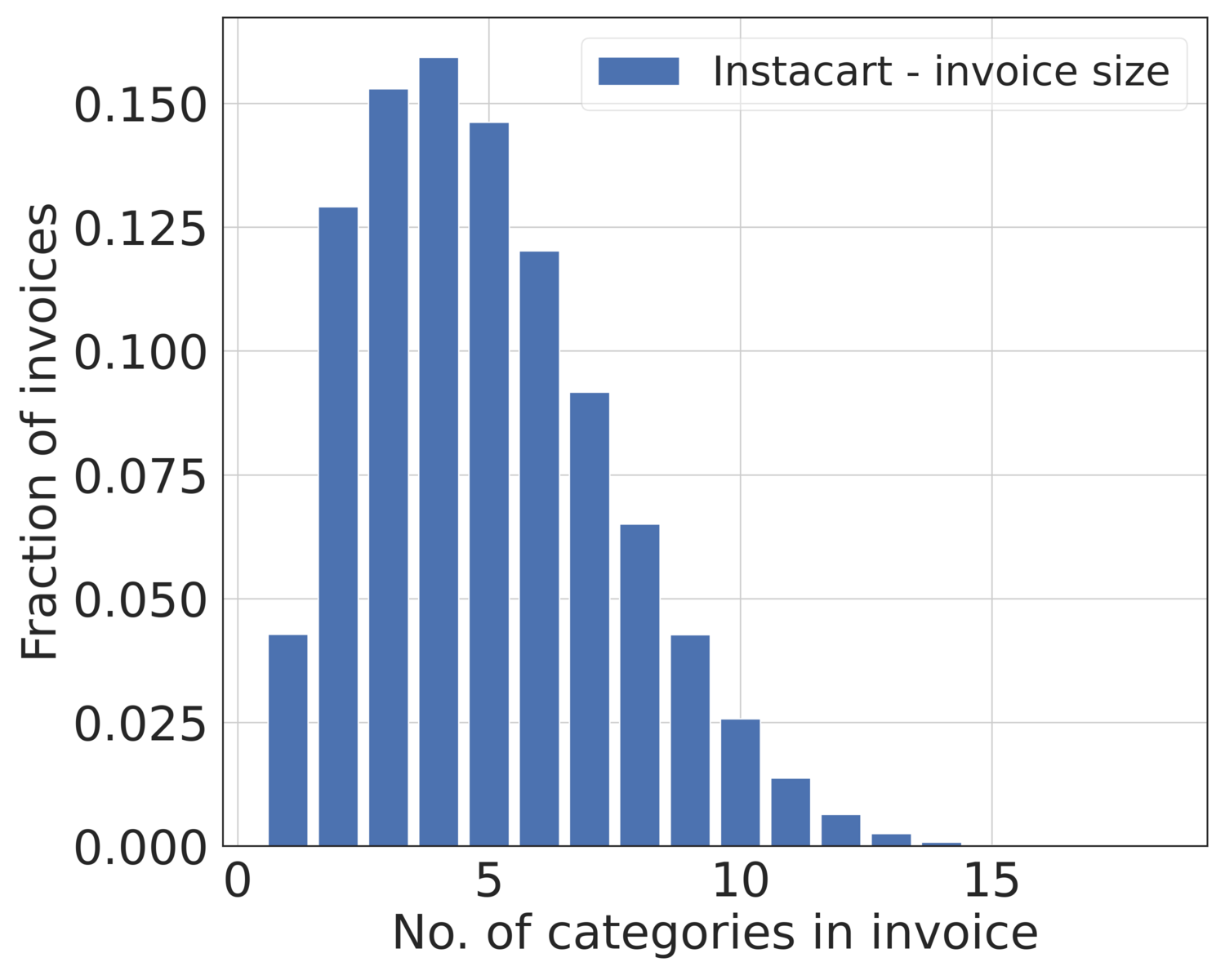}
        \caption{\label{fig:instacart_pdf_categories}} 
    \end{subfigure}\hfill
    \begin{subfigure}[b]{.25\linewidth}
        \centering
        \includegraphics[width=\linewidth]{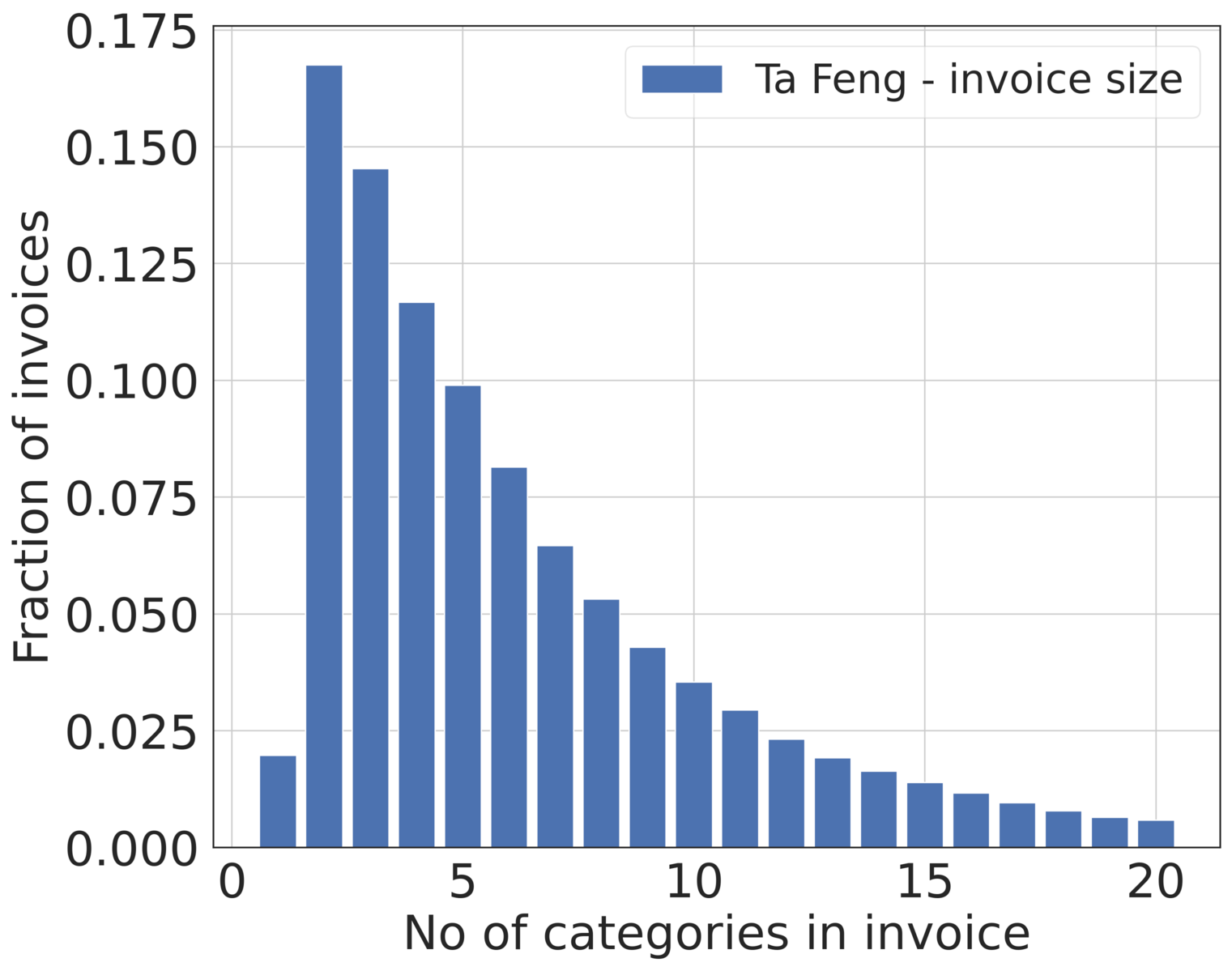}
        \caption{\label{fig:tafeng_pdf_categories}}
    \end{subfigure}\hfill
    \begin{subfigure}[b]{.25\linewidth}
        \centering
        \includegraphics[width=\linewidth]{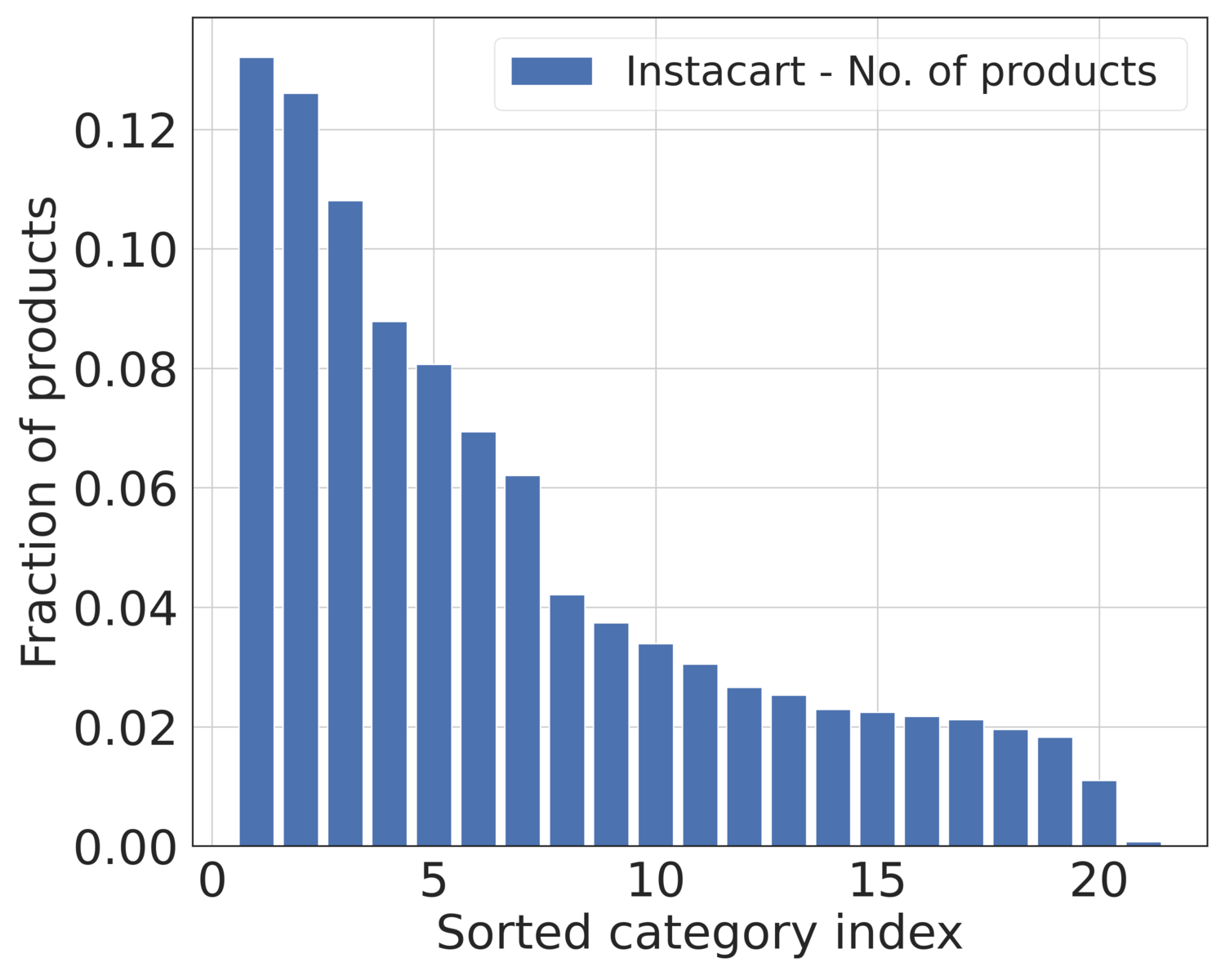}
        \caption{\label{fig:products_per_cat_instacart}}
    \end{subfigure}\hfill
    \begin{subfigure}[b]{.25\linewidth}
        \centering
        \includegraphics[width=\linewidth]{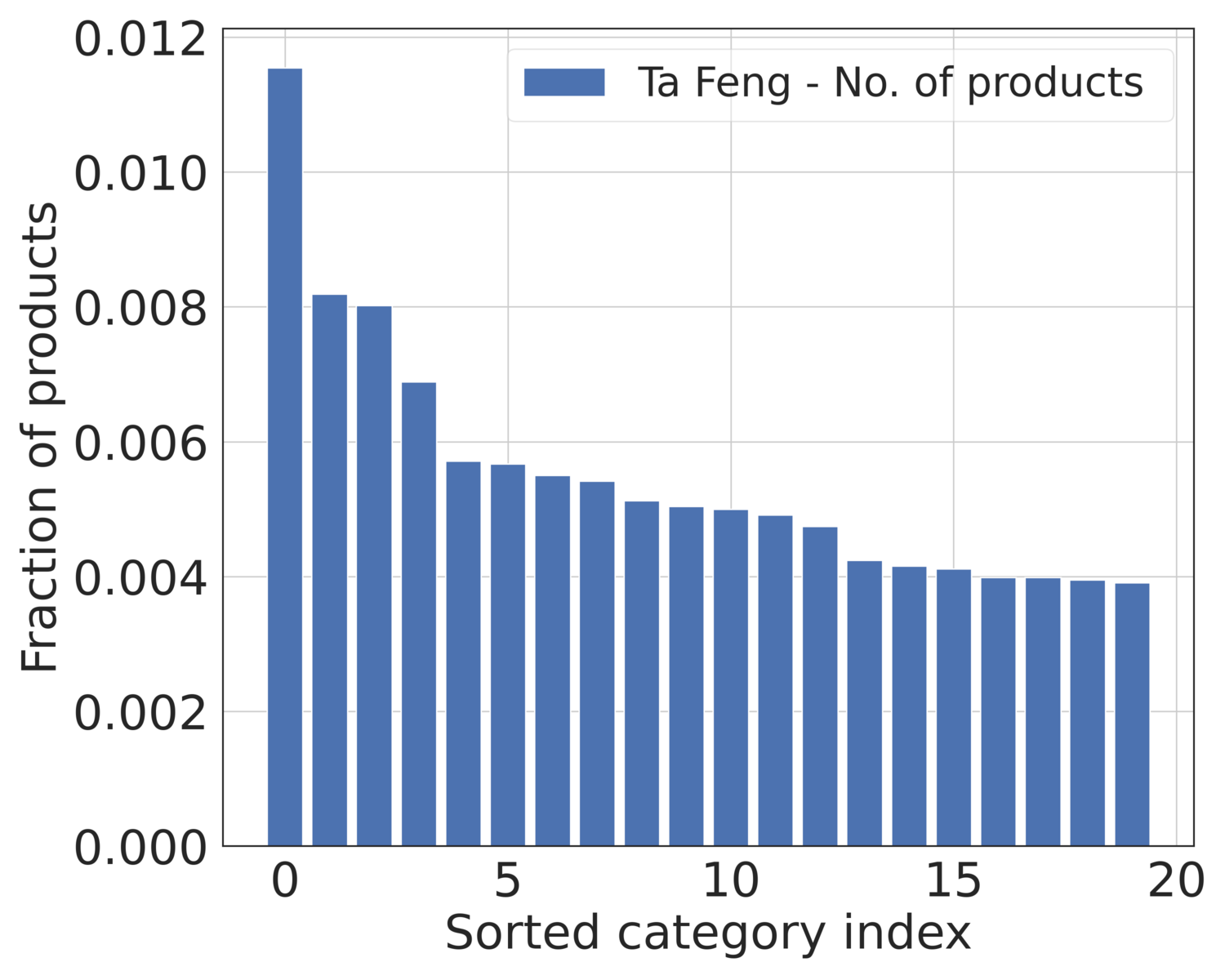}
        \caption{\label{fig:products_per_cat_tafeng}}
    \end{subfigure}\hfill
  \vfill
    \caption{Purchases typically involve multiple categories in a multi-purchase transaction.
    (\subref{fig:instacart_pdf_categories}): Histogram showing fraction of invoices that have a certain number of categories in the Instacart dataset.
    (\subref{fig:tafeng_pdf_categories}): Same as (\subref{fig:instacart_pdf_categories}) for the Ta Feng dataset.
    (\subref{fig:products_per_cat_instacart}): Histogram showing the proportion of unique products present in each of the top $20$ categories for the Instacart dataset.
    (\subref{fig:products_per_cat_tafeng}): Same as (\subref{fig:products_per_cat_instacart}) for the Ta Feng dataset.
    }
\end{figure}

Our second observation is that consumer purchases often tend to have products from multiple different categories in a given transaction. For instance, Figures \ref{fig:instacart_pdf_categories} and \ref{fig:tafeng_pdf_categories} show that the fraction of invoices that involve cross-category products are at least 95$\%$ of the multi-purchase transactions in two representative datasets. The proportion of products per category, which influence these fractions, are shown in Figures \ref{fig:products_per_cat_instacart} and \ref{fig:products_per_cat_tafeng}.

Single choice models cannot transparently capture these first-order patterns in the datasets discussed above. Thus, there is a need for developing suitable multi-choice models that can faithfully represent the observed empirical statistics above, and one such parsimonious attempt is discussed next. We will use the Instacart dataset as a case study to understand how this proposed multi-choice model captures data patterns in Section~\ref{sec:beta}. 

%% file: 2_model.tex
\section{The \bmvl~Choice Model For Multi-Purchase Behavior}
\label{sec:model}

The family of multi-purchase choice models that we formulate and study in this work, namely the \bmvlk{} family, has roots in the Marketing and the Spatial Statistics literature~\citep{russell2000analysis}. Models in this family describe the probability with which a customer purchases a bundle $\anychosenset$ (with $|\anychosenset| \leq K$) of unique products when the platform recommends a set $\offeredAsmt$ of products. Thus, the suffix K parameterizes these models by the maximum size of bundles that a customer can purchase (for instance, \rcm{} model captures purchases of bundles of size at most $2$). We start by specifying the conditional utility (a conditional random variable) of purchasing a product given the purchase decisions corresponding to all other products that were offered for the generic \bmvlk{} model as:
\begin{equation}
U(i | \{ X_j=x_j: j \in \offeredAsmt, j \neq i \} ) =  \left(\alpha_i + \sum_{j \in \offeredAsmt, j \neq i} \beta_{ij}x_j + \epsilon_i \right) \ind \left\{ \sum_{j \in \offeredAsmt, j \neq i} x_j < K  \right\},
\label{eqn:conditional-utility} 
\end{equation}
where $\alpha_i$ is a product specific parameter, parameters $\beta_{ij}$ capture interactions between product pairs $i$ and $j$, $\epsilon_i$ is a noise random variable distributed according to the Gumbel distribution, $\ind\{\}$ is the indicator function that evaluates to one (zero) if the inequality is true (false), and $X_j$ represent binary random variables that signify whether the customer purchased item $j$ or not ($x_j$ are the corresponding realizations) when $\offeredAsmt$ is offered. The $\beta_{ij}$ parameters are symmetric in the product pair, i.e., $\beta_{ij} = \beta_{ji}$. We can interpret from the above equation that the conditional utility of adding a product to a purchased bundle depends on its intrinsic value and its pairwise relationship with other purchases. For example, in grocery shopping, the utility of buying eggs for a customer who has already decided to buy flour and butter (which would let them bake a cake) as part of this bundle can be higher, as compared to buying eggs when they had decided to not purchase flour and butter. Further, even if the bundles are of size $K$, the effect on the utility by other products is pairwise, restricting the number of parameters to be of $O(\numberOfItems^2)$, where $\numberOfItems$ is the number of products.

The conditional probability of choosing $\anychosenset \subseteq \offeredAsmt$ given the above purchase behavior is:
\begin{align}
P(\anychosenset|\offeredAsmt) = \frac{  \exp{ \left( \sum_{i \in \anychosenset } \alpha_i +  \sum_{i, j \in \anychosenset  , i <j} \beta_{ij} \right)  }   }{v_0 + \sum_{ \anychosenset' \subseteq C, | \anychosenset'| \leq K} \exp{ \left( \sum_{i \in \anychosenset' } \alpha_i +  \sum_{i, j \in \anychosenset'  , i <j} \beta_{ij} \right)  }  },
\label{eqn:choice-prob-main}
\end{align}
where $v_0$ is a parameter corresponding to the no-purchase probability, and is introduced to be consistent with similar parameters appearing in several single choice models. This expression (see \appModelDerivation{} for additional details) lends itself to a high degree of interpretability (positive $\beta_{ij}$ implying complementary and the opposite implying substitutable relationships), and intuitively suggests that bundles with large positive values of the intrinsic parameters ($\alpha_i$s) and large positive pairwise affinity values ($\beta_{ij}$s) will have a higher probability of being purchased. We next show how this model can capture multi-purchase patterns using an example dataset, before relating it to extant models in the literature (Section~\ref{sec:bmvl-additional}) and developing optimization techniques for recommendation sets based on it (Section~\ref{sec:alg-rec-set}).

\subsection{Model-driven Insights: Case Study with the Instacart Dataset}
\label{sec:beta}

We now show how the estimated parameters of \bmvl-2 can be used to further understand purchase patterns using the Instacart dataset. We focus on products that belong to one of the following nine categories: frozen, bakery, produce, beverages, pantry, breakfast, dairy eggs, snacks, and deli (the few remaining categories are not interpretable). For computational reasons, we also filter to retain transactions of size $10$ or lower.

At the outset, we see that the two observations made in Section~\ref{sec:stats} are partially reflected by the estimated model, and the remaining gap can be attributed to \bmvl-2 not modeling more than 2 purchases in a transaction. Assuming that the no purchase probability is $0$ (since this is not observed in the data), the probability of a single purchase ($\sum_{i \in C} P(\{i\}|C)$) computed using \bmvl-2 is $\sim 55\%$, compared to an approximate estimate of $\sim 46\%$ from data (the ratio of single purchases versus at most two purchases as seen in Table~\ref{tab:product-percentages}, while not precise, gives us a rough handle of this). This is in contrast to a single choice model such as the MNL or the MMNL, which would simply be $100\%$ (since we assumed that the probability of no purchase is $0$). Similarly, the probability of having products from different categories, given that they were purchased ($\sum_{i,j \in C} P(\{i,j\}|C,\textrm{category}(i) \neq \textrm{category}(j))/\sum_{i,j \in C} P(\{i,j\}|C)$) computed using \bmvl-2 is $\sim83\%$, compared to $\sim95\%$ observed in Figure~\ref{fig:instacart_pdf_categories}. In contrast, the estimate using the MNL model, which breaks down multiple purchases as independent single purchases, is $\sim73\%$. Thus, \bmvl-2 better captures the empirical statistics of this dataset.

The $\beta_{ij}$ parameter in \bmvl-2 for a pair of products $i$ and $j$ can indicate the additional utility gained from buying these products as a bundle, which naturally lends itself to identifying pairs of products that are substitutes and complements — where purchasing one item makes the second item less or more attractive, respectively. In particular, the parameter $\beta_{ij}$ can be negative or positive. A negative value is indicative of the fact that the two products in question have negative additional utility due to being purchased together compared to being purchased separately on separate occasions. Similarly, a positive value is indicative of an additional positive utility. Also, some products may have low individual utility (a low $\alpha_i$) while some others may not, and the interaction parameter and the individual parameters together determine the purchase probability of bundles. 

Table~\ref{tab:bottom_top_10_beta} shows the composition of category pairs among the highest and lowest valued $\beta_{ij}$ pairs, respectively. Similarly, Table~\ref{tab:freq_beta_pct} shows which categories are prevalent at the highest and lowest percentiles of $\beta_{ij}$ values. Both these results reveal that the model is capturing natural complementary categories. It's important to note that the number of products in a category can impact the results since more products increase the likelihood of a $\beta_{ij}$ value being present. 

Note that if a product pair is never observed, then the corresponding $\beta_{ij}$ is $-\inf$, or equivalently, the probability of choosing the pair is $0$. Thus, of the $\beta_{ij}$s that were estimated (i.e., at least one transaction with the product pairs was observed), our model also shows that more than $95\%$ of them are positive, and this is true at each category level. This indicates a non-negligible tendency for products to be purchased in bundles. In other words, when product pairs are observed to be purchased together in the data, our model more often than not assigns a positive $\beta_{ij}$ value to them. However, there are approximately $0-5\%$ of these pairs where the $\beta_{ij}$ values are negative, as shown in the Table~\ref{tab:instacart_beta_percent_coop}. These negative $\beta_{ij}$ values indicate that the observed frequency of these products being purchased individually is higher than the frequency of them being purchased together (substitution). Finally, about $55-90\%$ of product pairs have $\beta_{ij}= -\inf$, which means these pairs are essentially unrelated to each other, and focusing on them for marketing/revenue management opportunities may not be effective. On the other hand, identifying the product pairs with positive and negative $\beta_{ij}$s can help with operational insights (such as improving the recommended sets shown, see Section~\ref{sec:alg-rec-set}).

Table~\ref{tab:instacart_beta_pct} shows that between $11-40\%$ of \emph{all possible} product pairs have a positive $\beta_{ij}$, implying a higher likelihood of being selected together compared to being selected independently. This means that customers tend to prefer buying these pairs as a bundle. For instance, the [Produce, Dairy Eggs] category pairs have $2.5$ times more co-purchased product pairs than the [Beverages, Breakfast] category pairs in the data, which approximately mirrors the proportion of positive $\beta_{ij}$s between these two. These findings demonstrate that \bmvl is able to partially capture co-purchases that could not be easily identified with single choice models. The nontrivial fraction of product pairs with positive beta values provides valuable insights for developing more effective recommendation set design (see Section~\ref{sec:alg-rec-set}), marketing and pricing strategies.

\begin{table}
\scriptsize
\centering
\begin{tabular}{|c|c||c|c|}
\hline
\textbf{Bottom Category Pairs} & \textbf{Percentage} & \textbf{Top Category Pairs} & \textbf{Percentage} \\ \hline
Produce, Dairy Eggs     & 14.74  &  Snacks, Dairy Eggs      & 6.87          \\ \hline
Produce, Beverages      & 12.07  &  Beverages, Dairy Eggs   & 6.57              \\ \hline
Produce, Snacks         & 8.85    & Beverages, Snacks       & 5.71                \\ \hline
Dairy Eggs, Snacks      & 7.81   & Frozen, Dairy Eggs      & 5.62              \\ \hline
Frozen, Produce         & 7.17      & Frozen, Snacks          & 5.42          \\ \hline
\end{tabular}
\caption{Composition of category pairs among the bottom and top $10\%$ of $\beta_{ij}$ values}
\label{tab:bottom_top_10_beta}
\end{table}

\begin{table}
\scriptsize
\begin{minipage}{.45\linewidth}
\centering
    \begin{tabular}{ |c|c|c|  }
     \hline
     \textbf{Category} & \textbf{Within-Category} & \textbf{Cross-Category} \\
     \hline
      Frozen & 0.21\% & 1.42\% \\
     \hline
      Bakery  & 0.65\%& 1.51\% \\
     \hline
     Produce & \textbf{3.74\%} &  \textbf{4.01\%}\\
     \hline
     Beverages & 1.34\% &  2.57\% \\
     \hline
      Pantry  & 0.17\% & \textbf{1.20\%} \\
     \hline
     Breakfast & \textbf{0.15\%}   & 1.24\% \\
     \hline
     Dairy Eggs & 2.28\% & 2.14\% \\
     \hline
     Snacks & 0.17\% &  1.43\%\\
     \hline
    Deli & 0.43\% & 1.42\% \\
    \hline
    \end{tabular}
    \captionof{table}{Percentage of estimated $\beta_{ij}$s that are negative within and across categories}
    \label{tab:instacart_beta_percent_coop}
\end{minipage} 
\hfill
\scriptsize
\begin{minipage}{.45\linewidth}
\centering
    \begin{tabular}{|c|c|c|}
    \hline
    \textbf{Category}  & \textbf{high} $\beta_{ij}$ & \textbf{low} $\beta_{ij}$ \\ \hline
    Frozen     & 13.13              & 8.57              \\ \hline
    Bakery     & 8.00               & 4.77              \\ \hline
    Produce    & 8.37               & \textbf{28.52}    \\ \hline
    Beverages  & 13.60              & 15.26             \\ \hline
    Pantry     & 10.91              & 6.73              \\ \hline
    Breakfast  & \textbf{5.61}      & \textbf{2.53}     \\ \hline
    Dairy Eggs & \textbf{17.51}     & 20.01             \\ \hline
    Snacks     & 16.82              & 9.80              \\ \hline
    Deli       & 6.05               & 3.81              \\ \hline
    \end{tabular}
    \captionof{table}{Frequency of each category among the top $10\%$ highest and lowest $\beta_{ij}$ values.}
    \label{tab:freq_beta_pct}
\end{minipage}
\end{table}

\begin{table}
\scriptsize
\centering
\begin{tabular}{|c|c|c|c|c|c|c|c|c|c|}
\hline
 &
  \textbf{Frozen} &
  \textbf{Bakery} &
  \textbf{Produce} &
  \textbf{Beverages} &
  \textbf{Pantry} &
  \textbf{Breakfast} &
  \textbf{Dairy Eggs} &
  \textbf{Snacks} &
  \textbf{Deli} \\ \hline
\textbf{Frozen}     & \textbf{0.14} & 0.21          & \textbf{0.33} & 0.21          & 0.20          & 0.19          & 0.26          & 0.19          & 0.23          \\ \hline
\textbf{Bakery}     & 0.21          & \textbf{0.11} & \textbf{0.35} & 0.20          & 0.21          & 0.21          & 0.28          & 0.19          & 0.24          \\ \hline
\textbf{Produce}    & 0.33          & 0.35          & \textbf{0.27} & 0.32          & 0.37          & 0.31          & \textbf{0.40} & 0.30          & 0.35          \\ \hline
\textbf{Beverages}  & 0.21          & 0.20          & \textbf{0.32} & \textbf{0.13} & 0.20          & 0.20          & 0.25          & 0.20          & 0.21          \\ \hline
\textbf{Pantry}     & 0.20          & 0.21          & \textbf{0.37} & 0.20          & \textbf{0.14} & 0.19          & 0.27          & 0.18          & 0.20          \\ \hline
\textbf{Breakfast}  & 0.19          & 0.21          & \textbf{0.31} & 0.20          & 0.19          & \textbf{0.15} & 0.27          & 0.20          & 0.19          \\ \hline
\textbf{Dairy Eggs} & 0.26          & 0.28          & \textbf{0.40} & 0.25          & 0.27          & 0.27          & \textbf{0.17} & 0.24          & 0.27          \\ \hline
\textbf{Snacks}     & 0.19          & 0.19          & \textbf{0.30} & 0.20          & 0.18          & 0.20          & 0.24          & \textbf{0.12} & 0.21          \\ \hline
\textbf{Deli}       & 0.23          & 0.24          & \textbf{0.35} & 0.21          & 0.20          & 0.19          & 0.27          & 0.21          & \textbf{0.14} \\ \hline
\end{tabular}
\caption{Percentage of positive $\beta_{ij}$s within and across categories}
\label{tab:instacart_beta_pct}
\end{table}

\begin{figure}
    \begin{subfigure}[b]{0.5\linewidth}
        \centering
        \includegraphics[width=\linewidth]{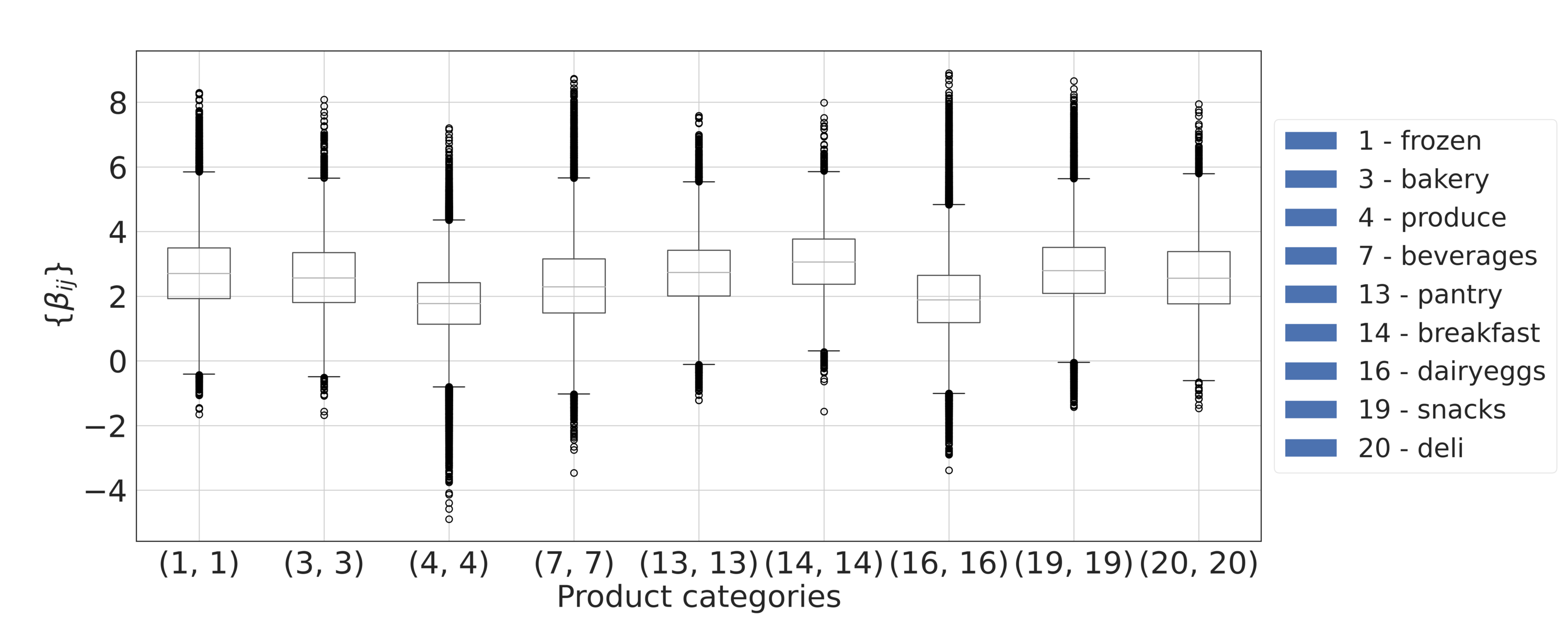}
        \caption{\label{fig:beta_within}}
    \end{subfigure}\hfill
     \begin{subfigure}[b]{.5\linewidth}
        \centering
        \includegraphics[width=\linewidth]{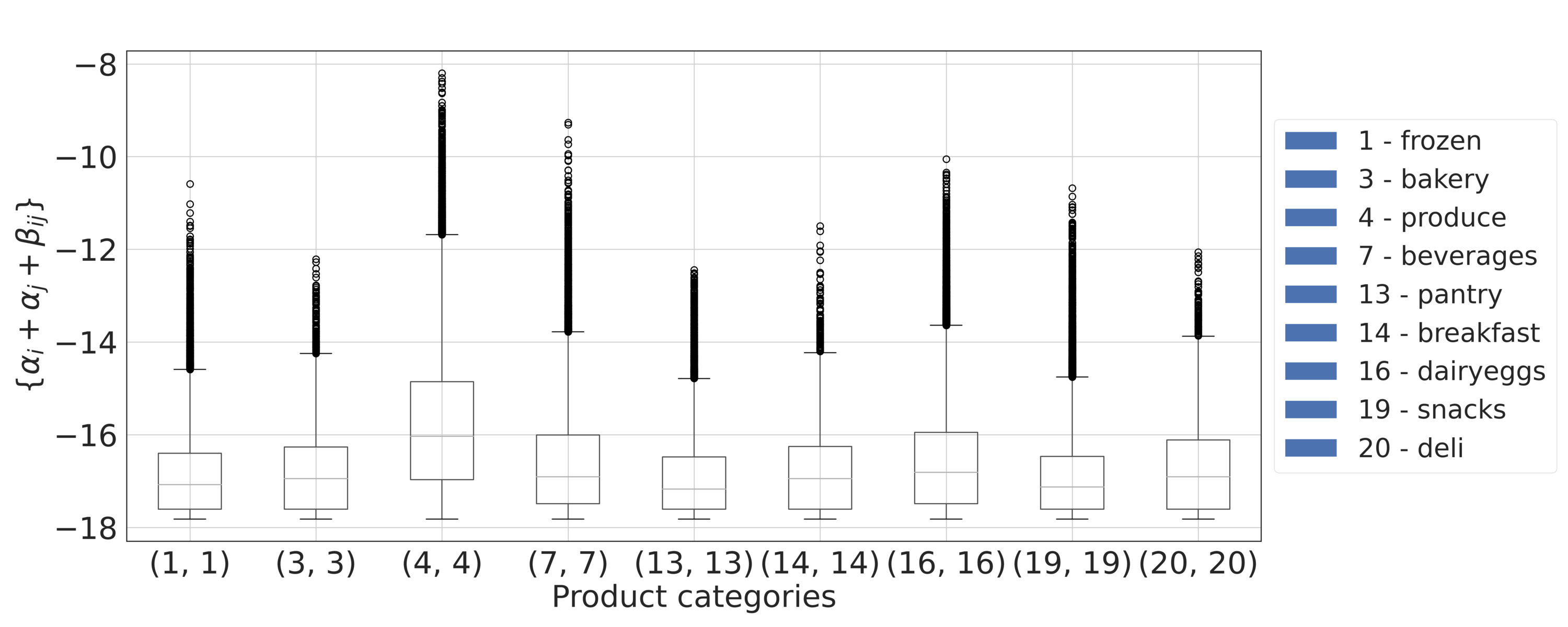}
        \caption{\label{fig:sum_within}}
    \end{subfigure}\vfill
    \begin{subfigure}[b]{.5\linewidth}
        \centering
        \includegraphics[width=\linewidth]{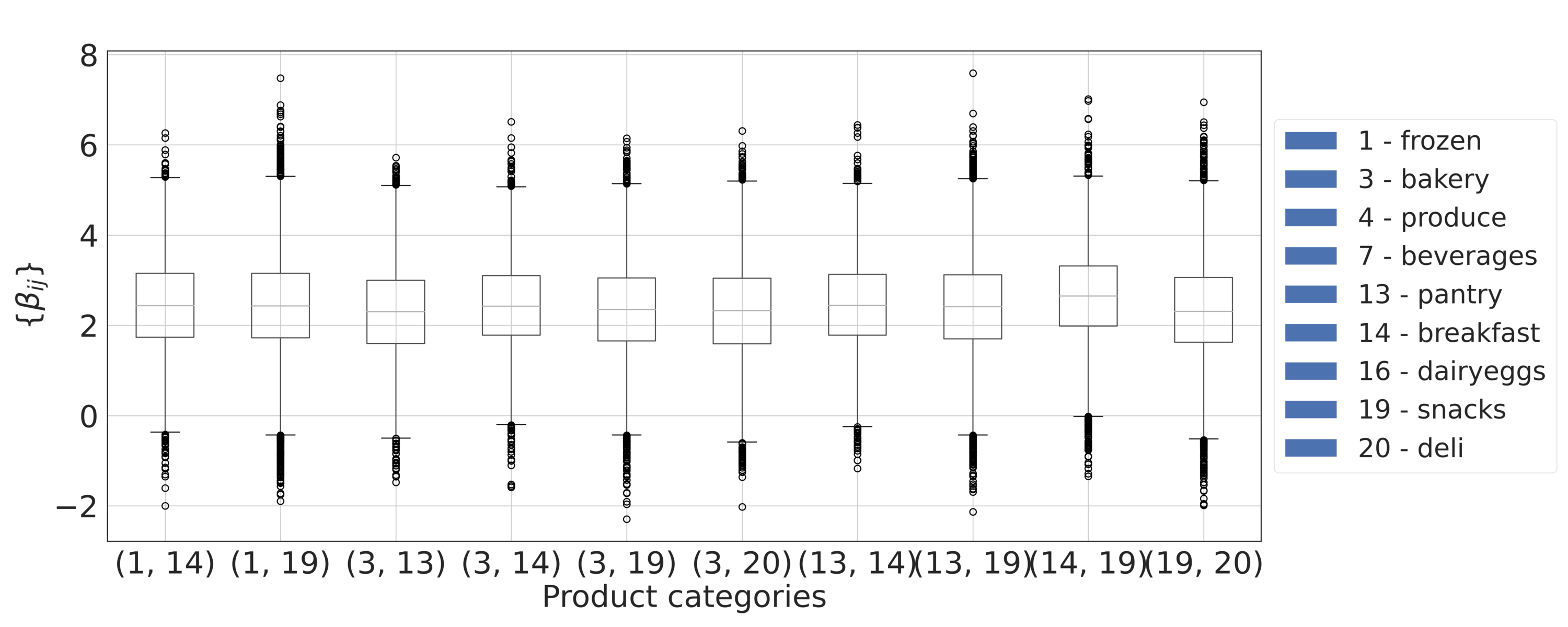}
        \caption{\label{fig:beta_across_positive}}
    \end{subfigure}\hfill
    \begin{subfigure}[b]{.5\linewidth}
        \centering
        \includegraphics[width=\linewidth]{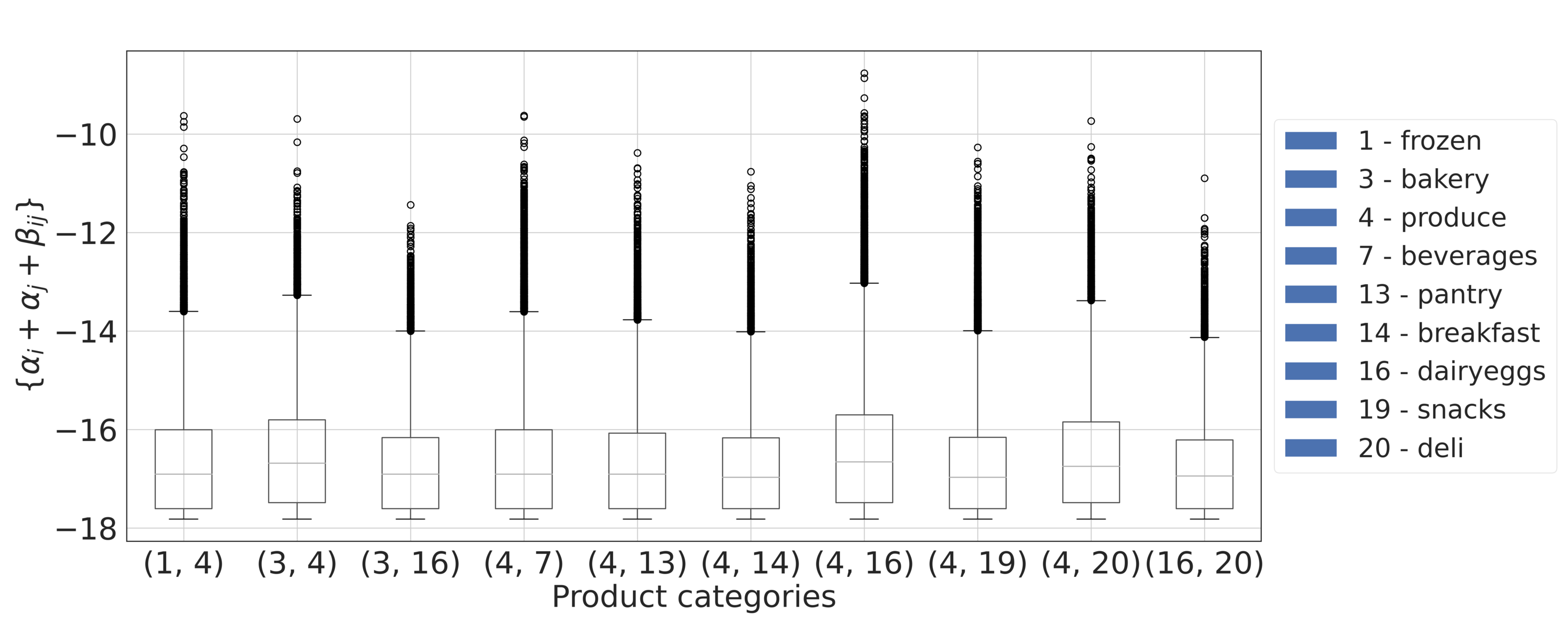}
        \caption{\label{fig:sum_across_positive}}
    \end{subfigure}\vfill
    \begin{subfigure}[b]{.5\linewidth}
        \centering
        \includegraphics[width=\linewidth]{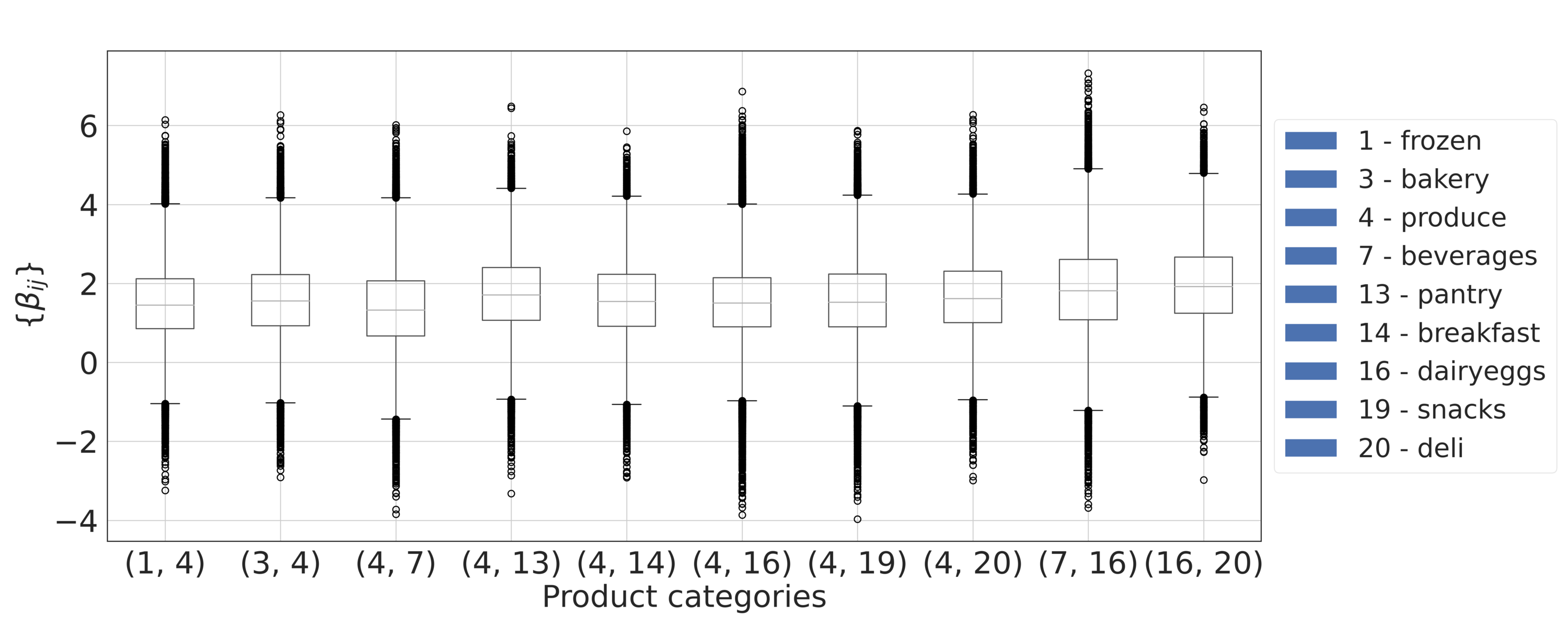}
        \caption{\label{fig:beta_across_negative}}
    \end{subfigure}\hfill
    \begin{subfigure}[b]{.5\linewidth}
        \centering
        \includegraphics[width=\linewidth]{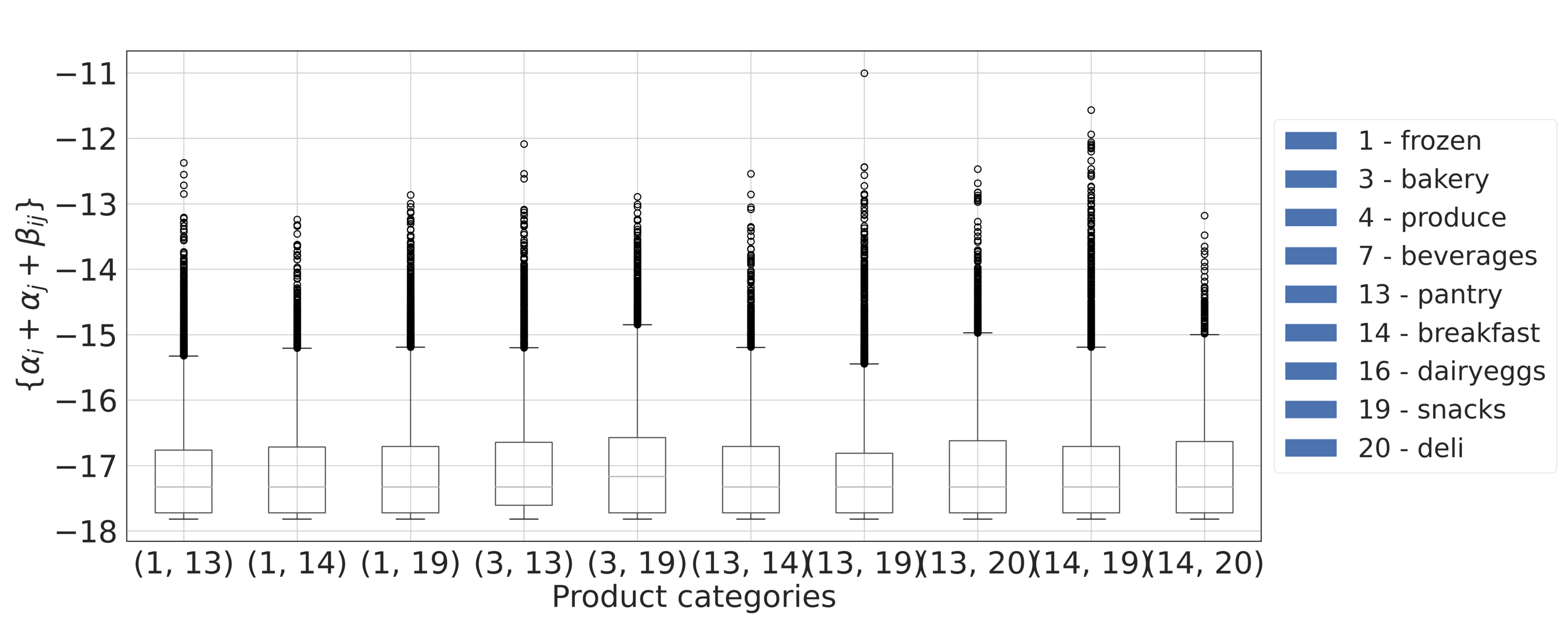}
        \caption{\label{fig:sum_across_negative}}
    \end{subfigure}\hfill
	\vfill
    \caption{ Estimation plots.
    (\subref{fig:beta_within}): $\{\beta_{ij}\}$ values for product pairs belonging to the same categories.
    (\subref{fig:sum_within}): $\{\alpha_i+\alpha_j+\beta_{ij}\}$ values for product pairs belonging to the same categories.
    (\subref{fig:beta_across_positive}): $\{\beta_{ij}\}$ values for select category pairs whose cross-category product pairs have high values.
    (\subref{fig:sum_across_positive}): $\{\alpha_i+\alpha_j+\beta_{ij}\}$ values for select category pairs whose cross-category product pairs have high values.
    (\subref{fig:beta_across_negative}):  $\{\beta_{ij}\}$ values for select category pairs whose cross-category product pairs have low values.
    (\subref{fig:sum_across_negative}):  $\{\alpha_i+\alpha_j+\beta_{ij}\}$ values for select category pairs whose cross-category product pairs have low values.
    }
\end{figure}

Figure~\ref{fig:beta_within} shows the distribution of $\beta_{ij}$ values via box plots for product pairs within the same category. We observe that purchases made within the \textit{produce} category have lower $\beta_{ij}$ values than others. This pattern can be attributed to the substitution type behavior, where consumers purchase say organic or regular items, but perhaps not both in any given transaction. For example, the product pairs (a) \textit{Organic Avocado} and \textit{Sweet Kale Salad Mix}, and (b) \textit{Clementines} and \textit{Organic Red Onion} have negative $\beta_{ij}$ values of $-4.58$ and $-4.13$ reflecting the substitution type behavior. Similarly, we observe that products from the \emph{snacks} category have higher $\beta_{ij}$ values than others. This pattern can be attributed to the complementary behavior where consumers purchase items in the \emph{snacks} category in combinations. For example, the product pairs (a) \textit{Chipotle Beef \& Pork Realstick} and \textit{Uncured Cracked Pepper Beef}, and (b) \textit{Double Chocolate Chip Protein Bar} and \textit{Protein Bar Chocolate Peanut Butter} have positive $\beta_{ij}$ values of $8.41$ and $8.22$, reflecting the complementary behavior. For a complete view, we also plot Figure~\ref{fig:sum_within}, which shows the distribution of $\{\alpha_i+\alpha_j+\beta_{ij}\}$ values, i.e., takes the intrinsic utilities into account.

Analogous to the above, Figures~\ref{fig:beta_across_positive} and ~\ref{fig:sum_across_positive} show the distribution of $\beta_{ij}$s and $\{\alpha_i+\alpha_j+\beta_{ij}\}$ values for category pairs whose product pairs have high values. In Figure~\ref{fig:beta_across_positive}, we observe cross-category product pairs that include the \textit{snacks} category have higher average $\beta_{ij}$ values than others, indicating a strong positive relationship between this category and others. But when the intrinsic parameters ($\alpha_i$s) are also considered as in Figure~\ref{fig:sum_across_positive}, cross-category product pairs that include the \textit{produce} category have higher total expected utilities in general. Figures~\ref{fig:beta_across_negative} and ~\ref{fig:sum_across_negative} show similar distributions for the bottom-most category pairs, and here we observe that cross-category product pairs that include the \textit{produce} category have lower $\beta_{ij}$s in general (and those that include \textit{pantry} have lower total expected utilities in general). This implies that even though products in the \emph{produce} category have weaker interactions with others, their intrinsic parameters ($\alpha_i$s) are high. Observations such as these and other similar ones can be readily deduced from the estimated \bmvl-2 model, and can help a decision maker make nuanced operational decisions (e.g., about the recommendation sets).

To summarize, using the Instacart dataset we have illustrated how the \bmvl-K family of models, specifically \bmvl-2, aligns well with real world data, and provides interpretable insights into product associations. It is evident that downstream benefits come from modeling these non-negligible proportion of pairwise positive/negative interactions when products are put together into a candidate recommendation set. In the Instacart dataset in particular, we observe that there is a non-negligible chance that the probability of a product pair getting purchased together is higher compared to purchasing them independently, and optimizing recommendation sets that take this key customer behavior into account can likely yield more expected revenue when compared to using models that don't have multi-purchase modeling capabilities.

\subsection{\bmvl-K: Key Properties, Estimation and Benchmarks}
\label{sec:bmvl-additional}

\noindent\textbf{Key Properties.} Given Equations~\ref{eqn:conditional-utility} and~\ref{eqn:choice-prob-main}, we make the following observations. First, we note that the model is not derived based on assigning a mean utility to every bundle, adding a random noise term and deriving purchase probabilities based on the utility maximizing bundle. Instead, the model assumes a form on the conditional utility of purchase of a product given the purchase decision on all other products and derives the only joint distribution consistent with this conditional probability distribution.

Second, while precursors to the \bmvlk \ family exist in the literature, our key contribution here is two fold: (a) we extend these precursors to the modeling of multiple products in addition to categories (a category is a collection of similar products), and (b) we are able to restrict the number of products purchased to a parameterized value $K$, and still obtain the \emph{softmax} (or sigmoidal) structure of the probability of purchase expression above. This restriction parametrized by $K$ and the softmax structure are both very appealing from an optimization perspective (see Section~\ref{sec:alg-rec-set}). For instance, the MNL single-choice model also has a similar form, which allows for the expected revenue maximizing set computation in polynomial time.  We achieve these two properties without making a restrictive assumption that is made for many other random utility models in the literature~\citep{benson2018discrete}: they assume independent Gumbel perturbations to the utilities of bundles, even if these bundles share products between them.

Third, with increasing values of $K$, we can model customers that purchase larger bundles (crucially, $K=1$ gives us back the popular MNL model). The model parameterization $(\alpha,\beta)$ remains the same for $K\geq 2$, although the data likelihood changes. The choice of $K$ is driven by the suitability of the model to data, and the tractability of the optimization problems (maximizing expected revenue or probability of purchase) downstream. For example, if most customers buy less than say $10$ products, although choosing $K$ as $10$ in the \bmvlk \ model suffices (see Figure~\ref{fig:walmart_tran_hist}), it may still be prudent to trade-off model fit with the computational tractability of optimization (by choosing a smaller value of $K$). As we show in Section~\ref{sec:experiments}, the choice of $K=2$ strikes a great balance for many real world datasets in terms of fit and in terms of scalability. In particular, we show that on one hand the revenue gains achieved by using the \rcm \ model over the MNL and MMNL models can be large, especially as the number of products grows. On the other hand, optimizing over \rcm{} is much more empirically tractable (more revenue gains in a small fraction of time) when compared to \bmvl-3.


\noindent\textbf{Estimation.} Any \bmvlk~ model can be estimated directly using maximum likelihood estimation (MLE). Let $\{S_l,\anyasmt_l\}_{l=1}^{m}$ be the dataset that potentially includes no purchase observations and $\max_{l}|S_l| \leq K$. Then the likelihood of observing the given data is:
\begin{equation}
\prob_{data} = \prod_{l=1}^m \prob( S_l | C_l \text{ was offered})
= \prod_{l=1}^m \left(\frac{  \exp{ \left( \sum_{i \in \anychosenset_l } \alpha_i +  \sum_{i, j \in \anychosenset_l  , i <j} \beta_{ij} \right)  }   }{v_0 + \sum_{ \anychosenset' \subseteq C_l, | \anychosenset'| \leq K} \exp{ \left( \sum_{i \in \anychosenset' } \alpha_i +  \sum_{i, j \in \anychosenset'  , i <j} \beta_{ij} \right)  }  }\right),
\end{equation}
which can be maximized numerically, using say first order methods, to get estimates of $\alpha_i$s and $\beta_{ij}$s. The estimation problem becomes easy when $K=2$ because one can optimize directly over $V_{\anychosenset} :=  \exp{ \left( \sum_{i \in \anychosenset } \alpha_i +  \sum_{i, j \in \anychosenset  , i <j} \beta_{ij} \right)  }$. 

\begin{lemma}\label{lemma:estimate}
Given values $V_\anychosenset$ for all bundles $\anychosenset$ of size $ \leq 2$, we can uniquely obtain the parameters $(\alpha, \beta)$ by solving the system of equations: $\log V_S = \sum_{i \in S} \alpha_i + \sum_{i, j \in S, j >i} \beta_{ij}$ for all $S$.
\end{lemma}

\textcolor{black}{The proof follows by showing that the linear transformation specified above between $\log V_S$ and $\alpha_i$s and $\beta_{ij}$s is invertible.}

Further, if the data is such that the same recommendation set is shown to consumers in each observation, then $V_{\anychosenset}$ can be estimated by simply counting and normalizing, which is a linear time computation. Because of the equivalence above, we will interchangeably use $V_\anychosenset$ as parameters as well when working with \rcm. 
\textcolor{black}{For estimating the \bmvlk\ model from data with transactions containing bundles of size greater than $K$, we propose splitting such transactions into multiple transactions each of which has a bundle of size at most $K$. This transformation allows computing the likelihood of the original transactions, and enables direct likelihood comparisons across models. The detailed procedure is described in \appEstimation.  }

\textcolor{black}{ If information about product features is also available, then the $\alpha$ and $\beta$ parameters can depend on these features as well. The mapping from such features to $\alpha, \beta$ values can be constrained by dimensionality/degrees-of-freedom so as to make it amenable to interpretability (such as being able to reason about what product features makes products complementary/substitutable). }

\noindent\textbf{Benchmark Models.} In addition to the MNL, we briefly discuss two models that are used for empirical comparisons later, namely the (multi-choice) MMC and the (single-choice) MMNL models.

\noindent\textit{The MMC model:} The Mixture Multi-Choice (MMC) model (\cite{benson2018discrete}) can be thought of as a two-stage mixture model. In the first step, the customer chooses the size of the bundle of products they are going to purchase. The size of this bundle is $m$ with probability proportional to $z_m$ for $1 \leq m \leq |C|$ where $C$ is the recommendation set. In the second step, the customer chooses an $m$-sized bundle $S$ by assigning random utilities $  \log V_S + \epsilon_{S} = \sum_{i \in S}\hat{\alpha}_i + \hat{\beta}_{S} + \epsilon_{S}$, where parameters $\hat{\alpha}_i$ represent intrinsic utility of product $i$, \ $\hat{\beta}_{S}$ represents additional utility from buying the products together as a bundle $S$, and $\epsilon_{S}$'s are i.i.d. Gumbel distributed random variables associated with every bundle $S$. The parameters $\hat{\beta}_{S}$ take non-zero values only for a small number of bundles $S$, leading to a sparse representation. To incorporate the possibility of not making any purchase in the second stage, the model can be extended to consider a utility of not making any purchase as $\log V_0^m + \epsilon_{\phi}^m$. Thus, the probability of choosing a bundle $S$ of size $m$  when a recommendation set $C$ is shown is given by $\frac{z_m}{z_1 + \cdots z_m} \frac{V_S}{ V_0^m + \sum_{S' \subset C s.t. |S'|=m} V_{S'} }$. The specific structure of the underlying generative process (i.e., the two stages described above) leads to computational intractability of optimization.

\noindent\textit{The MMNL model:}
The Mixture of MNLs (MMNL) model \citep{mendez2014branch} is a parametric extension of the MNL model. Its discrete variant (also referred to as latent class MNL or LC-MNL) is a flexible model that can be interpreted as a weighted average MNL with $K$ classes/consumer-types described by class proportions $\mu_1,.....,\mu_K$. The probability that product $j$ is chosen when recommendation set $C$ is offered is given by $ P(j|C) = \sum_{k=1}^{K} \mu_k \frac{\exp(v_{jk})}{\sum_{i \in C \cup \{0\}}\exp(v_{ik})} $
where $v_{jk}$ is the nominal utility that a consumer in segment $k$ gives to product $j$, and where $v_{0k}$ is the no purchase utility for class $k$. The class proportions satisfy: $\mu_k \geq 0, \sum_{k=1}^{K} \mu_k = 1$. Similar to the MNL, the probability of choosing a bundle $S$ of size $m$ when $C$ is shown is given by $ \prod_{j \in S} P(j|C)$.

\noindent\textbf{Estimation of Benchmark Models.} Similar to \bmvl-2, we attempt to fit the benchmark models based on maximizing their respective maximum likelihood (MLE) objectives, and when possible, use stochastic gradient descent with reasonable defaults for this. For MMNL, we end up using an improved conjugate gradient method~\citep{berbeglia2022comparative}. In the \ucm~model, the MLE maximization needs to be preceded by the calculation of the optimal sets which should receive corrections, which is an NP-Hard problem (see Proposition 3.2 in \cite{benson2018discrete}).

%% file: 3_algos.tex
\section{Algorithms for Revenue Maximization}
\label{sec:alg-rec-set}

\noindent\textbf{The Objective.} Maximizing revenue from a set of offered products is one of the key goals of retailers. 
Under the \bmvlk~ model,  the expected revenue of the platform if it offers recommendations $\offeredAsmt$ is given by $\revk(\offeredAsmt) = \frac{\sum_{S \subseteq \offeredAsmt, |S| \leq K} V_{S}r_{S}}{ v_0 + \sum_{S \subseteq \offeredAsmt, |S| \leq K} V_{S}}$, where $r_S = \sum_{i \in S} r_i$, and $r_i>0$ is the fixed and known revenue corresponding to product $i \in \prodUniverse$ and $\prodUniverse = \{ 1, 2, \cdots \numberOfItems \}$ is the set of all products. We assume that the products are ordered such that $r_1 \geq r_2 \cdots \geq r_\numberOfItems$.  As mentioned earlier, we will primarily focus on the \rcm{} model, whose expected revenue function can be written more explicitly as:
\begin{equation}
    \revtwo(\offeredAsmt) = \frac{ \sum_{i \in \offeredAsmt} r_i V_{\{i\}}  +  \sum_{i \in \offeredAsmt} \sum_{j \in \offeredAsmt, j >i } (r_i + r_j)V_{\{i,j\}}  }{v_0 +   \sum_{i \in \offeredAsmt} V_{\{i\}} + \sum_{i \in \offeredAsmt} \sum_{j \in \offeredAsmt , j >i } V_{\{i,j\}}} = \frac{ \sum_{i \in \prodUniverse} \sum_{j \in \prodUniverse} \hat{r}_{ij} \theta_{ij} x^{\offeredAsmt}_i x^{\offeredAsmt}_j  }{ v_0 + \sum_{i \in \prodUniverse} \sum_{j \in \prodUniverse}  \theta_{ij} x^{\offeredAsmt}_i x^{\offeredAsmt}_j},
\end{equation}
where $x^{\offeredAsmt}_i =1$  if $i \in \offeredAsmt$ else $0$,
$ \theta_{ij} =  \begin{cases} \frac{V_{\{ij\}}}{2} &  i \neq j \\
V_{ \{i \} } & i =j
\end{cases},  $ and $
 \hat{r}_{ij} =  \begin{cases} r_i + r_j &  i \neq j \\
r_i & i =j
\end{cases}
$.  Our objective is to find a recommendation set $\offeredAsmt^*$  that has the maximum expected revenue among all feasible recommendation sets (represented using collection $\feasibleAsmtSet$):
\begin{align} 
\label{eqn:AOproblem}
\max_{\anyasmt \in \feasibleAsmtSet} \revtwo(\anyasmt).
\end{align}
\vspace{-0.7cm}

\begin{theorem}
\label{thm:bmvl_hard} [Hardness Result for \rcm]
Under the \rcm \ model, the decision version of the unconstrained revenue optimization problem (\ref{eqn:AOproblem}) is NP-complete.
\end{theorem}

\noindent\textbf{A Noisy Binary Search Based Approach.} In the absence of polynomial time exact algorithms for the optimization problem (\ref{eqn:AOproblem}), we develop exact algorithms without any run-time guarantees, as well as create heuristic polynomial time algorithms, with improvements based on novel structural properties. These properties allow one to decide if a product is part of the optimal recommendation set in constant time, reducing the size of the instance that needs to be solved computationally (see \appStructural). Our algorithmic approach is based on a \emph{binary search outer loop}, and focuses on improving the computational speed of the comparison steps by \emph{exploiting the structure of our multi-purchase choice model} as well as by using scalable heuristics. In particular, the computationally expensive comparison step checks if the optimal revenue is greater than the specified threshold $\compthresh_j$. A key insight is that this calculation can be transformed as follows:
\begin{align} 
\label{eqn:comparison}
     \max_{\anyasmt \in \feasibleAsmtSet}  R(\anyasmt) & \geq \compthresh_j
\Longleftrightarrow   \max_{\anyasmt \in \feasibleAsmtSet} \sum_{i \in \prodUniverse} \sum_{j \in \prodUniverse} \theta_{ij}x_i^\anyasmt x_j^\anyasmt (\hat{r}_{ij} - \compthresh_j)  \geq \compthresh_j v_0. 
\end{align}
The optimization problem on the left hand side of the transformed comparison is a quadratic integer program (QIP) and is a special case of the well known Quadratic Unconstrained Binary Optimization (QUBO) problem that is known to be NP-hard~\citep{pardalos1992complexity}. We run multiple QUBO heuristics in parallel, and also use a noisy binary search variant~\citep{burnashev1974interval} for the outer loop to robustify our approach. Further, we are able to capture constraints on recommendation sets that lead to linear inequalities (e.g., capacity constraints) parsimoniously by solving modified QUBO instances (see \appLinearConstr). Note that the structural properties of the optimal solution do not hold for the constrained setting, and we will use a noisy binary search outer loop without any enhancements.

In each iteration of the proposed noisy binary search process, called \noisybinsearcheff, we prevent incorrect comparison step outcomes from misleading the search process and narrow the size of the interval in which $R(\offeredAsmt^*)$ lies using the aforementioned structural properties of the optimal solution. \appAlgFull{} provides additional details regarding \noisybinsearcheff, along with related algorithms, viz., \binsearch~(vanilla binary search, see \appAlgBS), \binsearcheff~(vanilla binary search with structural properties, see \appAlgBSE) and \noisybinsearch~(noisy binary search). The iterative nature of our approach helps in terminating the search for the optimal recommendations gracefully, especially under stringent timing requirements expected in applications. The binary search template can also be used to solve other \bmvlk{} models (the \textsc{Compare-Step} will be different), and Section \ref{sec:experiments} documents its use with the \tcm~model, for which we formulate mixed non-linear integer programs for the comparison steps, and solve them using \texttt{Bonmin} from COIN-OR~\citep{bonmin}.

\noindent\textbf{Benchmark Algorithms for \bmvl-2.} We also consider three benchmark methods that can be used in lieu of \noisybinsearcheff. We discuss the unconstrained setting first. The first benchmark is a mixed integer program (MIP) that builds on an earlier formulation for the MMNL model studied in~\cite{blanchet2016markov} (see \appAlgMIP{} in \appAlgFull) and produces an optimal solution.  The second benchmark is a generalization of the \adxopt~ algorithm~\cite{jagabathula2014assortment}, which we call \adxoptk.
This is a choice model agnostic greedy algorithm which in every iteration looks for a set of $L$ products whose addition/deletion/exchange will lead to recommendations with a higher revenue. Our third benchmark is a heuristic algorithm which chooses the revenue-ordered recommendation set with the highest revenue, and has a time complexity of $O(\numberOfItems^3)$. This heuristic is known to be optimal or close to optimal for several single-choice models~\citep{rusmevichientong2010assortment}. For the two heuristic benchmarks, we provide technical results that relate their solutions to an optimal solution in \appMainBenchmarks. For the constrained setting (specifically capacity constraints), the \adxopt~ and \adxopttwo~ heuristics can be modified so that in every greedy step, the best subset of products to add is chosen while ensuring that the capacity constraint is obeyed. Similarly, for the revenue-ordered heuristic, we can choose the best among all revenue-ordered sets of size less than the given capacity. Finally, the MIP can incorporate constraints in a straightforward manner.

\noindent\textbf{Benchmark Algorithms for Benchmark Models.} Finally, we note that to solve for the best/nearly optimal recommendation sets under the competing models, we use: (a) a revenue ordered approach (for unconstrained settings) and a linear programming formulation (for linearly constrained settings) for the MNL, (b) a computationally fast branch-and-cut algorithm~\citep{mendez2014branch} for both unconstrained/constrained settings for the MMNL (the problem is known to be NP-hard), and (c) a mixed integer program/MIP for both unconstrained/constrained settings for the MMC (see \appObj{} for the hardness result, the MIP is omitted due to space limitations).

%% file: 4_experiments.tex
\section{Experiments}
\label{sec:experiments}

We perform two sets of experiments. The goal in the first set is to validate the merits of the \rcm~model compared to other models on real data based on the empirical fit and the expected revenue obtained. In the second set, we benchmark the solution quality and computational times of the optimization approaches for \bmvl-2 (Section~\ref{sec:alg-rec-set} and \appAlgFull) in detail, to ascertain their suitability in the online recommendation setting. 

All results reported here are based on $50$ Monte Carlo runs for each configuration, unless otherwise noted. For computation times and the optimality gap metrics, we have plotted the median along with the $25^{\textrm{th}}$ and $75^{\textrm{th}}$ percentiles. We segment the number of products  $\numberOfItems$ into three regimes while discussing scalability trends, viz., (a) Small: $20-80$, (b) Medium: $100-400$, and (c) Large: $500-1500$ products. Some additional runs/variations of these experiments are documented in \appAddExp.

\subsection{\rcm~Model versus Other Single/Multi-Purchase Models}

We perform three sub-experiments using real world datasets: (a) we assess the fit of the \rcm~model versus others, (b) we assess the revenue gains achieved compared to other models, and (c) we assess the run-time gains across models.

\noindent\textbf{Empirical Fit.} We compare the \rcm~model with the \ucm{}, the MNL and the MMNL models using the datasets discussed in Section~\ref{sec:stats}. When estimating the \bmvl-2~and the \ucm~models, we transform all observations into ones that involve at most $K$-sized bundle purchases as described in \appEstimation. Similarly, we transform all observations into single purchases for the single choice models.

All datasets are split in the ratio $80:20$, with the former used for learning the parameters, and the latter for reporting out-of-sample log-likelihood fit. As these datasets do not contain information about customers leaving without making any purchase, it is not possible to estimate $v_0$. Consequently, we rely on domain knowledge to pick an appropriate value in our experiments, although we note that this has been observed to vary wildly from one application to another. While the \rcm{} and MNL have no tunable parameters, we need to pick the number of \emph{corrections} for the \ucm{} model and the number of mixture components for MMNL. For \ucm~increasing the number of corrections increases the flexibility of this model, but also increases the number of parameters. While adding such corrections, we pick bundles (namely the \textit{H-sets}) in descending order of their frequency of appearance in the datasets~\citep{benson2018discrete}. 

The fit of the models are tabulated for the Walmart Items dataset in Table \ref{tab:walmartdatasetresults} and for an additional five datasets in \appTabEstDataset{} in \appAddExp. The number of corrections in the \ucm~model have been reported as a percentage (between 0\% to 100\%) of the number of unique subsets of purchases of size greater than one that are observed. The number of estimated parameters are also reported for each model. The number of parameters in \rcm \ differs from \ucm \ with $100\%$ corrections because of the way these are estimated, which leads to different numbers of parameters getting estimated in each case. From these two tables, we infer that the \rcm~model provides the best fit  (i.e., the highest log-likelihood value) in all six real world datasets, when compared to the MNL, MMNL and the \ucm~models. The gap between the \rcm~and \ucm~models (the only two multi-purchase models among the four) when compared to the single-choice MNL and MMNL models is quite large (up to $1.5-2\times$ lower log-likelihoods for the latter in some cases), validating the necessity for multi-purchase choice modeling. In summary, for all six real world datasets, the \rcm~model is shown to be a viable and strong alternative to competitors (e.g., the \ucm~model) in capturing the rich multi-purchase behavior exhibited.

\begin{table}
\scriptsize
\centering
\begin{tabular}{ |c||c|c|c|  }
 \hline
 \textbf{Dataset} & \multicolumn{3}{c||}{\textbf{Walmart Items Dataset}} \\
 \hline
 \textbf{Model}& \textbf{Number of Parameters} & \textbf{Train Log-likelihood} & \textbf{Test Log-likelihood} \\
 \hline
 MNL    & 1075 & -160299 & -40526 \\
 \hline
 MMNL (5 components)   & 5380 & -158568 & -40039 \\
 \hline
 \ucm~(0 corrections)&   1075  & -168250   & -42494 \\
 \hline
 \ucm~(1\% corrections) & 1098 & -142794 &  -35985 \\
 \hline
 \ucm~(5\% corrections)    & 1194 & -138402 &  -35017 \\
 \hline
 \ucm~(20\% corrections)&   1551  & -135825 & -34871 \\
 \hline
 \ucm~(50\% corrections)& 2265 & -131954   & -35018 \\
 \hline
 \ucm~(100\% corrections)& 3456 & -125531 & -48916 \\
 \hline
 \rcm~& 3384 & -125410 & \textbf{-26518}\\
 \hline
\end{tabular}%
\caption{Log-likelihood values under different models for the Walmart Items dataset.}
\label{tab:walmartdatasetresults}
\end{table}

\noindent\textbf{Revenue Gains.} Even though the \rcm~model provides a good fit on real data, one of the primary goals of using such choice models in practice is to realize higher expected revenues. To start, we estimate parameters of the competing models using the Ta Feng and the UCI datasets because these also have price/revenue information. Next, we randomly sample without replacement a subset (of pre-specified sizes) of products, and define a choice model instance using their estimated parameter values.  As there is no information about customer interactions that did not lead to a sale in these datasets, we fix the probability of no-purchase when all products are displayed to 30\% (changing this probability to much higher and lower values while experimenting did not qualitatively change our conclusions), and estimate the $v_0$ (or equivalent) parameter accordingly for each instance. When doing comparisons, we also study the trends as a function of the model instance size (i.e., the number of products $\numberOfItems$).

\begin{figure}
    \begin{subfigure}[b]{.3\linewidth}
        \centering
        \includegraphics[width=\linewidth]{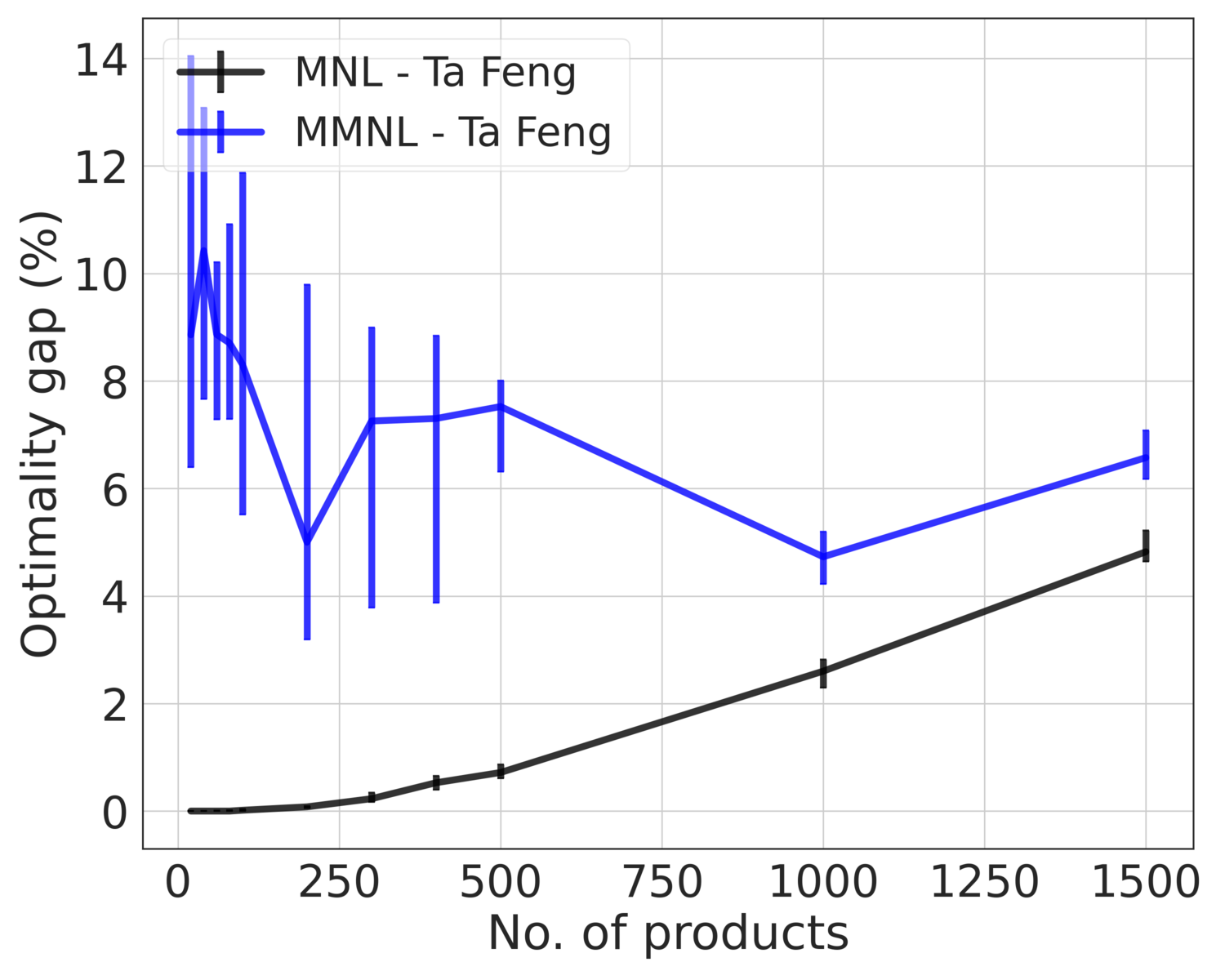}
        \caption{\label{fig:mmnl-gap-tafeng}}
    \end{subfigure}\hfill
    \begin{subfigure}[b]{.3\linewidth}
        \centering
        \includegraphics[width=\linewidth]{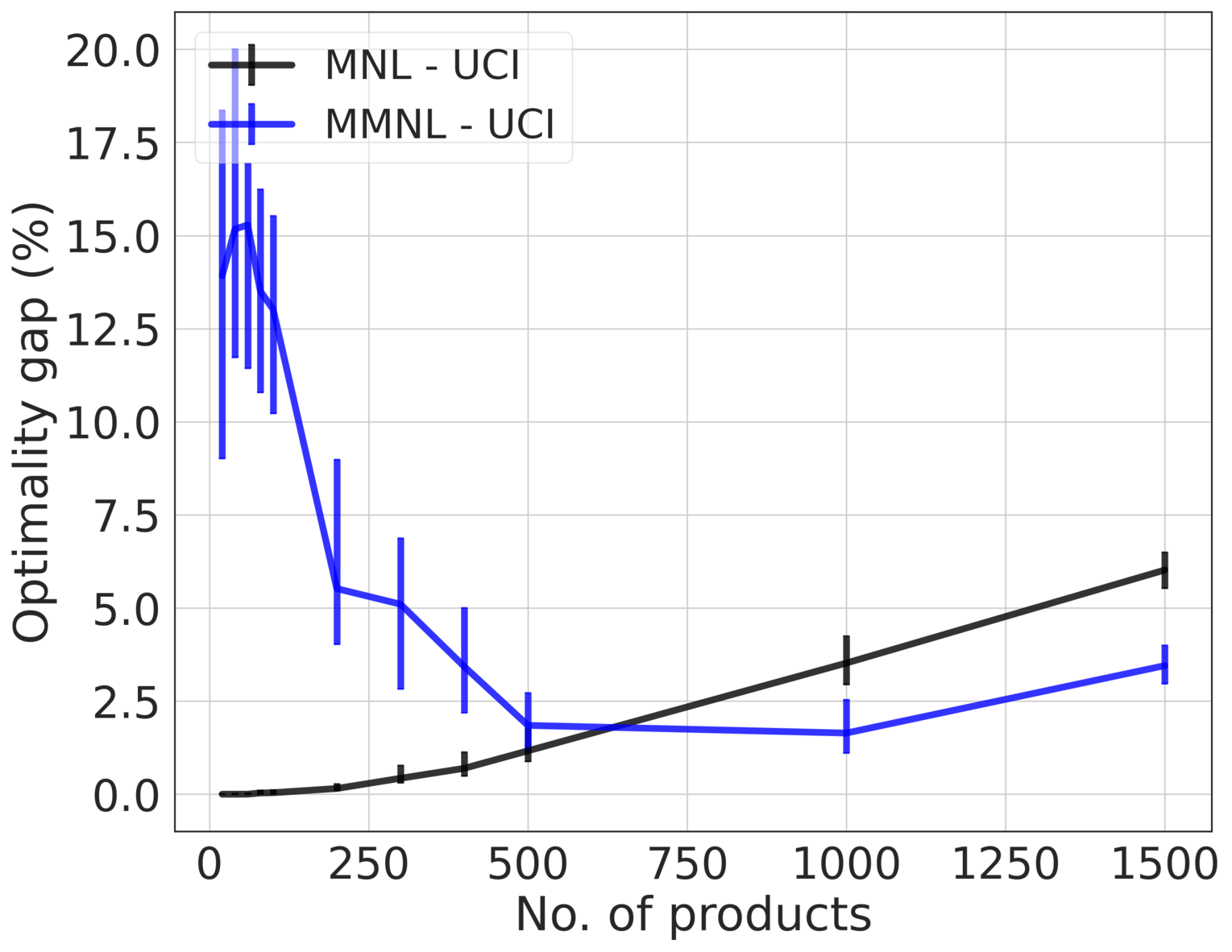}
        \caption{\label{fig:mmnl-gap-uci}}
    \end{subfigure}\hfill
    \begin{subfigure}[b]{.3\linewidth}
        \centering
        \includegraphics[width=\linewidth]{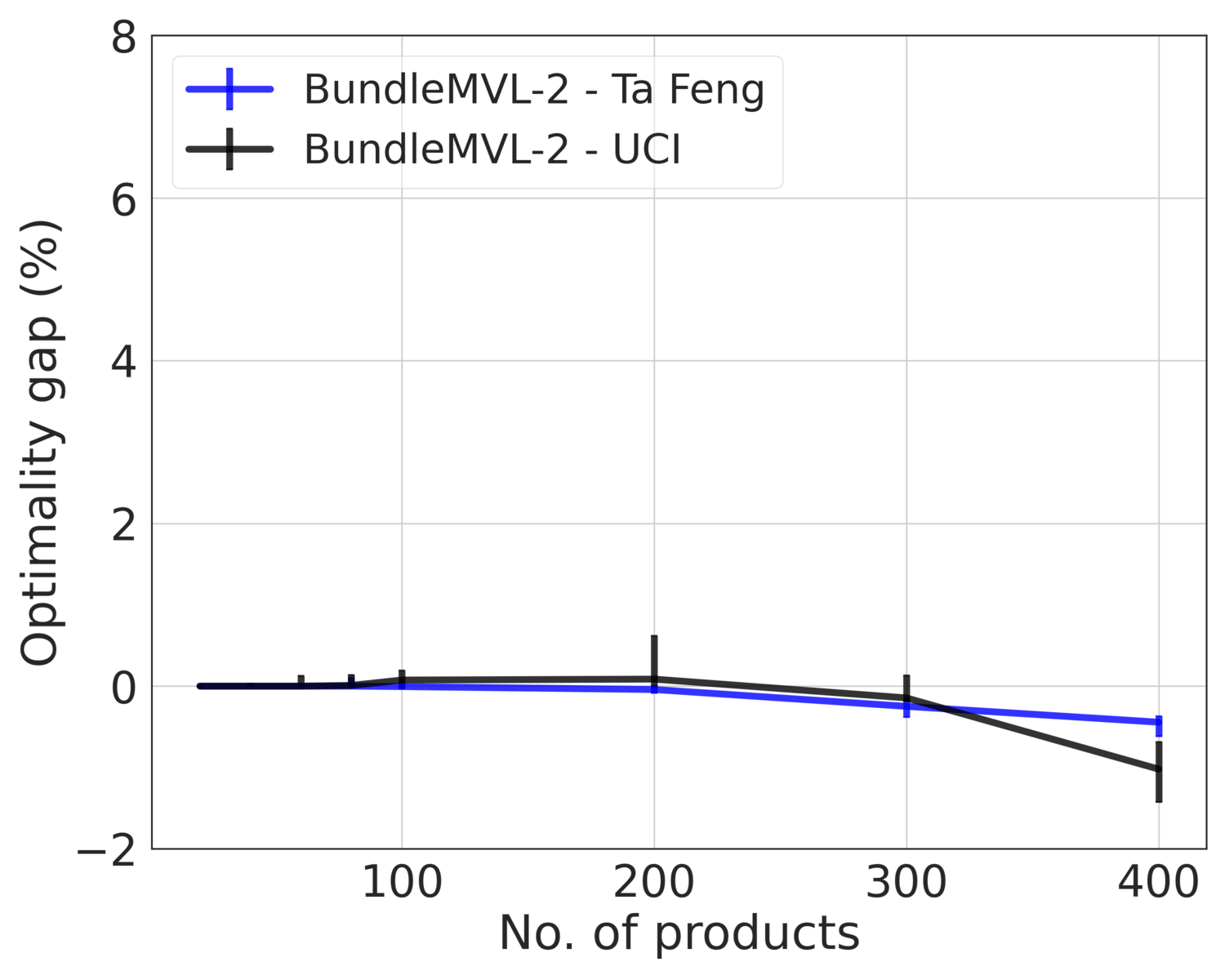}
         \caption{\label{fig:e5tcmrcmanalysis1}}
    \end{subfigure}\hfill
    \begin{subfigure}[b]{.3\linewidth}
        \centering
        \includegraphics[width=\linewidth]{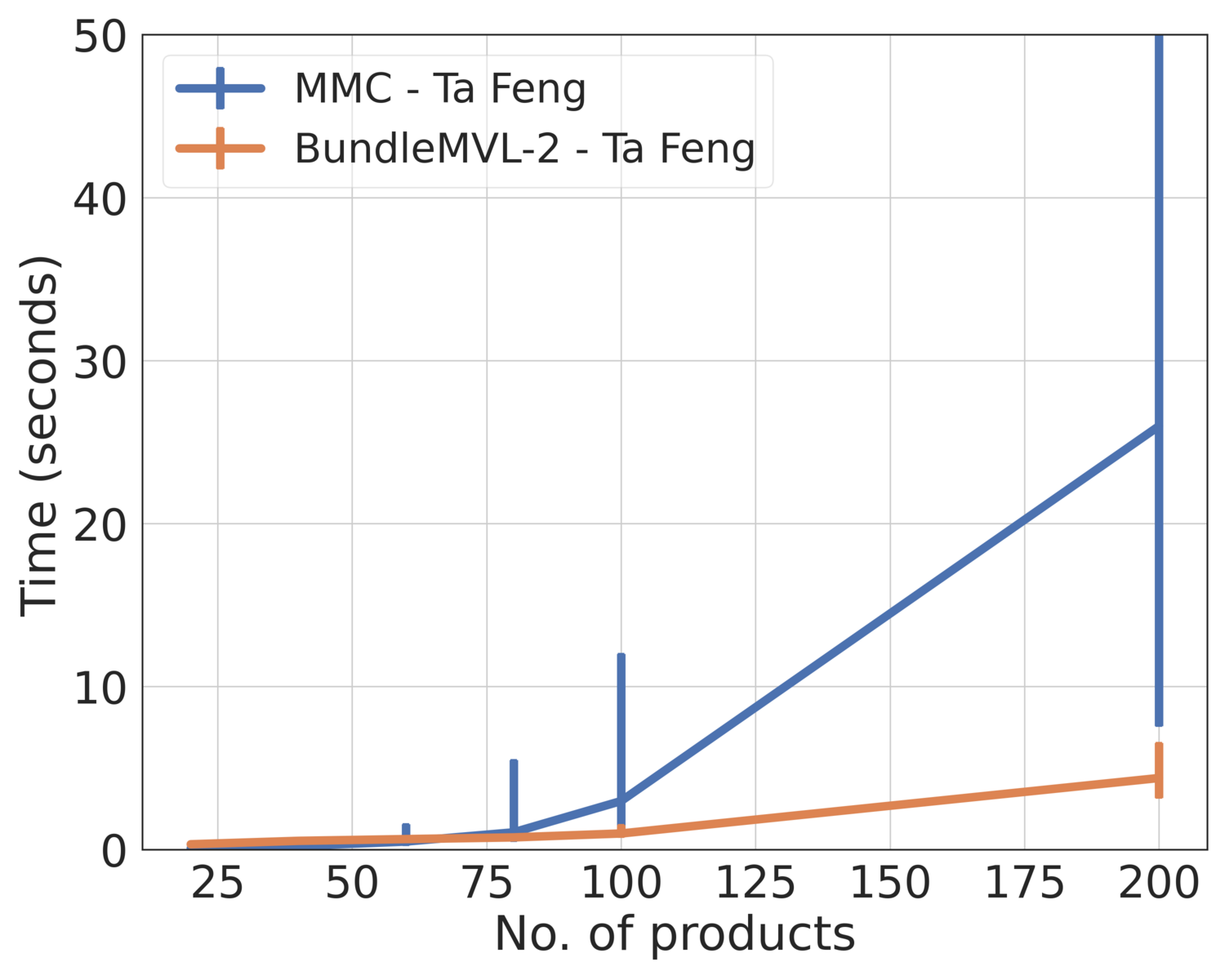}
        \caption{\label{fig:rcm_ucm_time_comparison}}
    \end{subfigure}\hfill
    \begin{subfigure}[b]{.3\linewidth}
        \centering
        \includegraphics[width=\linewidth]{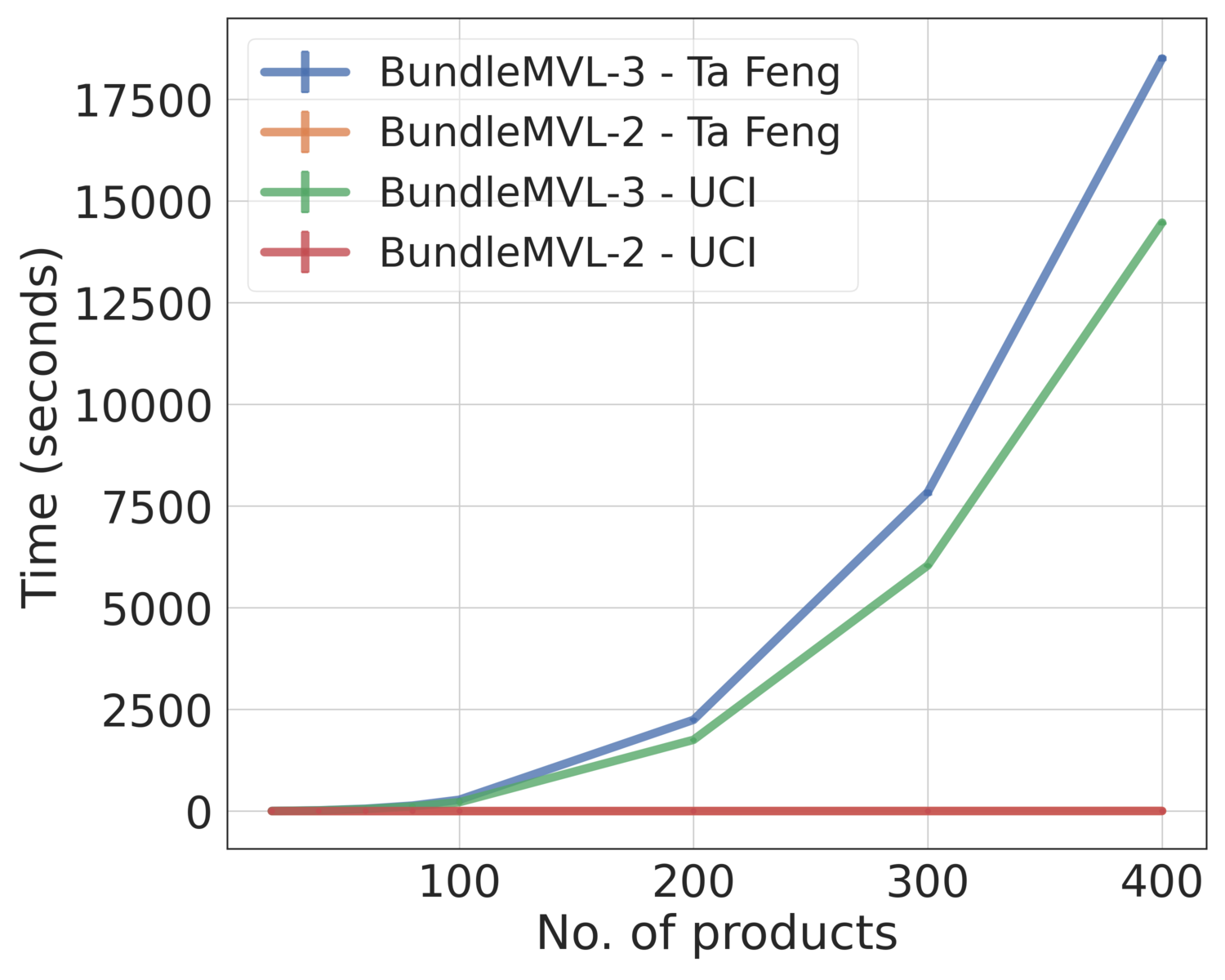}
        \caption{\label{fig:e5tcmrcmanalysis2}}
    \end{subfigure}\hfill
    \begin{subfigure}[b]{.3\linewidth}
        \centering
        \includegraphics[width=\linewidth]{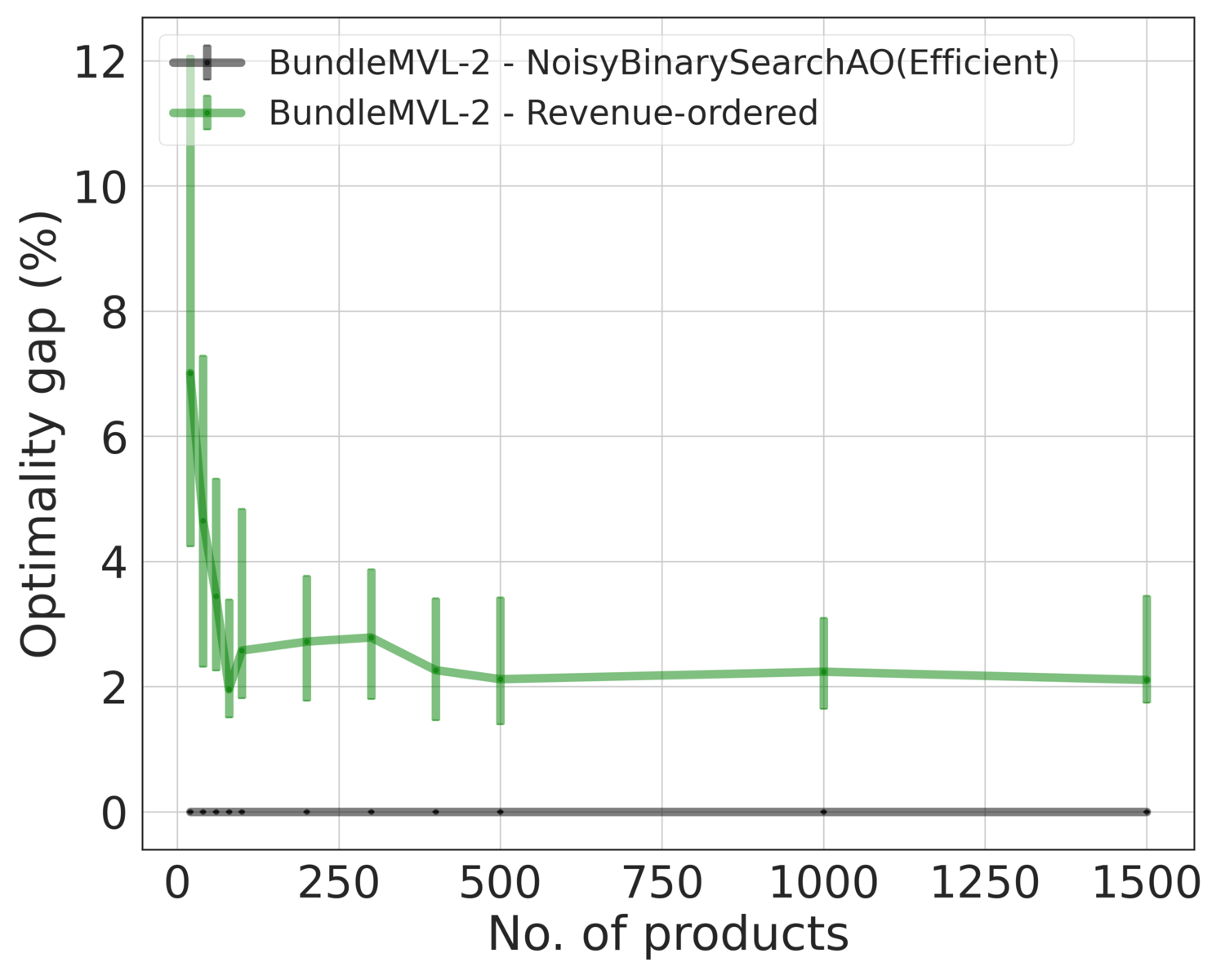}
        \caption{\label{fig:revenueorderedsuboptimalitygap}}
    \end{subfigure}\hfill
	\vfill
    \caption{ \bmvl-2 vs other models.
    (\subref{fig:mmnl-gap-tafeng}): Optimality gap of MNL and MMNL with \bmvl-2 as the ground truth using the Ta Feng dataset.
    (\subref{fig:mmnl-gap-uci}): Same as \subref{fig:mmnl-gap-tafeng}) using the UCI dataset.
    (\subref{fig:e5tcmrcmanalysis1}): Optimality gap for \bmvl-2 with \bmvl-3 as the ground truth.
    (\subref{fig:rcm_ucm_time_comparison}): Run-times (unconstrained setting) under \ucm~and \bmvl-2.    
    (\subref{fig:e5tcmrcmanalysis2}): Run-times (unconstrained setting) under \bmvl-2 and \bmvl-3 (\rcm~run-times overlap).
    (\subref{fig:revenueorderedsuboptimalitygap}): Optimality gap of the revenue-ordered heuristic on a synthetic dataset under \bmvl-2.
    }
\end{figure}

Figures~\ref{fig:mmnl-gap-tafeng},~\ref{fig:mmnl-gap-uci} and~\ref{fig:e5tcmrcmanalysis1} demonstrate the optimality gaps (the percentage difference from the optimal expected revenue under a ground truth model, lower is better). In Figures~\ref{fig:mmnl-gap-tafeng},~\ref{fig:mmnl-gap-uci}, \bmvl-2 is the ground truth. We observe that the recommendation sets produced using the MNL model lead to sub-optimal revenue, and the gap increases as the number of products increase, producing a $5-6\%$ relative loss at around $1500$ products. Similarly, the optimality gap for the MMNL starts from $9-15\%$ relative loss for $20-80$ products and reduces a bit and then slowly increases with the number of products, leading to $\sim3-6\%$ loss at around $1500$ products. We believe these improvements in expected revenue are due to a non-negligible percentage of product pairs having positive interaction terms $\beta_{ij}$s (we saw this for the Instacart dataset in Section~\ref{sec:beta}, but it holds true for Ta Feng and UCI Online Retail datasets as well), thus boosting the overall probability of purchase by a customer by a few percent. Since the expected revenue is a weighted aggregation of the probabilities of purchase of one or two bundles, we naturally see an improvement in the expected revenue as well.

In Figure~\ref{fig:e5tcmrcmanalysis1}, \bmvl-3 is the ground truth (note that the solver only produces a local optimum for \bmvl-3). We observe that the optimality gaps are similar to the gaps observed in the \rcm~ vs. MNL/MMNL comparison in the $20-400$ product range, showing that \bmvl-3 may not be needed for these datasets to achieve similar expected revenues. Beyond $400$ products, we are currently not able to solve for \bmvl-3 within reasonable run-times to assess optimality gaps. To summarize, modeling purchase of multiple products (such as by using \rcm) can lead to non-negligible revenue gains (especially, when the number of products is large), and if we can also improve the run-times of the optimization techniques involved, we can practically realize these gains in realistic applications.  For instance, since optimization over MNL (unconstrained) is linear time, it would be fruitful to reduce the computational complexity of unconstrained optimization over \rcm{} model as well, and we investigate this in detail in Section~\ref{sec:run-time-experiments} below.

We have also used a fairly general synthetic model (defined using independent utilities from co-purchasing any one or two products given any recommendation set) as the ground truth. Here, the number of parameters is exponential in the number of products, and cannot be estimated for the Ta Feng and UCI Online Retail datasets (because in these, we only observe purchases under a fixed recommendation set). Under this setting, the expected revenues obtained under the MNL and MMNL models were $15-30\%$ smaller in relative terms compared to \bmvl-2 when the number of products was varied from $20-200$. Optimizing over any instance of this ground truth model is intractable for the aforementioned product range, so we omit reporting the expected revenue gap between \bmvl-2 and the ground truth itself. 

\noindent\textbf{Run-time Gains.} Here we show how the run-times for optimizing under the \bmvl-2 model are competitive with other benchmark models. Complementary to this, in Section~\ref{sec:run-time-experiments}, we do an analysis of run-times of various approaches for optimizing under the proposed \bmvl-2 model. First we compare the times to solve for recommendation sets under \bmvl-2 and \ucm using the Ta Feng dataset. To keep the comparison fair, we solve the corresponding integer programs. Based on Figure \ref{fig:rcm_ucm_time_comparison}, we can conclude that the run-time is much faster under the \rcm~ model. 
This can be partially attributed to the difficulty in exploiting the \ucm~model's structure.

Next, we compare the run-times under \bmvl-2 and \bmvl-3 on both Ta Feng and UCI datasets. As seen in Figure~\ref{fig:e5tcmrcmanalysis2}, the time taken to optimize increases dramatically for \tcm{}. This likely stems from the difficulty in solving cubic integer optimization instances for \bmvl-3 as compared to quadratic integer optimization instances for \bmvl-2. Recalling Figure~\ref{fig:e5tcmrcmanalysis1}, we observe that the revenue gains are small/non-existent and likely do not outweigh the computational burden of optimizing over \tcm~ (and other \bmvlk \ models for $K>3$).

\subsection{Run-times for Computing \rcm{} based Recommendation Sets}\label{sec:run-time-experiments}

We now fix the \rcm{} model and benchmark the computation times and quality of solutions produced by various algorithms described in Section~\ref{sec:alg-rec-set} on the Ta Feng and UCI datasets. To get problem instances of different sizes, we use the same process as above. The choice of which QUBO heuristics to use in parallel during the comparison steps of binary search approaches is made a priori, after performing an extensive comparative study on synthetic data (\appAddExp). Table~\ref{tab:optimizationalgortihmsummary} summarizes the list of algorithms (for both constrained and unconstrained settings) that we consider in various product size regimes. The regimes in which each algorithm can be tested is decided based on the run-time and memory requirements of the algorithm. The MIP formulations are solved using CPLEX version 12.9.0 while the others are based on Python 3.7.4/Numpy 1.17, with some custom C++ code for the QUBO heuristics. We report two measures: (a) the fraction of times an algorithm generates a suboptimal solution, and (b) the median optimality gap over all Monte Carlo runs. The following two sub-experiments illustrate how scalable \noisybinsearcheff{} is compared to others for the Ta Feng dataset (similar trends are observed for the UCI dataset, see \appAddExp):
\begin{enumerate}
    \item Benchmarking  algorithms for the \rcm~model (unconstrained setting), and
    \item Benchmarking  algorithms for the \rcm~model (constrained setting).
\end{enumerate}

\begin{table}
\scriptsize
\centering
\begin{tabular}{ |c|c|c|c|c|  }
 \hline
 \textbf{Optimization Algorithm}& \textbf{Applicable Choice Model(s)} & \textbf{Type}& \textbf{Product Size Regime} \\ \hline
 \binsearcheff & \rcm & $\epsilon$-optimal & Small, Medium, Large \\ \hline
 \noisybinsearcheff & \rcm & Heuristic & Small, Medium, Large\\ \hline
 Mixed Integer Programming & \rcm & Exact & Small, Medium \\\hline
 Mixed Integer Programming & MMNL & Exact & Small, Medium, Large \\\hline
 Mixed Integer Programming & \ucm & Exact & Small \\\hline
 Mixed Integer Non-Linear Programming & \bmvl-3 & Heuristic & Small, Medium \\\hline
 Revenue Ordered & Any & Heuristic & Small, Medium, Large \\ \hline
 \adxopt/\adxopttwo& Any & Heuristic & Small \\\hline
\end{tabular}
\caption{Summary of algorithms and the underlying multi-choice models.}
\label{tab:optimizationalgortihmsummary}
\end{table}

\begin{figure}
    \begin{subfigure}[b]{.3\linewidth}
        \centering
        \includegraphics[width=\linewidth]{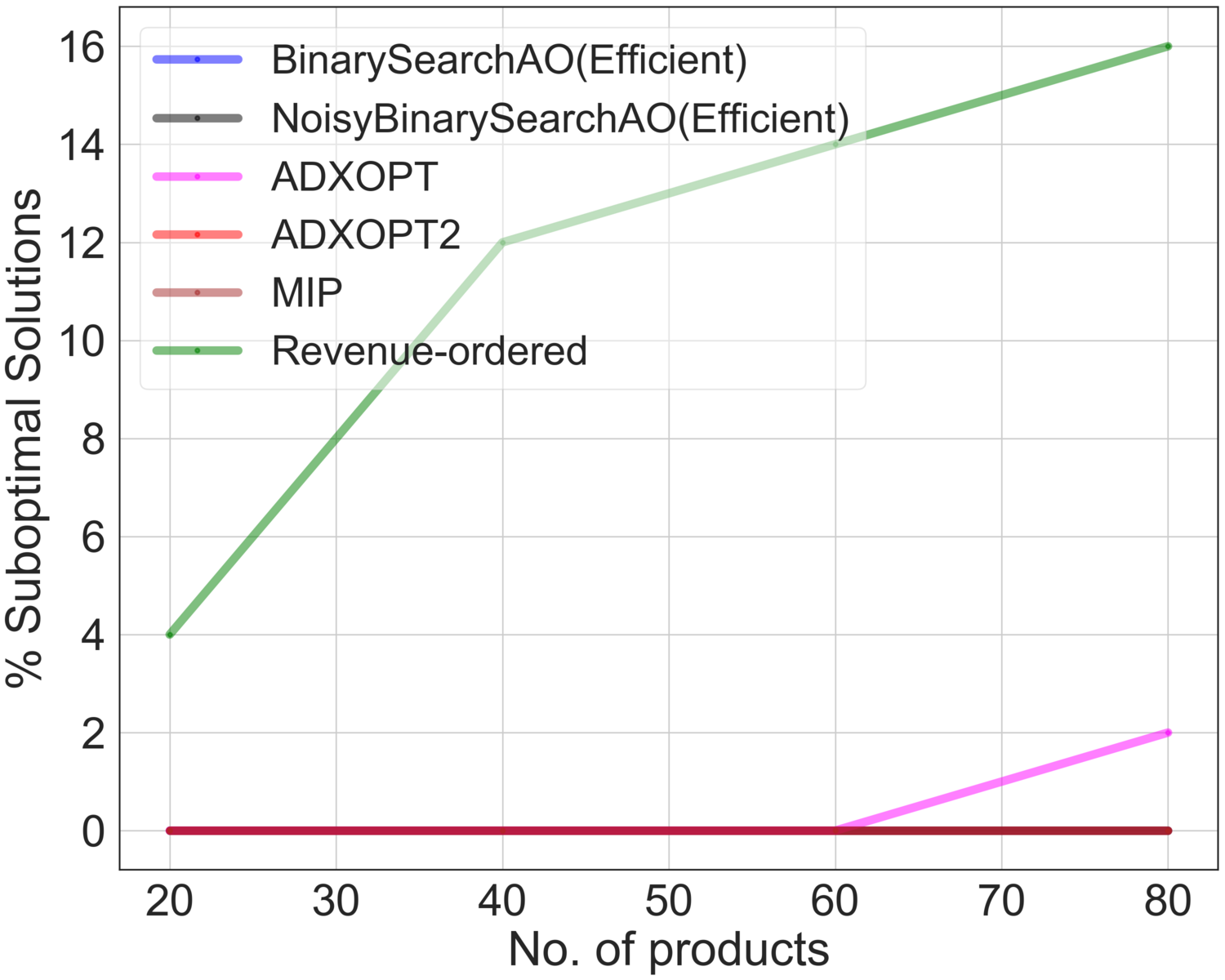}
        \caption{\label{fig:e1suboptimalityfraction1}}
    \end{subfigure}\hfill
    \begin{subfigure}[b]{.3\linewidth}
        \centering
        \includegraphics[width=\linewidth]{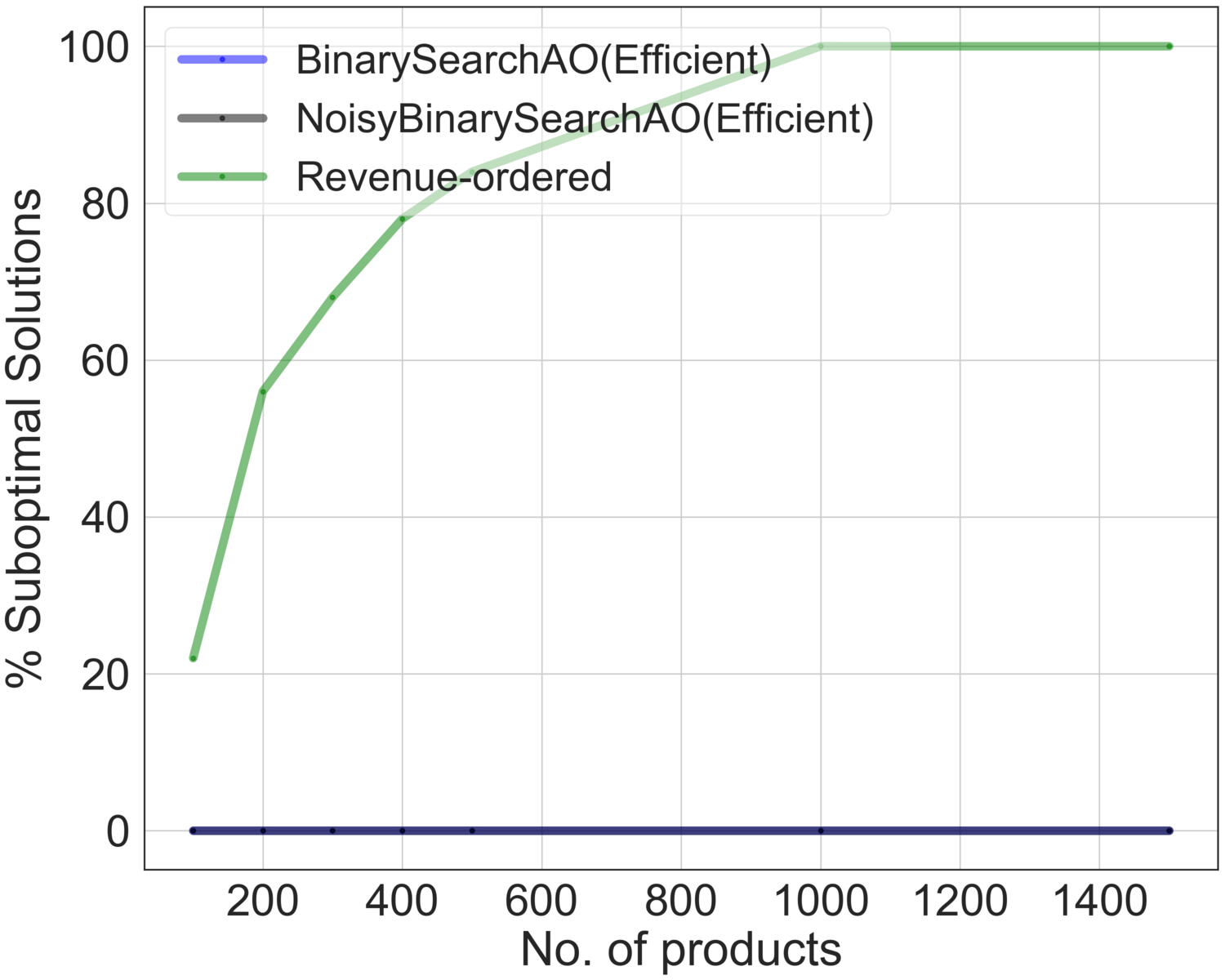}
        \caption{\label{fig:e1suboptimalityfraction2}}
    \end{subfigure}\hfill
    \begin{subfigure}[b]{.3\linewidth}
        \centering
        \includegraphics[width=\linewidth]{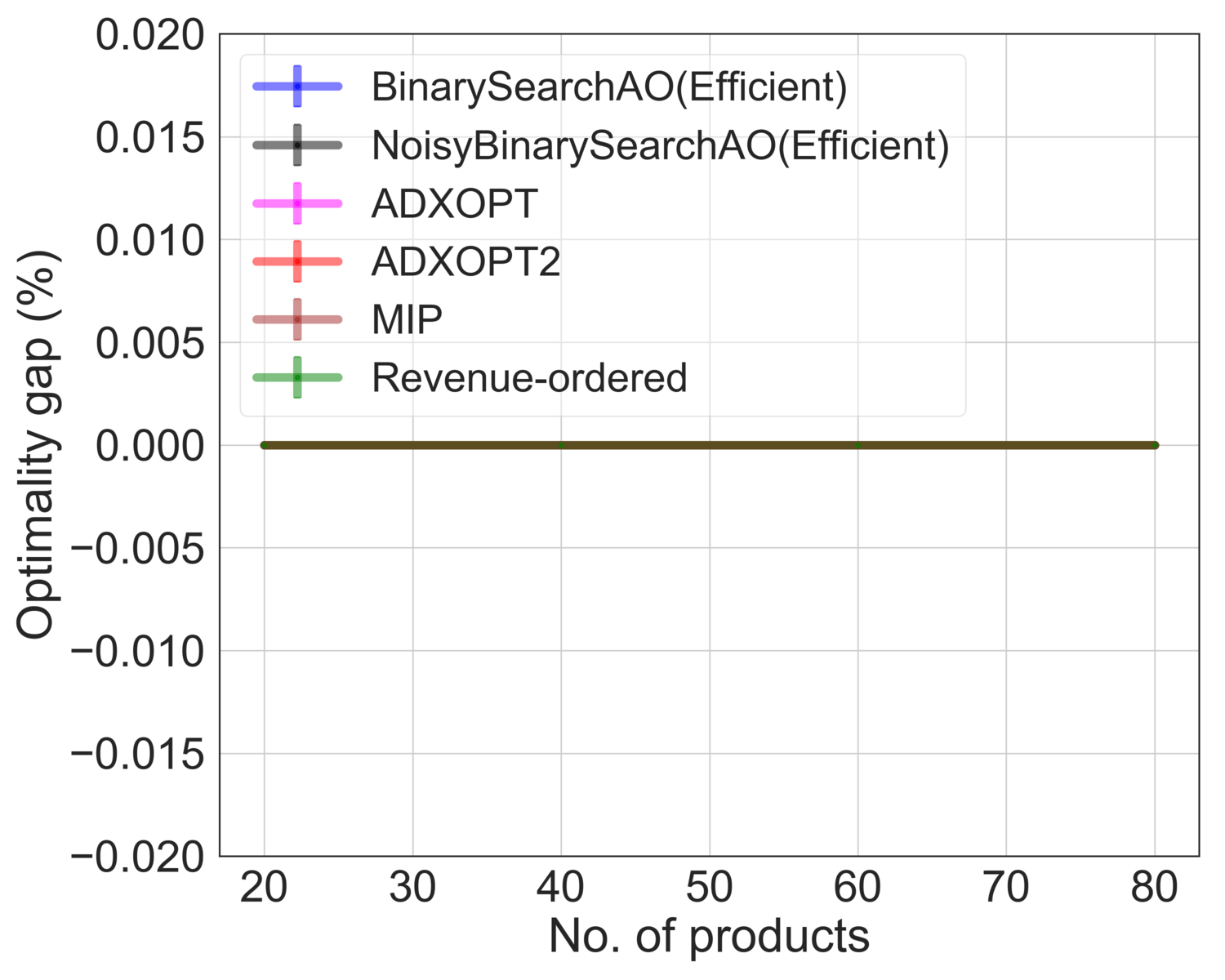}
         \caption{\label{fig:e1suboptimalitygap1}}
    \end{subfigure}\vfill
    \begin{subfigure}[b]{.3\linewidth}
        \centering
        \includegraphics[width=\linewidth]{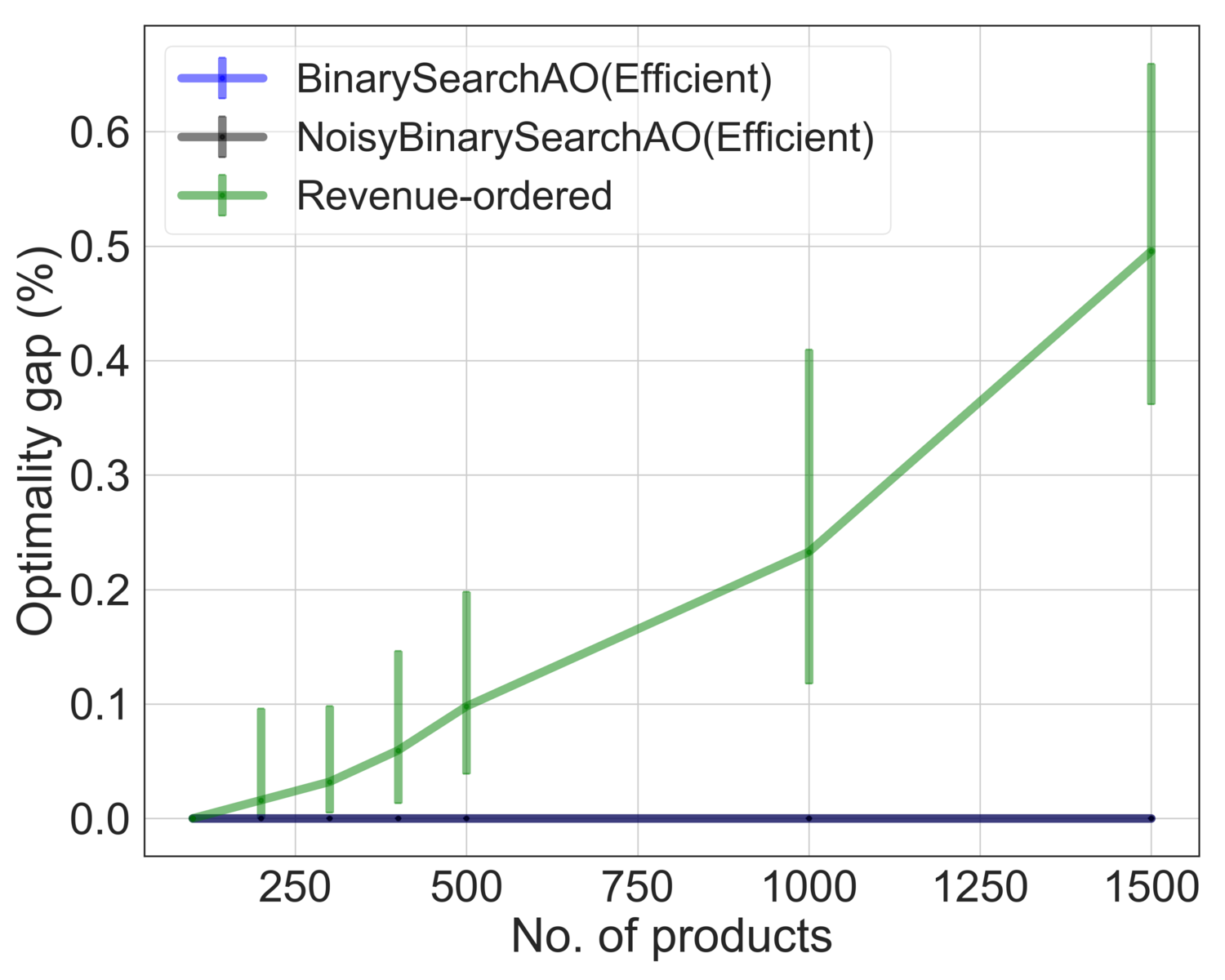}
        \caption{\label{fig:e1suboptimalitygap2}}
    \end{subfigure}\hfill
    \begin{subfigure}[b]{.3\linewidth}
        \centering
        \includegraphics[width=\linewidth]{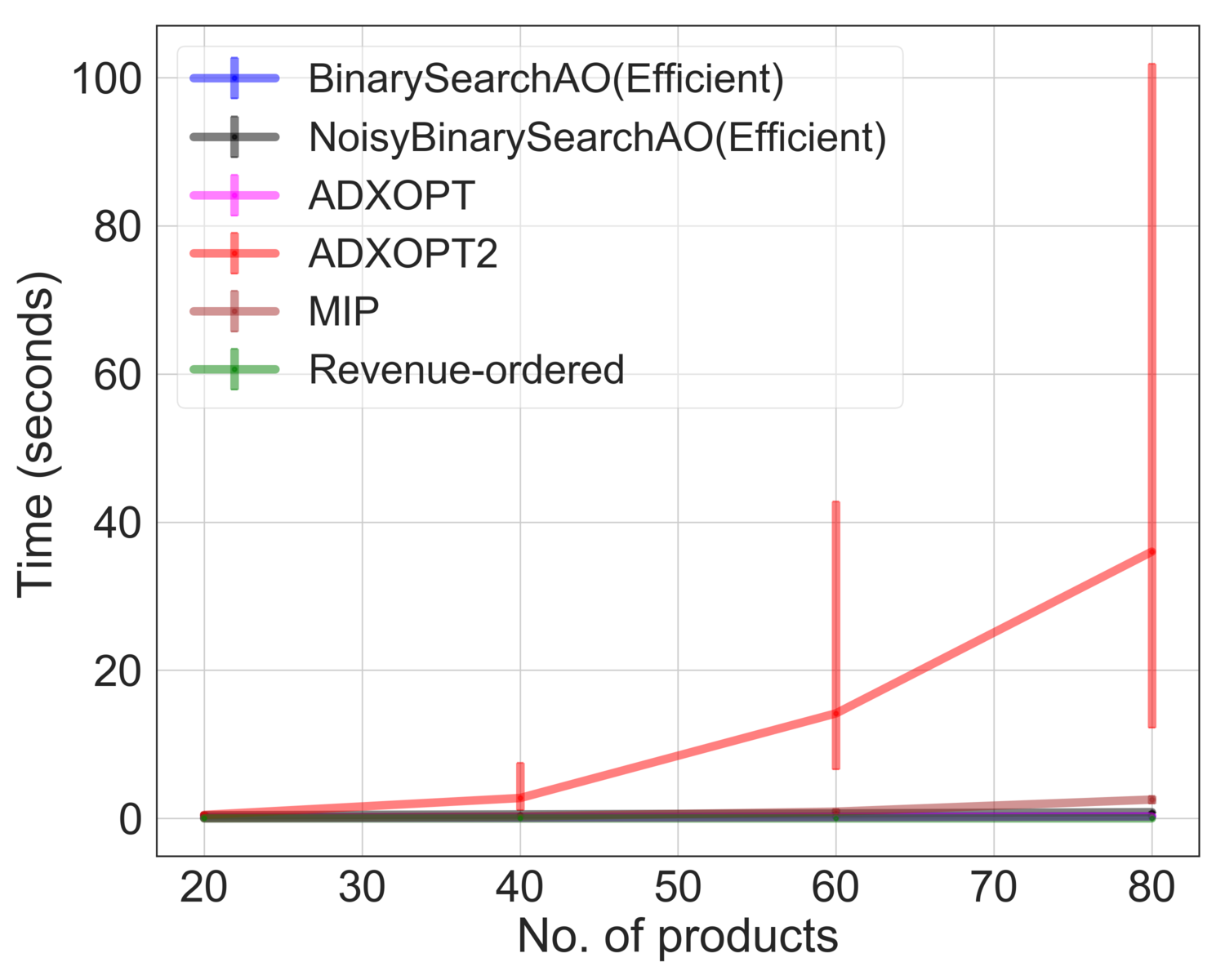}
        \caption{\label{fig:e1timeanalysis1}}
    \end{subfigure}\hfill
    \begin{subfigure}[b]{.3\linewidth}
        \centering
        \includegraphics[width=\linewidth]{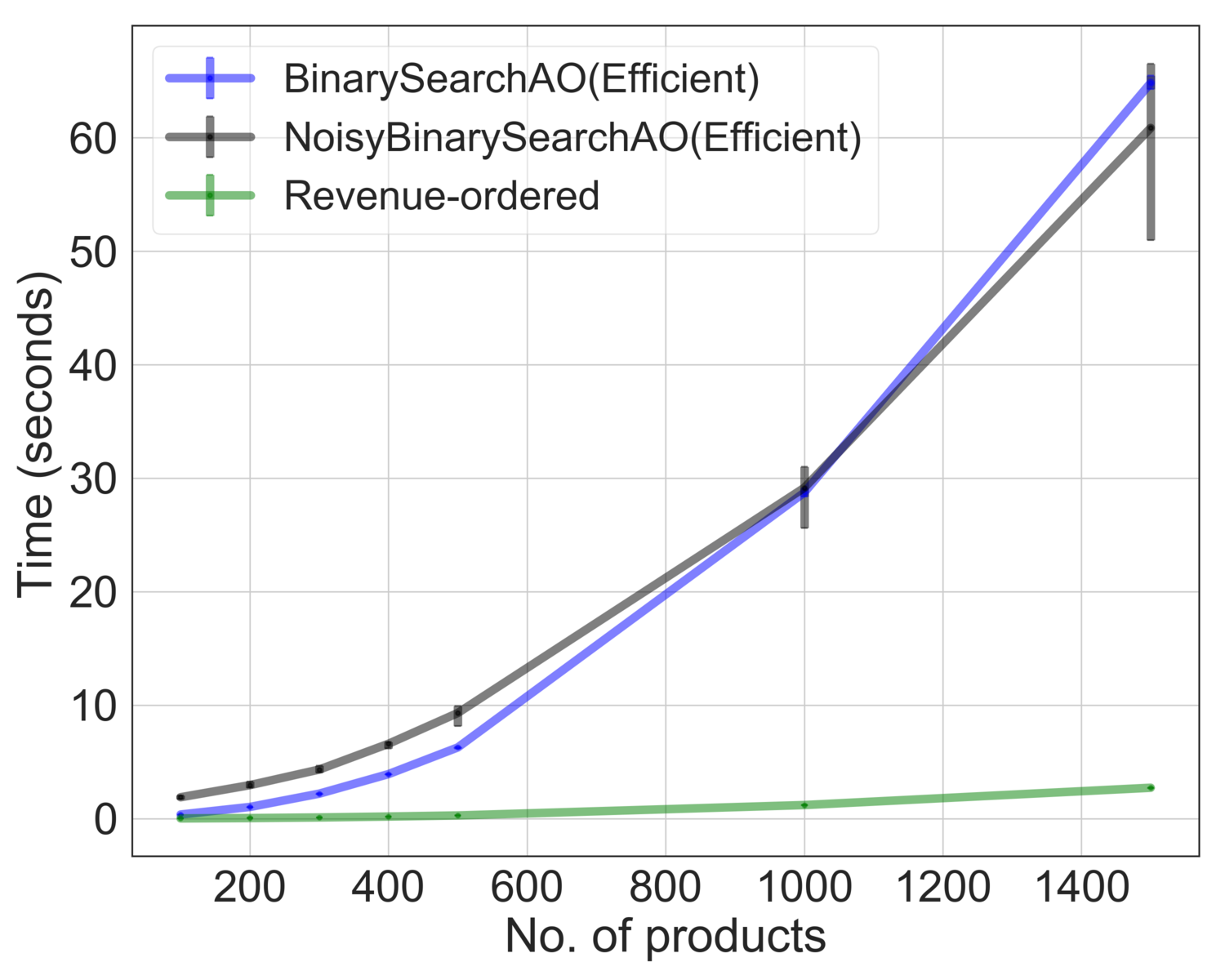}
        \caption{\label{fig:e1timeanalysis2}}
    \end{subfigure}\vfill
    \caption{Optimality gaps (\subref{fig:e1suboptimalityfraction1}-\subref{fig:e1suboptimalitygap2}) and run-times (\subref{fig:e1timeanalysis1}-\subref{fig:e1timeanalysis2}) in the unconstrained setting under \bmvl-2 using the Ta Feng dataset. The gaps are with respect to the best among the considered algorithms. 
    \label{fig:e1tafenganalysis}}
\end{figure}

\noindent\textbf{Benchmarking algorithms for the \rcm~model (unconstrained setting).}  Figures ~\ref{fig:e1suboptimalityfraction1},~\ref{fig:e1suboptimalityfraction2} and ~\ref{fig:e1suboptimalitygap1},~\ref{fig:e1suboptimalitygap2} show the fraction of times suboptimal solutions were returned and the optimality gap of various algorithms respectively. In Figure~\ref{fig:e1suboptimalityfraction1}, \binsearcheff, \noisybinsearcheff, \adxopttwo~and \adxopt~return no suboptimal solution when the number of products $\numberOfItems \leq 60$. In Figure~\ref{fig:e1suboptimalityfraction2}, the revenue-ordered heuristic trails in terms of obtaining optimal solutions. In Figure~\ref{fig:e1suboptimalitygap1}, we see that all algorithms have a $0$ median optimality gap. In Figure~\ref{fig:e1suboptimalitygap2}, \binsearcheff~and \noisybinsearcheff~curves overlap and are very close to $0$. Finally, we see that the gaps for the revenue-ordered heuristic are small in the context of this dataset.

From these plots, we note that \noisybinsearcheff \ has good performance on all instance sizes. It converges to the optimal solution for more than $90\%$ of the instances for all regimes of product sizes. We also note that the other heuristic approaches - the revenue-ordered heuristic, \adxopt \ and \adxopttwo \ - also output close to optimal solutions, even though the revenue-ordered heuristic fails to return an optimal solution most of the time. We drop \binsearch~and \noisybinsearch from this experiment, because we observe that utilizing structural properties (\appStructural) improves solution quality and leads to faster computations due to the number of products to optimize over decreasing in subsequent \textsc{Compare-Step}s.

Figures ~\ref{fig:e1timeanalysis1} and ~\ref{fig:e1timeanalysis2} show the run-times of the algorithms. In Figure~\ref{fig:e1timeanalysis1}  \adxopt, MIP, \binsearch~and the revenue-ordered heuristic's curves are close to zero, and they tend to increase for larger sizes. In Figure~\ref{fig:e1timeanalysis2}, we see that the revenue-ordered heuristic is much faster to compute as expected. The greedy algorithms (\adxopt \ and \adxopttwo) and the MIP solver don't scale well as the number of products increases, with the former having a large variance in run-times. Consequently, we limit their evaluation to the small regime (where the number of products is less than $100$). Among the binary search based approaches, \noisybinsearcheff{} becomes slightly faster in comparison to exact binary search as the number of products increases. Also, as expected, the revenue-ordered heuristic takes the least time to run. Overall, the binary search and revenue-ordered algorithms scale well as the number of products increases when compared to other approaches. Between binary search approaches and the revenue-ordered heuristic, the former have lower optimality gaps (with direct implications on expected revenue). 

While one could make a case about trading-off sub-optimality with speed in favor of the revenue-ordered heuristic, we observe that the heuristic's performance is very sensitive to the datasets. 
In the Ta Feng and UCI datasets, the proportion of estimated $\beta_{ij}$s is $< 10\%$, which explains the efficacy of the heuristic based on our analysis (see \appThmSmallVij{} in \appMainBenchmarks). Motivated by this observation, we generate an additional synthetic dataset by first dividing products into two groups of equal size - high priced products and low priced products, and assuming/imposing that users are very unlikely to buy two high priced products. To capture this,  $V_{i,j}$ parameters are sampled from Beta$(1,10)$ distribution when products $i$ and $j$ both belong to the high priced group, and the rest of the $V_{i,j}$s are sampled from Beta$(10,1)$. Additionally, the prices ($r_i$s) are sampled from Beta$(2,10)$ and the $V_{i}$s are sampled from Beta$(1,1)$. Figure \ref{fig:revenueorderedsuboptimalitygap} shows that the 
gap for the revenue-ordered heuristic on instances of this type is non-negligible across a broad range,
supporting our hypothesis.

\begin{figure}
	\begin{subfigure}[b]{.3\linewidth}
		\centering
		\includegraphics[width=\linewidth]{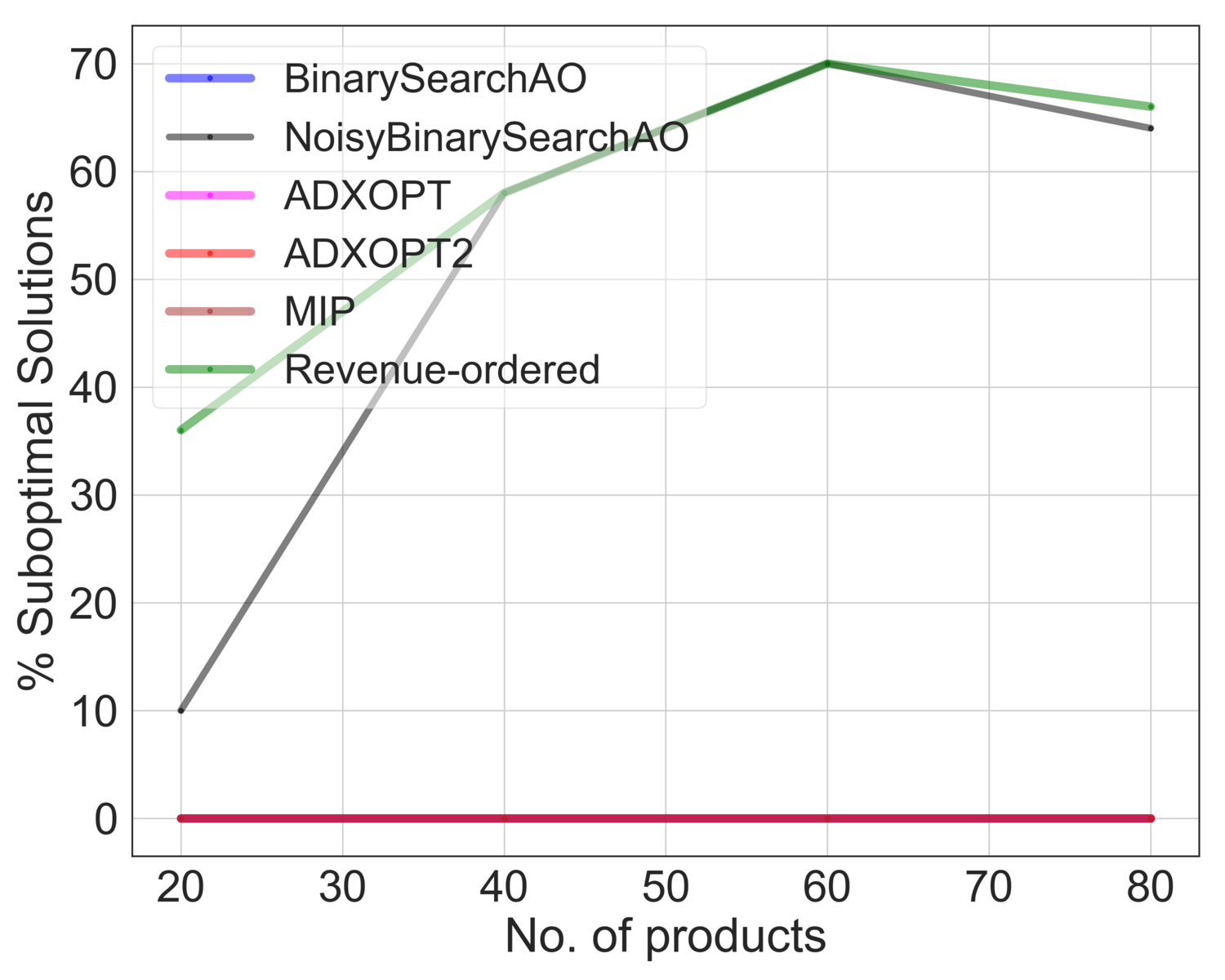}
		\caption{\label{fig1a}}
	\end{subfigure}\hfill
	\begin{subfigure}[b]{.3\linewidth}
		\centering
		\includegraphics[width=\linewidth]{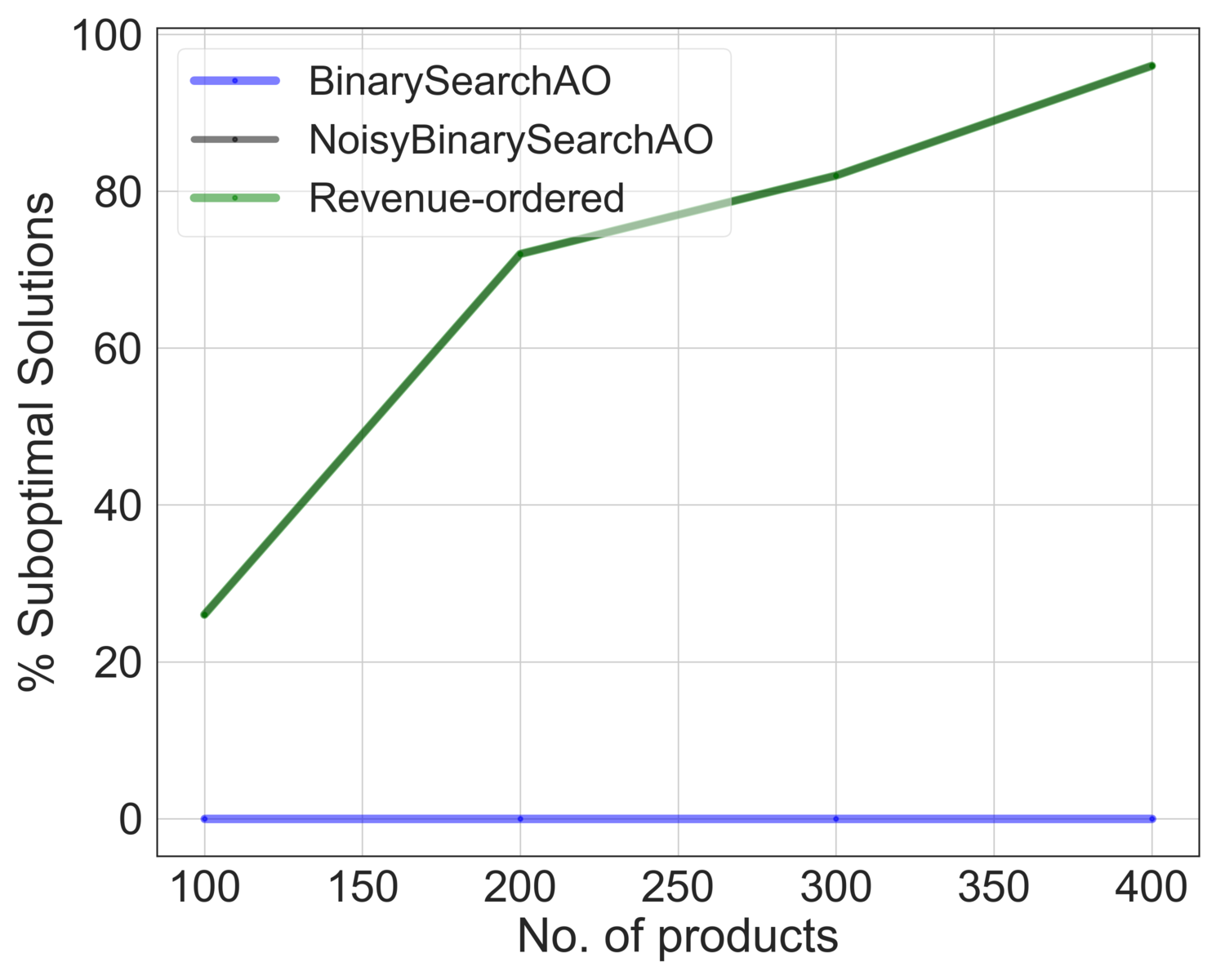}
		\caption{\label{fig1b}}
	\end{subfigure}\hfill
	\begin{subfigure}[b]{.3\linewidth}
		\centering
		\includegraphics[width=\linewidth]{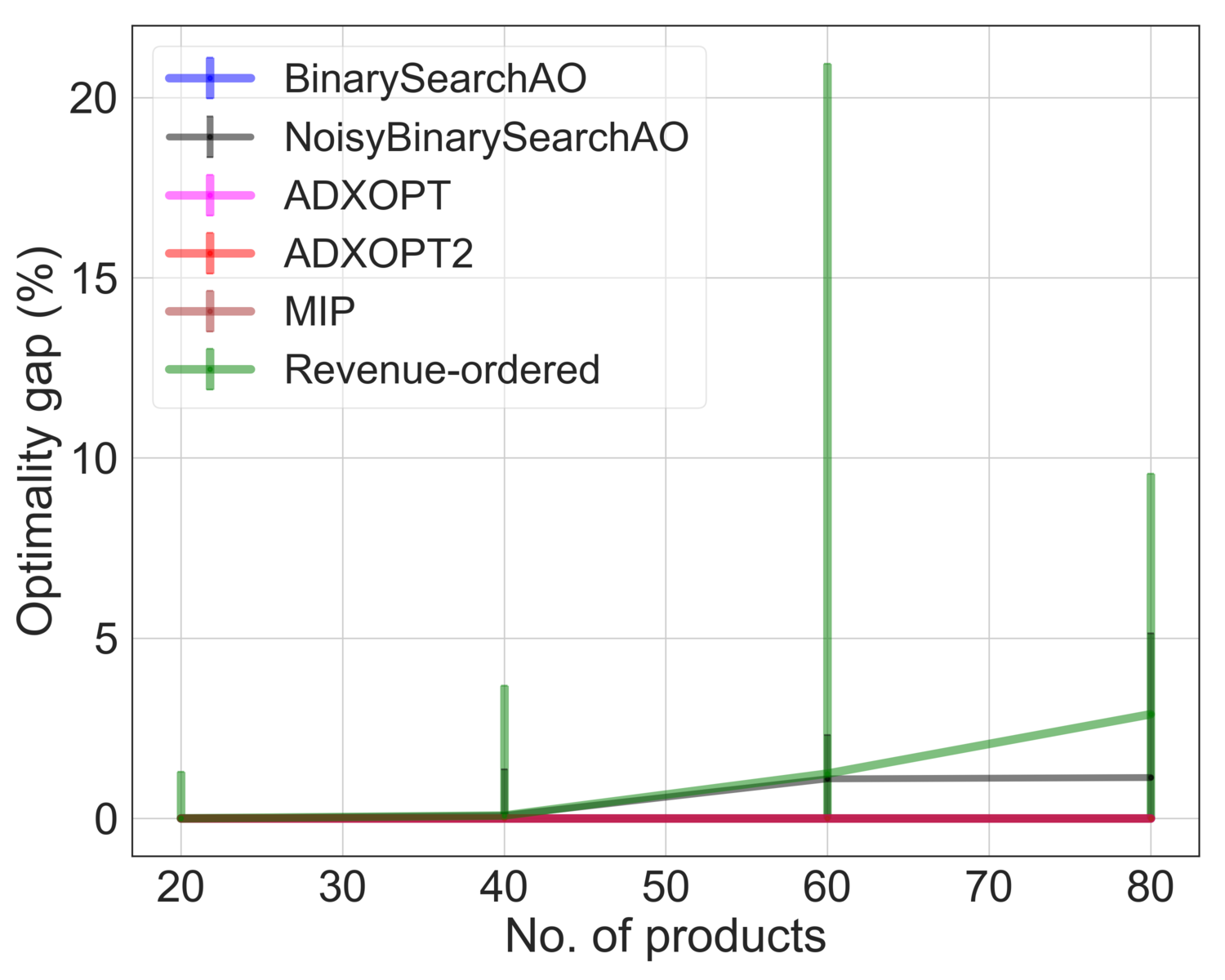}
		 \caption{\label{fig1c}}
	\end{subfigure}\vfill
	\begin{subfigure}[b]{.3\linewidth}
		\centering
		\includegraphics[width=\linewidth]{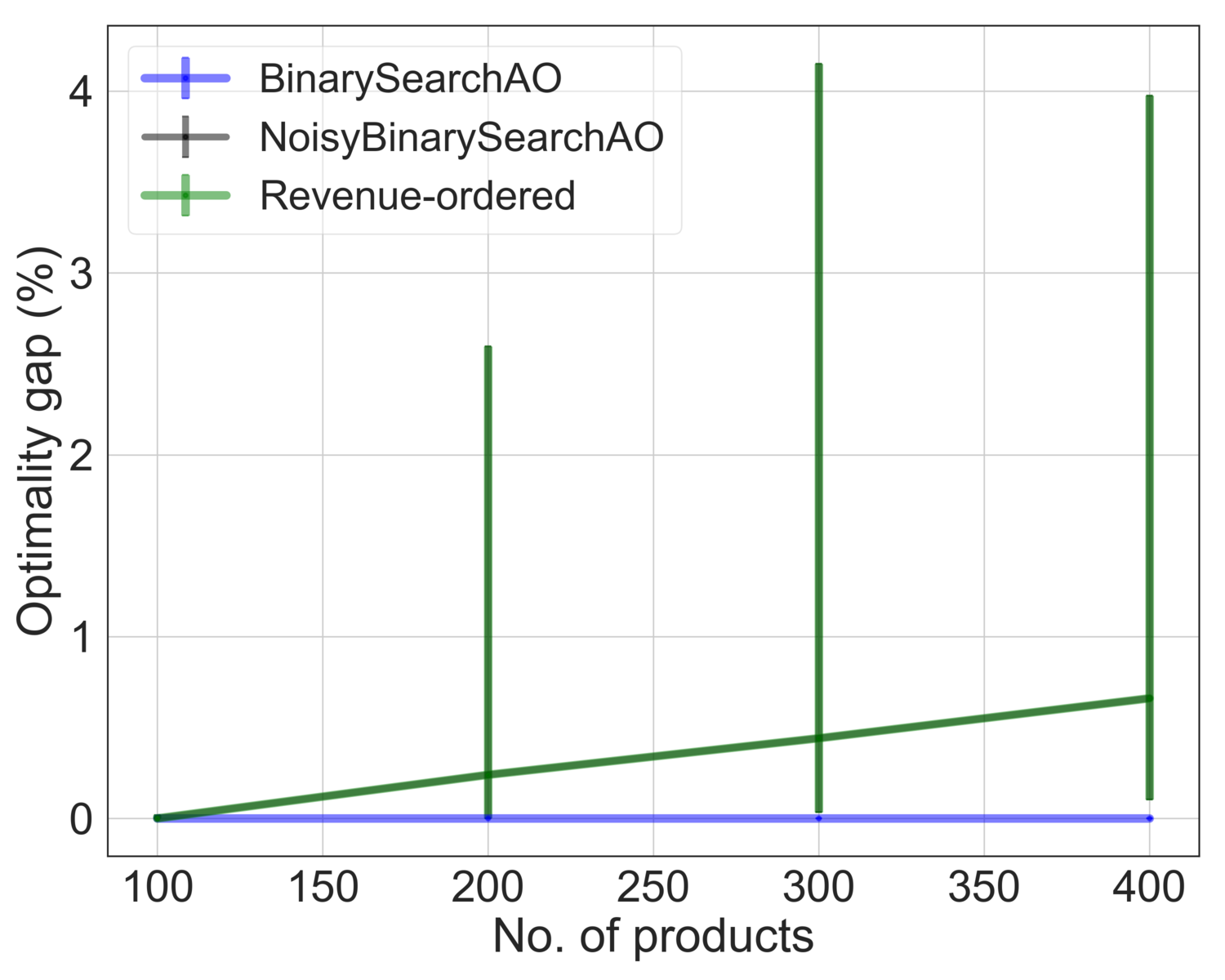}
		\caption{\label{fig1d}}
	\end{subfigure}\hfill
	\begin{subfigure}[b]{.3\linewidth}
		\centering
		\includegraphics[width=\linewidth]{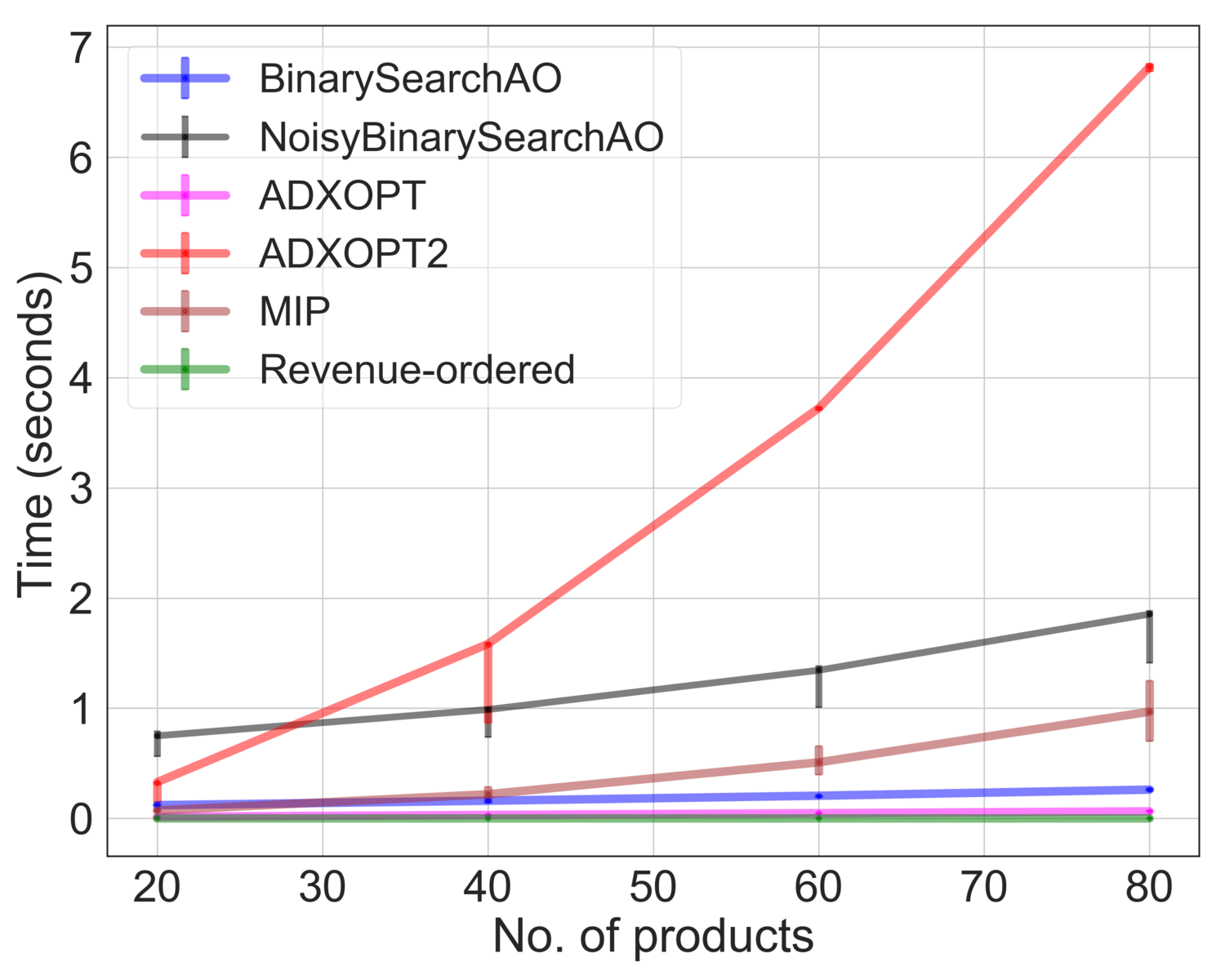}
		\caption{\label{fig1e}}
	\end{subfigure}\hfill
	\begin{subfigure}[b]{.3\linewidth}
		\centering
		\includegraphics[width=\linewidth]{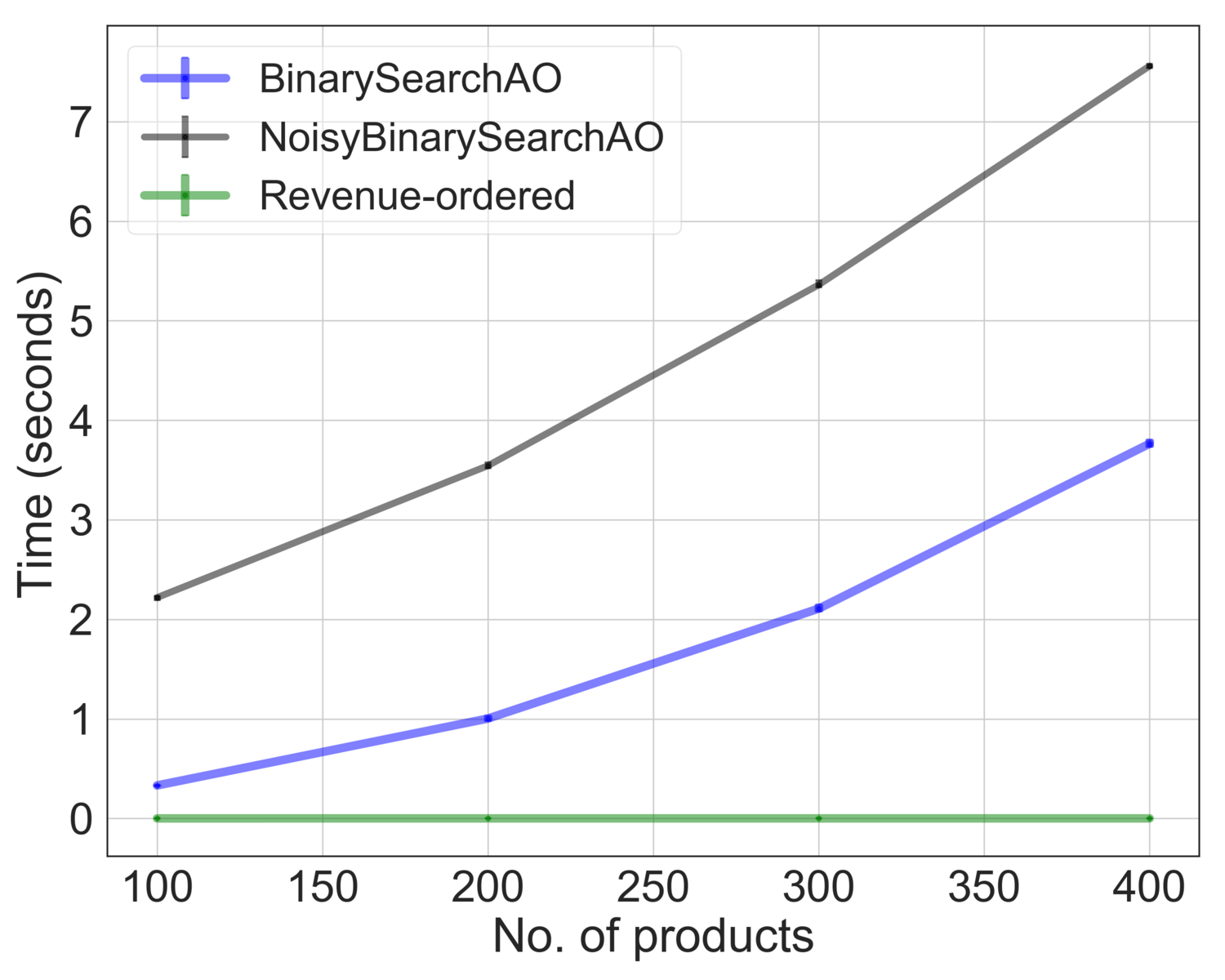}
		\caption{\label{fig1f}}
	\end{subfigure}\vfill
	\caption{Optimality gaps (\subref{fig1a}-\subref{fig1d}) and run-times (\subref{fig1e}-\subref{fig1f}) in the capacity constrained setting under \bmvl-2 using the Ta Feng dataset. The gaps are with respect to the best among the considered algorithms.
		\label{fig:e3tafenganalysis}}
\end{figure}

\noindent\textbf{Benchmarking algorithms for the \rcm~model (constrained setting).} In this experiment, we evaluate the performance of algorithms when there is a constraint on the size of the feasible recommendation set. We focus our evaluation on small and medium regimes for the number of products, and fix the capacity constraint to $5$ and $20$ for these two regimes respectively. 
The performance of different algorithms is presented in Figure \ref{fig:e3tafenganalysis}. In Figure~\ref{fig1a}, \binsearch, MIP and \adxopt~ get optimal solution in most runs, thus the curves overlap near $0$. In Figure~\ref{fig1b} and Figure~\ref{fig1d}, we observe that \noisybinsearch, which uses the revenue-ordered solution as a lower bound, generates no gain on top of it. Thus, the curves overlap for these two as well. In Figure~\ref{fig1c}, \adxopt, \adxopttwo, MIP and \binsearch~curves overlap as the median optimality gap is $0$ in the small product size regime. Revenue-ordered heuristic's gap improves from Figure~\ref{fig1c} to Figure~\ref{fig1d} due to an increase in the capacity constraint. Most importantly, we see that \noisybinsearch~(structural properties of the optimal solution we derived no longer hold) with QUBO heuristics does worse in terms of solution quality compared to \binsearch. This is likely due to the reformulated QUBO instances becoming dense instances (the matrix in the objective becomes dense). In the unconstrained case, the proportion of non-zero $\beta_{ij}$s are relatively small and we don't have to do any reformulation, leading to sparse instances. 

Further, there is no significant improvement in solution quality using various algorithms compared to the simple revenue-ordered heuristic for this dataset. Due to the capacity constraint, the greedy algorithms \adxopt \ and \adxopttwo \ also run much faster (when compared to their performance in the unconstrained setting). The revenue-ordered heuristic has a constant time performance as it only needs to evaluate the first $\capconst$ revenue-ordered sets (where $\capconst$ is the capacity constraint), irrespective of the total number of products. Overall, we conclude that the current shortlisted QUBO heuristics (\appAddExp) are not competitive for the constrained setting when compared to other approaches. Nonetheless, the binary search approach with exact computation at each comparison step (\binsearch) works well. It also scales well in the studied product size regimes. We will need newer heuristics/strategies to improve the run-time in order to compete with the benchmark approaches (revenue-ordered, \adxopt{} or \adxopttwo{}) in larger product size regimes.

%% file: 5_discussion.tex
\section{Discussion}\label{sec:discussion}

While the focus of the paper has been on proposing a parsimonious model for multi-purchase behavior in retail and other online settings followed by algorithmic solutions for maximizing revenues, there are a few key managerial insights that become apparent. Firstly, the investigations in this paper support the importance of modeling richer consumer behavior models going beyond what has been done previously. Although operational decisions based on these richer models may pose challenges, we have shown fairly extensively that scalable near real-time computation of revenue maximizing recommendation sets is entirely feasible, even if the problem is NP-hard in theory. Not only is this computation fast in practice, but the expected revenue gains to be had with richer models make them well worth the effort. Second, practitioners should not be afraid of the increase in complexity with these models. For instance, the proposed models in this work are parametric and interpretable, and only depend quadratically on the number of products, a feature common with many single-choice models such as the Markov chain choice model, the paired combinatorial logit model and others. 

Third, managers and practitioners can easily trade-off model complexity with increase in optimization times with the models and algorithms proposed here. The richer the model, the better it represents consumer purchase behavior, while at the same time increasing the operational aspects (such as run-times). In fact, the iterative nature of our approach allows one to control this trade-off in a fine grained manner: they can choose less number of iteration steps to solve for candidate recommendation sets, which may not be optimal but useful to improve the user experience and revenues compared to other alternatives. Further, \rcm{} model from the \bmvlk{} family seems to be rich enough to provide tangible gains when compared to single-choice models. 


Finally, a key insight useful for decision makers and practitioners that emerged from our experiments is the remarkable effectiveness of provably suboptimal heuristics on real world data. While theory gives limited insights on the performance of heuristics such as the revenue-ordered heuristic or \adxoptk, we observe that they were fairly competitive for the two real world datasets. This supports the notion that one should rely on real world data related to the specific problem at hand while making algorithmic as well as modeling choices. The interpretability and simplicity of these heuristics are an added bonus in this regard.

%% file: 6_conclusion.tex
\section{Conclusion}
\label{sec:conclusion}

In this work, we evaluated the effectiveness of multi-purchase choice behavior captured via the proposed \bmvlk{} family on improving revenues, and compared it to other state-of-the-art models. We made one of the first contributions towards using multi-purchase models for expected revenue maximization.
Although the revenue gains were non-negligible, the optimization problems were harder than those for single-choice models, and we designed scalable algorithms that can allow a practitioner to realize these gains in demanding applications such as e-commerce platforms.

%% file: 90_appendix.tex
\setcounter{page}{1}

%% file: 91_model_and_estimation.tex
\section{Model: Choice Probabilities, Relation to MNL and Estimation}
\label{sec:model-choice-mnl-estimation}

\subsection{Derivation of the Choice Probabilities for \bmvl-K}
\label{sec:model-derivation}

We briefly discuss how the analytical form of the choice probability for the \bmvl-K model shown in Equation~\ref{eqn:choice-prob-main} is obtained.
If the consumer has made a decision for each of the other products, then they will purchase product $i$ only if the conditional utility defined in Equation~\ref{eqn:conditional-utility} exceeds the threshold value $0$. The conditional probability of buying product $i \in \offeredAsmt $ can be computed as: 
\begin{equation*} 
\prob\left(X_i=1 | \{ X_j=x_j: j \in \offeredAsmt, j \neq i \}\right) =  \frac{\exp(\alpha_i + \sum_{j \in \offeredAsmt, j \neq i} \beta_{ij}x_j)}{\exp(\alpha_i + \sum_{j \in \offeredAsmt, j \neq i} \beta_{ij}x_j) + 1} \ind \left\{ \sum_{j \in \offeredAsmt, j \neq i} x_j < K  \right\}. 
\end{equation*}
\textcolor{black}{This form of the probability is due to the noise being Gumbel distributed}. Also, note that the above probability is non-zero only when the number of products already purchased is strictly less than $K$. These conditional probabilities can be combined using Besag's characterization theorem~\citep{besag1974spatial} to get a consistent joint probability distribution of purchase of bundles. Let $\phi$ be the empty bundle signifying a no-purchase event. As per Besag's theorem, for any $\mathbf{x} = (x_1, \cdots x_\numberOfItems)$ such that $P(\mathbf{x}) > 0 $, we have $ \frac{ \prob(\mathbf{x} )}{\prob(\phi)} = \prod_{i \in \offeredAsmt}   \frac{ g^i(x_i)}{g^i(0)}$, where $g^i(1) + g^i(0) = 1$ and:
\begin{gather*}
g^i(1)  =   \frac{\exp(\alpha_i + \sum_{j \in \offeredAsmt, j < i} \beta_{ij}x_j)}{\exp(\alpha_i + \sum_{j \in \offeredAsmt, j < i} \beta_{ij}x_j) + 1} \ind \left\{ \sum_{j \in \offeredAsmt, j < i} x_j < K  \right\}.
\end{gather*}

Thus, $\frac{\prob(\mathbf{x})}{\prob(\phi)} = 
\exp\left( \sum_{i \in \offeredAsmt} \alpha_i + \sum_{i \in \offeredAsmt} \sum_{j \in \offeredAsmt, j < i} \beta_{ij}x_j \right)  \text{if } \ind \{ \sum_{j \in \offeredAsmt, j < i} x_j < K  \}  =1$ and $0 $ otherwise. Using the fact that the sum of probability of purchase over all bundles is  one, we get the probability of purchase of a bundle $\anychosenset$ given $\offeredAsmt$ as: $\prob(\anychosenset | \offeredAsmt) = \frac{V_\anychosenset}{1 + \sum_{\anychosenset' \subseteq \offeredAsmt , |\anychosenset'| \leq K  }V_{\anychosenset'}}$,
where $V_\anychosenset = \exp\left( \sum_{i \in \offeredAsmt} \alpha_i x_i + \sum_{i \in \offeredAsmt} \sum_{j \in \offeredAsmt, j < i} \beta_{ij} x_i x_j \right)  \ \forall \ \anychosenset \subseteq \offeredAsmt, \ |\anychosenset| \leq K $, and $x_j = 1$  if $j \in \anychosenset$ and zero otherwise. Note that one can extend our model representation power by including parameters involving three or more products as well, because as long as we ensure that the conditional probability of purchasing a product does not depend on the order of previous purchases, Besag's theorem can still be used to derive an analogous multi-purchase model. To be consistent with the literature on single-choice models, we introduce another parameter $v_0$ corresponding to the no-purchase probability by scaling each $V_\anychosenset$ by $\frac{1}{v_0}V_\anychosenset$. One can interpret this parameter as the result of comparing the conditional utilities to a non-zero threshold.

\subsection{Comparison of the Likelihood Expressions of MNL and \bmvl-2}\label{subsec:likelihood-compare}

The likelihood difference of MNL and \bmvl-2 is from two sources: (1) $\beta_{ij}$s (the interactions between product pairs) and (2) the way a purchase with greater than two products is broken up into multiple observations involving a single purchase (for MNL) and at most two item purchases (for \bmvl-2). Even in the  case when $\beta_{ij}$s are all zero, the likelihood of observing a purchase is not the same under MNL and \bmvl-2 because the denominator of the conditional probabilities will be different. Note that only when the probability of choosing a product pair is the same as the product of the probability of choosing each individually, the corresponding $\beta_{ij}$ will be $0$ (this can be rare). When the probability of choosing the product pair itself is $0$ (e.g., completely unrelated products), the corresponding $\beta_{ij} = -\inf$ (this is common). 

Next, we show that the likelihood of a purchase under \bmvl-1 (which is MNL) is not the same as \bmvl-2 when $\{\beta_{ij}\}$s are all zero via an illustrative example. Let $(a,b,c,d)$ be the recommendation set and $(a,b,c)$ be the bundle purchase. Under MNL, we consider this as three singleton choices/purchases given the same recommendation set. Thus, the likelihood of observing this example is as follows. 
 \begin{align*}
\prob_{(a,b,c)|(a,b,c,d)}^{MNL} &= P(a|(a,b,c,d))\times P(b|(a,b,c,d))\times P(c|(a,b,c,d)),\\
&= \frac{v_a}{v_0 + v_a + v_b + v_c + v_d}\times \frac{v_b}{v_0 + v_a + v_b + v_c + v_d} \times \frac{v_c}{v_0 + v_a + v_b + v_c + v_d},\\
&= \frac{v_a\times v_b\times v_c}{(v_0 + v_a + v_b + v_c + v_d)^3}.
\end{align*}
    
We next compute the likelihood of the same observation under \bmvl-2, when $\beta_{i,j}$s are all zeros. Let $\exp(\alpha_i) = v_i$ for $i=\{a,b,c,d\}$ and $\exp(\alpha_i + \alpha_j) = v_{ij}$ for $i,j \in \{a,b,c,d\}$. Further, let
    \begin{align*}
    D &= v_0+ v_a + v_b + v_c + v_d + v_{ab} + v_{ac} + v_{ad} + v_{bc} + v_{bd} + v_{cd}.
    \end{align*}
    Since $K=2$ for \bmvl-2, we consider four likely generative processes for this observation. Given the recommendation set $(a,b,c,d)$, the customer could have purchased: (i) $a$ and $(b,c)$ separately, or (ii) $b$  and $(a,c)$ separately, or (iii) $c$ and $(a,b)$ separately, or (iv) each of $a$, $b$ and $c$ separately. Lets represent each of these events using their case numbers (e.g., the first event is represented as (i)). Under each event, the choice probabilities are shown below:
    \begin{align*}
    \prob_{(a)(b,c)|(a,b,c,d)}^{\bmvl-2} &= \frac{v_a}{D}\times \frac{v_{bc}}{D} = \frac{v_{a} v_{b}v_{c}}{D^2},
    \quad \prob_{(b)(a,c)|(a,b,c,d)}^{\bmvl-2} = \frac{v_b}{D}\times \frac{v_{ac}}{D} = \frac{v_{a} v_{b}v_{c}}{D^2},\\
    \prob_{(c)(a,b)|(a,b,c,d)}^{\bmvl-2} &= \frac{v_c}{D}\times \frac{v_{ab}}{D} = \frac{v_{a} v_{b}v_{c}}{D^2}, \textrm{ and }
    \prob_{(a)(b)(c)|(a,b,c,d)}^{\bmvl-2} = \frac{v_a}{D}\times \frac{v_{b}}{D}\times \frac{v_{c}}{D} = \frac{v_{a} v_{b}v_{c}}{D^3}.
    \end{align*}
    They can be added together to get the likelihood of the observation under $\bmvl$-2 as:
    \begin{align*}
    \prob_{(a,b,c)|(a,b,c,d)}^{\bmvl-2} &= P(i)\times \prob_{(a)(b,c)|(a,b,c,d)}^{\bmvl-2}
    + P(ii)\times \prob_{(b)(a,c)|(a,b,c,d)}^{\bmvl-2}\\
    & + P(iii) \times \prob_{(c)(a,b)|(a,b,c,d)}^{\bmvl-2}
    + P(iv) \times \prob_{(a)(b)(c)|(a,b,c,d)}^{\bmvl-2}.
    \end{align*}

Clearly, the above expression will not be the same as the likelihood of MNL. 

\subsection{Data Augmentation before MLE for the \bmvlk~Model}
\label{sec:estimation}

\begin{minipage}[t]{.5\columnwidth}
\begin{algorithm}[H]
\caption{Dataset Pre-processing}
\begin{algorithmic}
\label{alg:est}
\REQUIRE{ Purchased bundles $\tilde{S}_1, \cdots \tilde{S}_{\tilde{m}}$. }
\STATE $\mathcal{S} \gets [ \ ]$, weights $ \gets  [ \ ], \ l  \gets1$.
\WHILE{ $l \leq \tilde{m}$}
    \IF{$|\tilde{S}_l| \leq K $}
        \STATE $\mathcal{S}$.append($\tilde{S}_l$) \& weights.append($1$).
    \ELSE
        \STATE Find the set of partitions: 
        \STATE $Q_l =\{\left( A_1, \cdots A_{t} \right)  \text{ s.t.}$ (1) $ \cup_{j=1}^t A_j = \tilde{S}_i$, 
        \STATE (2) $ |A_j| \leq K \ \forall \ 1 \leq j \leq t;$ and
        \STATE (3) $  A_1, A_2 , \cdots A_{t} \ \text{are pairwise disjoint}\}$.
        \FOR{ each $A_i$ in $\left( A_1, \cdots A_{t} \right)$ in $Q_l$}
            \STATE $\mathcal{S}$.append($A_i$) \&  weights.append($\frac{1}{|Q_l|}$).
        \ENDFOR
    \ENDIF
    $l \gets l + 1$.
\ENDWHILE
\RETURN{$\mathcal{S}$, weights}
\end{algorithmic}
\end{algorithm}
\end{minipage}
\begin{minipage}[t]{.48\textwidth}
\begin{algorithm}[H]
\caption{MIP Formulation  for recommendation set optimization (\rcm~ model)}
\label{alg:mip}
\begin{align*}
    \max &  \sum_{i \in \prodUniverse }\sum_{j \in \prodUniverse }  \hat{r}_{ij}  p_{ij} \\
    \text{s.t. } 
    & p_{ij} \leq x_{ij}\  \forall \ i,j \in \prodUniverse,  \\
    & p_{ij} \leq \frac{V_{\{i,j\}}}{v_0} p_{00} \ \forall \ i, j \in \prodUniverse  \\
    & p_{ij} +  \frac{V_{\{i,j\}}}{v_0}\left(1-x_{ij}\right) \geq  \frac{V_{\{i,j\}}}{v_0}p_{00} \ \forall \ i,j \in \prodUniverse\\
&    x_{ii} + x_{jj} -1 \leq x_{ij}  \leq \min(x_{ii},x_{jj}) \ \forall \ i,j \in \prodUniverse \\
   &p_{00} +  \sum_{i \in \prodUniverse}\sum_{j \in \prodUniverse}p_{ij} = 1 \\
  &  x_{ij}  \in \{0,1\} \ \forall \ i, j \ \in \prodUniverse \\
  & p_{ij} \geq 0 \ \forall \ i,j \in \prodUniverse \\
  & p_{00} \geq 0.
\end{align*}
\vspace{.08cm}
\end{algorithm}
\end{minipage}

If we want to estimate a \bmvlk \ model where $K$ is smaller than the size of some purchased bundles in the dataset, then we can pre-process these observations. In particular, we first partition the purchased bundle, which is of size larger than $K$, into subsets of size at most $K$, and augment each such subset as an additional observation. We consider all possible such partitions and assign them equal probability weights (see Algorithm~\ref{alg:est}). The MLE objective is also updated to incorporate these importance weights. In other words, the probability of purchasing bundle $\tilde{S}_l$ from offer set $C_l$ when $\tilde{S}_l > K$ can be written as: $\textrm{Prob}(\tilde{S}_l | C_l) = \frac{1}{|Q_l|}\sum_{\left( A_1, \cdots A_{t} \right) \in Q_l} \prod_{k=1}^{t} \textrm{Prob} (A_k|C_l)$, where each set $A_k$ satisfies $|A_k| \leq K$ needed for the model. Note that in the generative process assumed for this pre-processing step, each partitioning of the original $\tilde{S}_l$ is assigned equal likelihood. This is a choice that was made to balance simplicity while addressing the model-dataset incompatibility issue. The choice of weights do not change the nature of tractability of estimation. For instance, when the offered set $C_l$ is the same for each observation $l$, then the estimation problem remains convex even if the weights are assumed unequal. The estimates themselves would change. Changing the generative process to include latent variables could be another direction that we could take to improve on the current process, but we defer this to future work.


%% file: 92_hardness.tex
\vspace{.5cm} 

\section{The Revenue Maximization Problem: Hardness and Structural Results}
\label{sec:objective}



\subsection{Hardness of Unconstrained Optimization for \bmvl-2}\label{sec:hardness}

The revenue functions $\revtwo(\anyasmt)$ and $\revk(\anyasmt)$ have a form similar to the expected revenue of recommendations under the MNL model. For the MNL model, it is known that the unconstrained revenue optimization problem can be solved in linear time as the optimal set is a revenue-ordered set, i.e., it only contains the $l$ highest priced products for some $l \in \mathbb{Z}_{+}$. But for the \rcm \ model, we assert that this does not hold.

The proof of Theorem~\ref{thm:bmvl_hard} below follows from a reduction of the well-known MAXCUT problem.  While the existence of a polynomial time approximation algorithm for problem (\ref{eqn:AOproblem}) is an open question, we believe it is inapproximable because it  is similar to the quadratic knapsack problem with additional constraints on the coefficients (that can be both positive and negative). And it is known that the quadratic knapsack problem (defined on an edge-series parallel graph) is hard to approximate.

\proof{Proof of Theorem \ref{thm:bmvl_hard}.} Consider the decision version of the unconstrained \rcm \ optimization problem: 
\begin{align*}
 \max_{\anyasmt \in 2^\prodUniverse }  \revtwo(\anyasmt)  \geq \compthresh &
\Longleftrightarrow   \max_{\anyasmt \in  2^\prodUniverse } \sum_{i \in \prodUniverse} \sum_{j \in \prodUniverse} \theta_{ij}x_i^\anyasmt x_j^\anyasmt (\hat{r}_{ij} - \compthresh)  \geq \compthresh v_0 & \textsc{(Compare-Step)} \end{align*}

We will show that this decision version of the revenue optimization problem under the \rcm~ model is NP-complete by a reduction from \maxcut \ to this problem. Without loss of generality, we can assume that revenues of all products is less than $\compthresh$ (if not, then these products will be in the recommendation set corresponding to the solution of the optimization problem). Consider a graph $G$ with nodes $\{ 1, \cdots m \}$. We obtain a modified graph $G'$ by removing all the self-loops in $G$.  Let the adjacency matrix  of $G'$ be $A'$. Let $\mathbf{d} = (d_1, \cdots d_m)$ denote a $m$-dimensional vector with the $i$-th entry being the degree of node $i$ in the graph $G'$. Consider the following $(m+1) \times (m+1)$ matrix $Q = \bigl( \begin{smallmatrix}
    0 & \mathbf{d/2} \\
    \mathbf{d/2} & -A'
    \end{smallmatrix}\bigr)$ with a generic entry $q_{i,j}$ (in the $i$-th row and the $j$-th column). We index the entries of this matrix starting from 0 and the nodes of the graph starting from 1. Hence, for $i>0$, the $i$-th column of the $Q$ matrix corresponds to the $i$-th node in the graph $G'$. Consider the optimization problem:
\begin{equation}
        \label{eq:quad_opt}
         \argmax_{\anyasmt \in  2^\prodUniverse } \sum_{ 0 \leq i \leq m } \sum_{ 0 \leq j \leq m} q_{i,j} x_i^\anyasmt x_j^\anyasmt, 
\end{equation}
with solution $\anyasmt^*$. This optimization problem is equivalent to the \maxcut \ problem on the graph $G$ as shown below. 
    Note that the only positive values $q_{i,j}$  are either in the first row or the first column, hence $x_0^{C^*} = 1$. Now, 
   \ifisVerbose{  
    $ \begin{aligned}[t]
    \sum_{0 \leq i \leq m } \sum_{0 \leq j \leq m} q_{i,j} x_i^{\anyasmt^*} x_j^{\anyasmt^*} &=         2 \sum_{0 \leq i \leq m} \sum_{0 \leq j \leq m, \ j>i} q_{i,j} x_i^{\anyasmt^*} x_j^{\anyasmt^*} \\
    & = 2 \sum_{1 \leq j \leq m} q_{0,j} x_j^{\anyasmt^*} +  2 \sum_{1 \leq i \leq m} \sum_{1 \leq j \leq m, \ j>i} q_{i,j} x_i^{\anyasmt^*} x_j^{\anyasmt^*} \\
    &= 2\sum_{j \in \anyasmt^*} \frac{d_j}{2} - 2 E_{G'}\left(\Tilde{\anyasmt}^*, \Tilde{\anyasmt}^*\right) \\
    &=  E_{G'}\left(\Tilde{\anyasmt}^*, \{1, \cdots m \} \backslash \{ \Tilde{\anyasmt}^* \} \right) \\
    &=  E_{G}\left(\Tilde{\anyasmt}^*, \{1, \cdots m \} \backslash \{ \Tilde{\anyasmt}^* \} \right) 
    \end{aligned} $
} \else{
    $ \begin{aligned}[t]
\sum_{0 \leq i \leq m } \sum_{0 \leq j \leq m} q_{i,j} x_i^{\anyasmt^*} x_j^{\anyasmt^*} &=    2 \sum_{1 \leq j \leq m} q_{0,j} x_j^{\anyasmt^*} +  2 \sum_{1 \leq i \leq m} \sum_{1 \leq j \leq m, \ j>i} q_{i,j} x_i^{\anyasmt^*} x_j^{\anyasmt^*} \\
&= 2\sum_{j \in \anyasmt^*} \frac{d_j}{2} - 2 E_{G'}\left(\Tilde{\anyasmt}^*, \Tilde{\anyasmt}^*\right) 
  \hspace{5mm} =  E_{G}\left(\Tilde{\anyasmt}^*, \{1, \cdots m \} \backslash \{ \Tilde{\anyasmt}^* \} \right), 
\end{aligned} $
} \fi

where $E_{G'}(\anyasmt, \anyasmt')$ represents the number of edges between the set of nodes $\anyasmt$ and $\anyasmt'$ in the graph $G'$, and $ \Tilde{\anyasmt}^* = \anyasmt^*\backslash\{0\}$. The final expression equals the number of edges across the cut $\left(\Tilde{\anyasmt}^*, \{1, \cdots m \} \backslash \{ \Tilde{\anyasmt}^* \} \right)$ in the graph $G$. 
    
We can also see that problem \eqref{eq:quad_opt} can be transformed into  an equivalent \textsc{Compare-step} problem as follows: choose numbers $\compthresh, r_0, r_1, \cdots r_m$ such that $r_0 > \compthresh$ and $ \compthresh/2 > r_1 > r_2 \cdots r_m >0$. Let $  \theta_{ij} = \frac{q_{ij}}{r_i + r_j - \compthresh}, \ \ 0 \leq i, j \leq m$. Thus, the problem of finding the maximum cut on any graph can be transformed to the optimization problem in the \textsc{Compare-Step} of a \rcm \ optimization problem. 
Moreover, given a  solution of the \textsc{Compare-Step} optimization problem, the maximum cut of the corresponding graph is evident. Hence, the decision problem and subsequently the \rcm \ optimization problem are NP-complete. $\blacksquare$

\endproof

\subsection{Hardness of Unconstrained Optimization for \ucm}\label{sec:mmc-hardness}


\begin{theorem} \label{thm:mmc_hard} [Hardness Result for \ucm]
	The decision version of the unconstrained revenue optimization problem under the \ucm~ model (with number of allowed purchases $\leq 2$) is NP-complete.
\end{theorem}


\proof{Proof of Theorem \ref{thm:mmc_hard}.} The unconstrained revenue optimization under the \ucm~ model is given by the following problem:
\begin{align*}
\tag{\ucmao}
    \arg\max_{\anyasmt \in 2^\prodUniverse}  z_1 \sum_{i \in \anyasmt} r_i \frac{V_{\{i\}}}{V_{\{0\}} + \sum_{k \in \anyasmt} V_{\{k\}} } + z_2  \sum_{i, j \in \anyasmt, i < j}( r_i + r_j) \frac{V_{\{i,j\}}}{V_{\{0,0\}} + \sum_{k,l \in \anyasmt} V_{\{k,l\}} },
\end{align*}
where $z_1, z_2$ are the probability of purchasing one and two products respectively such that $z_1 + z_2 =1$. Although \cite{benson2018discrete} do not model the no purchase option, one way to incorporate the no-purchase option is to assign a utility to the no-purchase option for each size of the subset that can be chosen. The no-purchase option is then treated as an alternative which is always present irrespective of the recommendation set and is represented with the parameters $V_{\{0\}}$ and $V_{\{0,0\}} $. To establish the NP-completeness of \ucmao, we use the fact that revenue optimization under the mixture of two multinomial logits (\mmnlao) is NP-complete~\citep{rusmevichientong2010assortment}. This optimization problem is:
\begin{align*}
    \tag{\mmnlao}
    \arg\max_{\anyasmt \in 2^\prodUniverse} \alpha_1 \sum_{i \in \anyasmt} s_i \frac{v_i^1}{v^1_0 + \sum_{j \in \anyasmt} v^1_j} +  \alpha_2 \sum_{i \in \anyasmt} s_i \frac{v_i^2}{v^2_0 + \sum_{j \in \anyasmt} v^2_j},
\end{align*}
where the revenues of products are given by $s_1, \cdots , s_\numberOfItems$ with $s_i \in \mathbb{Z_+}  \ \forall \ i \in [\numberOfItems]$, the preference weights are $ \left( v_0^g, v_1^g, \cdots v_\numberOfItems^g \right)$ with $v_i^g \in \mathbb{Z_+} \forall \ i \in [\numberOfItems], \ g=1,2$, and the probability of a customer belonging to each of the mixture compoenents is $\left(\alpha_1 ,\alpha_2\right) \text{with } \alpha_g \in \mathbb{Q_+}, \ g = 1, 2$ and $\alpha_1 + \alpha_2 =1$. We claim that there is a reduction from a \mmnlao \ instance to a \ucmao \ instance and prove this in two steps:
\begin{enumerate}
    \item  Transform an instance of \mmnlao \ to an instance of \ucmao:
    Given an instance of \mmnlao, we define an instance of \ucmao \ by including an additional (\numberOfItems +1)-th product, which we refer to as a \emph{snowball}.
    The revenues of the products in the transformed instance are same as their original revenue and the snowball has zero revenue i.e. $r_i = s_i \ \forall i \in [\numberOfItems]$ and  $r_{\numberOfItems +1 } = 0$. 
    The probability of purchasing one and two products is equal to the probability of belonging to each group i.e. $z_g = \alpha_g , g =1, 2.$
    The preference weights in the transformed instance are given as: 
    \begin{enumerate}
        \item $V_{i} = v^1_i \ \forall \ i \in \{ 0, 1, \cdots, \numberOfItems \} $  and $V_{\numberOfItems +1 } = 0$
        \item \ifisVerbose { $V_{\{i,j\}} = \begin{cases}
            v^2_0  & \text{if } i=0, j=0  \\
            v^2_i & \text{if } j= \numberOfItems +1, i \in [\numberOfItems]   \\
            v^2_j & \text{if } i= \numberOfItems +1, j \in [\numberOfItems]   \\
            0 & \text{else}.
            \end{cases}$
        } \else { $V_{\{i,j\}} =  v^2_0   \text{ if } i=0, j=0; 
        v^2_i  \text{ if } j= \numberOfItems +1, i \in [\numberOfItems];
        v^2_j  \text { if } i= \numberOfItems +1, j \in [\numberOfItems];
        0  \text{ else.}$
    } \fi 
    \end{enumerate}
    
    \item Given a solution $S^*$ of the above instance of \ucmao \, obtain a solution for the original instance of \mmnlao: this can be done using Lemma \ref{lem:ucmao_mmnla0}.
\end{enumerate} 

Thus, any instance of the \mmnlao \  can be reduced to an instance of \ucmao, proving that \ucmao \ is also NP-complete. $\blacksquare$
    
\endproof
    
\begin{lemma} \label{lem:ucmao_mmnla0}
If $S^*$ is the solution for the above instance of \ucmao, then $S^*\backslash \{ \numberOfItems + 1\}$ is optimal for the \mmnlao \ problem. 
\end{lemma}
    
\proof{Proof.} It is easy to see  that $R_{\ucm}(S) \leq R_{\ucm}(S \cup \{\numberOfItems + 1\} ) \ \forall \ S \in 2^\prodUniverse $. Thus, without loss of generality, $\numberOfItems + 1 \in S^*$. We define $ R_{MMNL}(S) = \alpha_1 \sum_{i \in S} s_i \frac{v_i^1}{v^1_0 + \sum_{j \in S} v^1_j} +  \alpha_2 \sum_{i \in S} s_i \frac{v_i^2}{v^2_0 + \sum_{j \in S} v^2_j}.$
Thus, with the parameters specified as above, $R_{MMNL}(S) = R_{\ucm}(S \cup \{n+1\} ), \ \forall \ S \in 2^\prodUniverse $. Assume $ \hat{S} \neq S^*\backslash \{ \numberOfItems + 1\}$ is the solution  for \mmnlao. Thus, $ R_{\ucm}(\hat{S} \cup \{  n+1 \} ) =  R_{MMNL}(\hat{S}) > R_{MMNL}(S^*) = R_{\ucm}(S^*)$ contradicting the assumption that $S^*$ is the solution to \ucmao.    $\blacksquare$
\endproof

\subsection{Structural Properties of the Optimal Unconstrained Solution for \bmvl-2}\label{sec:structural}

Under the setting $\feasibleAsmtSet = 2^{\prodUniverse}$, i.e., for the unconstrained optimization problem  (\ref{eqn:AOproblem}), we prove the following structural properties satisfied by the optimal solution $\offeredAsmt^*$: 

\begin{lemma} \label{lem:shdbelong} For all products $i \in \prodUniverse$ that are not in any optimal recommendation set $\offeredAsmt^*, \ r_i \leq \revtwo(\offeredAsmt^*)$. Equivalently, $\offeredAsmt^*_u \subseteq \offeredAsmt^*$, where $\offeredAsmt^*_u = \{i: r_i > \revtwo(\offeredAsmt^*) \}$.
\end{lemma}

\begin{lemma} \label{lem:shdnotbelong} Let $\offeredAsmt^*$ be optimal. For every $i \in \offeredAsmt^*, \ \exists \ j(i) \in \offeredAsmt^*$, where $j(i) \neq i$ and $r_i + r_{j(i)} \geq \revtwo(\offeredAsmt^*)$. 
\end{lemma}
\textit{Remark}: If $\offeredAsmt^*$ is an optimal recommendation set of the smallest cardinality, then $\forall \ i \in \offeredAsmt^*, \ \exists \ j(i) \in \offeredAsmt^*$, where $j(i) \neq i$ and $r_i + r_{j(i)} > \revtwo(\offeredAsmt^*)$.

\begin{lemma} \label{lem:monotone} Let the $i$-th revenue-ordered recommendation set be defined as $A_{i}= \{1, 2, \cdots i \}, i \in \prodUniverse$. Then, the revenue of revenue-ordered recommendation sets increases monotonically as long as the price of all the products in the revenue-ordered recommendation set is greater than $\revtwo(\offeredAsmt^*)$, i.e., $\revtwo(A_1) \leq \revtwo(A_2) \cdots \leq \revtwo(A_k)$ where $r_k > \revtwo(\offeredAsmt^*) \geq r_{k+1}$.
\end{lemma}

Lemma~\ref{lem:shdbelong} says that if a product's revenue is greater than the optimal revenue, then it belongs to the optimal recommendation set. Lemma~\ref{lem:shdnotbelong} suggests that a product that is in the optimal recommendation set has a corresponding pair that also belongs to the optimal set such that the sum of their revenues is greater than the expected revenue of the set. Finally, Lemma~\ref{lem:monotone} suggests that the objective function has a partial monotonicity property. Overall, these three properties can help us narrow the search for the optimal recommendations. For instance, if an algorithm keeps an estimate of an upper bound on the optimal revenue, then this can help prune the search space based on Lemma~\ref{lem:shdbelong}.
We start by decomposing the revenue function.

\noindent \begin{lemma}
For two sets $\offeredAsmt$ and $\offeredAsmt'$ such that $\offeredAsmt \cap \offeredAsmt'$ is the empty set, we have:
\begin{equation}
\label{eqn:revDecomp}
    \revtwo(\offeredAsmt \cup \offeredAsmt')=  \alpha \revtwo(\offeredAsmt) + (1-\alpha) T(\offeredAsmt, \offeredAsmt'), 
\end{equation}
where $\alpha = \frac{ v_0 + \sum_{i \in \offeredAsmt} \sum_{j \in \offeredAsmt}  \theta_{ij}}{ v_0 + \sum_{i \in \offeredAsmt \cup \offeredAsmt'} \sum_{j \in \offeredAsmt \cup \offeredAsmt'}  \theta_{ij} }$ is a value between 0 and 1, 
and the function 
$T(\offeredAsmt, \offeredAsmt')$ is defined as  \\ $T(\offeredAsmt, \offeredAsmt') = \frac{\sum_{i \in \offeredAsmt'} \sum_{j \in \offeredAsmt'} \hat{r}_{ij} \theta_{ij} + 2\sum_{i \in \offeredAsmt} \sum_{j \in \offeredAsmt'} \hat{r}_{ij} \theta_{ij} }{\sum_{i \in \offeredAsmt'} \sum_{j \in \offeredAsmt'}  \theta_{ij} + 2\sum_{i \in \offeredAsmt} \sum_{j \in \offeredAsmt'}  \theta_{ij}} $. 
\end{lemma}
\proof{Proof.}
\begin{align*}
    \revtwo(\offeredAsmt \cup \offeredAsmt')&= \frac{ \sum_{i \in \offeredAsmt \cup \offeredAsmt'} \sum_{j \in \offeredAsmt \cup \offeredAsmt'} \hat{r}_{ij} \theta_{ij}   }{ v_0 + \sum_{i \in \offeredAsmt \cup \offeredAsmt'} \sum_{j \in \offeredAsmt \cup \offeredAsmt'}  \theta_{ij} } \\
    &=   \left(\frac{\sum_{i \in \offeredAsmt} \sum_{j \in \offeredAsmt} \hat{r}_{ij} \theta_{ij}}{ v_0 + \sum_{i \in \offeredAsmt} \sum_{j \in \offeredAsmt}  \theta_{ij}} \right) \left(\frac{ v_0 + \sum_{i \in \offeredAsmt} \sum_{j \in \offeredAsmt}  \theta_{ij}}{ v_0 + \sum_{i \in \offeredAsmt \cup \offeredAsmt'} \sum_{j \in \offeredAsmt \cup \offeredAsmt'}  \theta_{ij} }\right)  + \\
    & \left(\frac{\sum_{i \in \offeredAsmt'} \sum_{j \in \offeredAsmt'} \hat{r}_{ij} \theta_{ij} + 2\sum_{i \in \offeredAsmt} \sum_{j \in \offeredAsmt'} \hat{r}_{ij} \theta_{ij} }{\sum_{i \in \offeredAsmt'} \sum_{j \in \offeredAsmt'}  \theta_{ij} + 2\sum_{i \in \offeredAsmt} \sum_{j \in \offeredAsmt'}  \theta_{ij}} \right) \left(\frac{ \sum_{i \in \offeredAsmt'} \sum_{j \in \offeredAsmt'}  \theta_{ij} + 2\sum_{i \in \offeredAsmt} \sum_{j \in \offeredAsmt'}  \theta_{ij}}{ v_0 + \sum_{i \in \offeredAsmt \cup \offeredAsmt'} \sum_{j \in \offeredAsmt \cup \offeredAsmt'}  \theta_{ij} }\right) \\
    &= \alpha \revtwo(\offeredAsmt) + (1-\alpha) T(\offeredAsmt, \offeredAsmt') \quad\quad\blacksquare
\end{align*}
\endproof

Given the above decomposition, the proofs of the Lemmas~\ref{lem:shdbelong}-\ref{lem:monotone} are provided below.

\proof{Proof of Lemma \ref{lem:shdbelong}.} 
Let $i \notin \offeredAsmt^*$ such that $r_i > \revtwo(\offeredAsmt^*)$.
We know, $\revtwo(\offeredAsmt^* \cup i) = \alpha \revtwo(\offeredAsmt^*) + (1-\alpha) T(\offeredAsmt^*, \{i\}) $
for some $0 \leq \alpha \leq 1$.
As $r_i > \revtwo(\offeredAsmt^*)$, we have $T(\offeredAsmt^*, \{i\}) \geq r_i > \revtwo(\offeredAsmt^*)$.
As $\revtwo(\offeredAsmt^* \cup i)$ is a convex combination of $\revtwo(\offeredAsmt^*)$ and $T(\offeredAsmt^*, \{i\})$, it is greater than $\revtwo(\offeredAsmt^*)$, contradicting the optimality of the recommendation set $\offeredAsmt^*$. 
$\blacksquare$
\endproof

\proof{Proof of Lemma \ref{lem:shdnotbelong}.}
Suppose $\exists \ i \in \offeredAsmt^*$ such that $r_i + r_j < \revtwo(\offeredAsmt^*) \ \forall \ j \in \offeredAsmt^* \backslash i$.
This implies that $T(\offeredAsmt^* \backslash i, \{i\}) < \revtwo(\offeredAsmt^*).$
We know that $\revtwo(\offeredAsmt^*)$ is a convex combination of $\revtwo(\offeredAsmt^* \backslash i )$ and $T(\offeredAsmt^* \backslash i, \{i\})$.
Thus, $\revtwo(\offeredAsmt^* \backslash i) > \revtwo(\offeredAsmt^*)$ contradicting the optimality of the recommendation set $\offeredAsmt^*$. 
$\blacksquare$
\endproof

\proof{Proof of Lemma \ref{lem:monotone}.}
Using (\ref{eqn:revDecomp}), we can decompose the revenue of the revenue-ordered recommendation set $A_m$ as 
$\revtwo(A_m) =  \alpha \revtwo(A_{m-1}) + (1-\alpha)T(A_{m-1}, \{m\}),$
for some $0 \leq \alpha \leq 1$. Also, note that $T(A_{m-1}, \{m\}) \geq r_m$.
Further, $r_m > \revtwo(\offeredAsmt^*)$ for $m \leq k$. 
Thus, $T(A_{m-1}, \{m\}) > \revtwo(\offeredAsmt ^*) \geq \revtwo(A_{m-1})$ for $m \leq k$.
But $\revtwo(A_m)$ is a convex combination of $T(A_{m-1}, \{m\})$ and $\revtwo(A_{m-1})$. Thus, $\revtwo(A_m) \geq \revtwo(A_{m-1})$ for $m\leq k$.
$\blacksquare$
\endproof

%% file: 93_algorithms.tex
\section{Revenue Maximizing Recommendations: Additional Details} 
\label{sec:algorithms-full}


\subsection{Binary Search with Efficient Comparisons}\label{sec:algo-efficient}

For any given tolerance $\epsilon >0$, our initial algorithm (Alg.~\ref{alg:bin_srch_outline}) gives an $\epsilon$-optimal solution, i.e., a solution within $\epsilon$ of the optimal value. In each iteration of the search process, we narrow the size of the interval in which $R(\offeredAsmt^*)$ lies as outlined in \binsearch~(Alg.~\ref{alg:bin_srch_outline}). The upper bound on $R(\offeredAsmt^*)$ is initialized as the maximum revenue possible from a bundle of two products i.e. $r_1 + r_2$. The optimal recommendation set is arbitrarily initialized as $\{ 1\}$ and is of relevance only when the optimal revenue is less than $\epsilon$, in which case all recommendation sets have revenue within $\epsilon$ of the optimal recommendation set.


\begin{figure}
\centering
\begin{minipage}[t]{.45\textwidth}
\begin{algorithm}[H]
\caption{\binsearch}
\label{alg:bin_srch_outline}
\begin{algorithmic}[1] 
\REQUIRE{ Parameters $\{r_{i}\}_{i=1}^{n}$, $\{\theta_{ij}\}_{i=1, j=1}^{n}$, tolerance level $\epsilon > 0$, and feasible sets $\feasibleAsmtSet $. } \\
\STATE{$L_1 = 0, U_1 = r_1 +r_2 , j=1,$ and $ \offeredAsmt^* = \{1 \} $.}\\ 
\WHILE{ $U_j - L_j > \epsilon$} 
\STATE{ $\compthresh_j = (L_j + U_j)/2 $.} \\ \label{alg:compare-step}
\IF{$\compthresh_j \leq  \max_{C \in \feasibleAsmtSet } R(C)$ } 
\STATE{$L_{j+1} =  \compthresh_j, U_{j+1} = U_j$.}
\STATE{Pick any $\offeredAsmt^* \in \{ C \in \feasibleAsmtSet :R(C) \geq \compthresh_j \}$.}
\ELSE 
\STATE{$L_{j+1} = L_j , U_{j+1} = \compthresh_j$.} 
\ENDIF 
\STATE{Increment $j$ by 1.}
\ENDWHILE
\RETURN{$ \offeredAsmt^*$}
\vspace{1.42in}
\end{algorithmic}
\end{algorithm}
\end{minipage}
\hspace{.1cm}
\begin{minipage}[t]{.45\textwidth}
\begin{algorithm}[H]
\caption{\binsearcheff}
\label{alg:bin_srch_outline_eff}
\begin{algorithmic}[1] 
\REQUIRE{ Parameters $\{r_{i}\}_{i=1}^{n}$,  $\{\theta_{ij}\}_{i=1, j=1}^{n}$, tolerance level $\epsilon > 0$, and feasible sets $\feasibleAsmtSet $.} \\
\STATE{$U_1 = r_1 +r_2 , j=1, i=1$ and $ \offeredAsmt^* = \{1 \} $.}\\ 
\STATE{\textbf{while} $r_{i+1}  \geq r_{i}$ \textbf{do}  Increment $i$ by 1.}
\STATE{$L_1 = r_{i+1}.$}
\WHILE{ $U_j - L_j > \epsilon$} 
\STATE{ $\compthresh_j = (L_j + U_j)/2 $.} \\
\IF{$\compthresh_j \leq  \max_{C \in \feasibleAsmtSet } R(C)$ \label{alg:compare-step-again}} 
\STATE{$L_{j+1} =  \compthresh_j, U_{j+1} = U_j$.}
\STATE{Pick a $\offeredAsmt^*$ such that $R(\offeredAsmt^*) \geq \compthresh_j, \overline{B} \subset \offeredAsmt^*, $ and $\ \underline{B} \cap \offeredAsmt^* = \phi $; where $ \overline{B} = \{ i: r_i > U \},$ and $ \underline{B} = \{ i: r_i + r_1 < L \} $.}
\ELSE 
\STATE{$L_{j+1} = L_j , U_{j+1} = \compthresh_j$.} 
\ENDIF 
\STATE{Increment $j$ by 1.}
\ENDWHILE
\RETURN{$ \offeredAsmt^*$}
\end{algorithmic}
\end{algorithm}
\end{minipage}
\end{figure}

\noindent\textbf{Using Structural Properties of $\offeredAsmt^*$.} 
\binsearch~can be made more efficient by using the properties of the optimal recommendation set derived earlier (Lemma~\ref{lem:shdbelong}-\ref{lem:monotone}). 
At the cost of additional pre-processing (which is small), we can start with a lower bound greater than 0 based on Lemma~\ref{lem:monotone}. Let $l$ be the maximum index $i$ such that $R(A_1) \leq R(A_2) \cdots \leq R(A_i)$. Then, from  Lemma~\ref{lem:monotone}, we know that $l \geq k$ (see the definition of $k$ in the Lemma). Thus, $r_k \geq R(\offeredAsmt^*) \geq r_{k+1} \geq r_{l+1}$. Hence, at the beginning of the binary search, the lower bound $L$ can be set as $r_{l+1}$. Lemmas \ref{lem:shdbelong} and \ref{lem:shdnotbelong} can be used to make the comparison step faster. In particular, when we have an upper bound $U$ on $R(\offeredAsmt^*)$, then using Lemma \ref{lem:shdbelong}, we know that all products with revenue greater than $U$ should belong to the optimal recommendation set. Similarly, with a lower bound $L$ on the revenue of the optimal recommendation set, we know that all products $i$ such that $r_i + r_1 < L$, cannot belong to the optimal recommendation set. These observations predetermine the fate of some products, reducing the problem size (sometimes significantly as seen in our experiments) in the comparison step (\ref{eqn:comparison}).
\binsearcheff{} incorporates these properties as shown in Algorithm \ref{alg:bin_srch_outline_eff}.

\noindent\textbf{QUBO Heuristics and Noisy Binary Search.} 
Though the QUBO problem is an NP-hard problem~\citep{pardalos1992complexity} as discussed before, there has been ample research in heuristic algorithms that return high quality solutions in extremely reasonable computation times~\citep{dunning2018works}. This makes it appealing to use these approximate algorithms in solving the \textsc{Compare-Step}. Nonetheless, solving problem (\ref{eqn:comparison}) approximately can potentially lead to narrowing down on an incorrect interval in the binary search outer loop. We take two steps to alleviate this issue.
Firstly, for each QUBO problem, we run multiple QUBO heuristics in parallel. 
The binary search interval will have an incorrect update only if all the heuristics lead to an incorrect answer for the \textsc{Compare-Step}. 

Secondly, we further robustify the binary search outer loop by using a noisy binary search variant~\citep{burnashev1974interval}. Here one maintains a distribution over the unknown $R(\offeredAsmt^*)$, and each comparison (with the median of the current distribution) is used to obtain an updated distribution using Bayes rule. This prevents incorrect comparison step outcomes from easily misleading the search process. For any specified tolerance level and a probability value with which the given solution needs to lie within the tolerance level, the number of iterations required for the noisy binary search still stays logarithmic. We refer to this algorithm as \noisybinsearch \ and its version which uses the structural properties as \noisybinsearcheff~(we omit its description due to space constraints).

\subsection{Optimization with Linear Constraints}\label{sec:linear-constr}

Constraints on the feasible recommendation sets are common in practice. For example, there can be a \emph{cardinality} constraint on the maximum number of products in a recommendation set due to webpage/screen size limits in the online e-commerce setting. Business rules and obligations such as ensuring sufficient representation from various sub-groups of products, or the requirement to maintain a precedence order among products can also be formulated as linear constraints \citep{davis2013assortment,sinha2017optimizing}.  We can incorporate linear constraints in the following way. For a general linear constraint set $Dx = e $, the \textsc{Compare-Step} becomes
\begin{align*} 
 \max_{\{x \in \{0,1\}^\numberOfItems: Dx = e\}} \sum_{i \in \prodUniverse} \sum_{j \in \prodUniverse} \theta_{ij}x_i x_j (\hat{r}_{ij} - \compthresh)  \geq \compthresh v_0. 
\end{align*}

This is a quadratic binary optimization problem with constraints that can be relaxed to get $$ \max_{x \in \{0,1\}^\numberOfItems} \sum_{i \in \prodUniverse} \sum_{j \in \prodUniverse} \theta_{ij}x_i x_j (\hat{r}_{ij} - \compthresh) +   \lambda(Dx - e)'(Dx  - e), $$ for a suitably large $\lambda < 0$. In this form, the aforementioned QUBO solvers can be used directly. 

If we have an inequality constraint, then as long as the components of $D$ and $e$ are integral, a similar transformation can be done with an appropriate number of slack variables \citep{glover2018tutorial}. For instance, suppose we want to ensure that the size of the recommendation set is at most $e \in \mathbb{Z}_{+}$, i.e., we have the constraint $\mathbf{1}'x \leq e$ (here, $\mathbf{1}$ is the vector of all ones). Then, using slack variables $s_1,...,s_{e}$, we get the following readily solvable QUBO instance:
\begin{equation}
\label{eq:cap_QUBO}
\max_{x \in \{0,1\}^\numberOfItems} \sum_{i \in \prodUniverse} \sum_{j \in \prodUniverse} \theta_{ij}x_i x_j (\hat{r}_{ij} - \compthresh) + \lambda(\mathbf{1}'x + \sum_{i=1}^{e}s_i - e)'(\mathbf{1}'x + \sum_{i=1}^{e}s_i - e). 
\end{equation}

\subsection{Guarantees for Benchmark Algorithms}\label{sec:main-benchmarks}



\begin{remark}
The MIP formulation builds on an earlier formulation for the \emph{mixture of MNLs} model studied in~\cite{blanchet2016markov}, and is described in Algorithm ~\ref{alg:mip}. Linear constraints can be incorporated into the formulation as well.
\end{remark}

 
\begin{lemma}
    \label{lem:adxoptSoln}
    Let $\widehat{\offeredAsmt}$ be the solution returned by  \adxopttwo~ using the \rcm~ model. Then, $ \widehat{\offeredAsmt} \supset \offeredAsmt^*_u $, where $\offeredAsmt^*_u = \{i: r_i > \revtwo(\offeredAsmt^*) \}$.
\end{lemma}

\proof{Proof of Lemma \ref{lem:adxoptSoln}.}
Using the arguments in proof of Lemma \ref{lem:shdbelong} and the local optimality of $\widehat{\offeredAsmt}$, we have $ \widehat{\offeredAsmt}_u \subset \widehat{\offeredAsmt} $, where $ \widehat{\offeredAsmt}_u = \{i: r_i > \revtwo(\widehat{\offeredAsmt}) \}.$
As $\revtwo(\widehat{\offeredAsmt}) \leq \revtwo(\offeredAsmt^*)$, $ \offeredAsmt^*_u \subset \widehat{\offeredAsmt}_u \subset \widehat{\offeredAsmt}$.
$\blacksquare$
\endproof

\begin{remark} The (worst-case) time complexity of the \adxopttwo \ algorithm is $O(\numberOfItems^7)$, which is prohibitive for medium to large instances as observed in our experiments.\end{remark}


\begin{assumption} \label{assume:val_consc} Model parameters for value conscious customers, i.e., those who prefer cheaper product(s) but derive higher value (product of price and utility) from higher priced product(s), satisfy:
    \begin{enumerate}[(a)] 
        \item $\Vi \leq \Vj$ and $ \Vik \leq \Vjk \ \forall i<j,  i\neq k, j\neq k$ and $ i, j, k \in \prodUniverse$
        \item $r_i \Vi \geq r_j \Vj $ and $\left( r_i +r_k \right)\Vik \geq \left( r_j +r_k \right)\Vjk \ \forall i < j,  i\neq k, j\neq k$ and $ i, j, k \in \prodUniverse$.
    \end{enumerate}
\end{assumption}

\begin{theorem} \label{thm:valCons}
    For value conscious customers, the revenue-ordered heuristic produces an optimal recommendation set for the optimization problem $\max_{ |\offeredAsmt| \leq \capconst } \revtwo(\offeredAsmt) $ for any $d >0$.
\end{theorem}

\proof{Proof of Theorem \ref{thm:valCons}.} Let $\offeredAsmt^*$ be an optimal recommendation set for the above problem. We will construct a revenue-ordered  recommendation set which has revenue  $\revtwo(\offeredAsmt^*)$. If  $\offeredAsmt^*$  is a revenue-ordered recommendation set, then the theorem is trivially true. Hence, we focus on the case when $\offeredAsmt^*$ is not a revenue-ordered recommendation set. Then, there exists products $l, \ m$ such that $l \in \offeredAsmt^*$ and $  m \notin \offeredAsmt^*$ for some $1 \leq m <l\leq \numberOfItems $.

\noindent Let $\Tilde{\offeredAsmt} =  \{ m\} \cup  \offeredAsmt^*   \backslash \{l\} $. Thus, we have
\ifisVerbose{
\begin{align*}
\revtwo(\Tilde{\offeredAsmt})    &=   \frac{ \sum_{i \in \Tilde{\offeredAsmt}} r_i \Vi   +  \sum_{i \in \Tilde{\offeredAsmt}} \sum_{j \in \Tilde{\offeredAsmt}, j >i } (r_i + r_j)\Vij  }{v_0 +   \sum_{i \in \Tilde{\offeredAsmt}} \Vi  + \sum_{i \in \Tilde{\offeredAsmt}} \sum_{j \in \Tilde{\offeredAsmt} , j >i } \Vij} \\ 
&=  \frac{ \sum_{i \in \offeredAsmt^* \backslash \{l\}} r_i \Vi   +  \sum_{i \in \offeredAsmt^* \backslash \{l\}} \sum_{j \in \offeredAsmt^* \backslash \{l\}, j >i } (r_i + r_j)\Vij  + r_m \Vm + \sum_{i \in \offeredAsmt^* \backslash \{l\}, i \neq m}  (r_i + r_m)\Vim   }{v_0 +   \sum_{i \in \Tilde{\offeredAsmt}} \Vi  +  \sum_{i \in \offeredAsmt^* \backslash \{l\}} \sum_{j \in \offeredAsmt^* \backslash \{l\}, j >i } \Vij + \Vm + \sum_{i \in \offeredAsmt^* \backslash \{l\}, i \neq m}  \Vim } \\
 &\geq \frac{ \sum_{i \in \offeredAsmt^* \backslash \{l\}} r_i \Vi   +  \sum_{i \in \offeredAsmt^* \backslash \{l\}} \sum_{j \in \offeredAsmt^* \backslash \{l\}, j >i } (r_i + r_j)\Vij  + r_l \Vl + \sum_{i \in \offeredAsmt^* \backslash \{l\}, i \neq l}  (r_i + r_l)\Vil   }{v_0 +   \sum_{i \in \Tilde{\offeredAsmt}} \Vi  +  \sum_{i \in \offeredAsmt^* \backslash \{l\}} \sum_{j \in \offeredAsmt^* \backslash \{l\}, j >i } \Vij + \Vl + \sum_{i \in \offeredAsmt^* \backslash \{l\}, i \neq l}  \Vil } \\ 
 &= \revtwo(\offeredAsmt^*).
\end{align*}
} \else{
    \begin{align*}
    \revtwo(\Tilde{\offeredAsmt})    
    &=  \frac{ \sum_{i \in \offeredAsmt^* \backslash \{l\}} r_i \Vi   +  \sum_{i \in \offeredAsmt^* \backslash \{l\}} \sum_{j \in \offeredAsmt^* \backslash \{l\}, j >i } (r_i + r_j)\Vij  + r_m \Vm + \sum_{i \in \offeredAsmt^* \backslash \{l\}, i \neq m}  (r_i + r_m)\Vim   }{v_0 +   \sum_{i \in \Tilde{\offeredAsmt}} \Vi  +  \sum_{i \in \offeredAsmt^* \backslash \{l\}} \sum_{j \in \offeredAsmt^* \backslash \{l\}, j >i } \Vij + \Vm + \sum_{i \in \offeredAsmt^* \backslash \{l\}, i \neq m}  \Vim } \\
    &\geq \revtwo(\offeredAsmt^*).
    \end{align*} } \fi
As $\offeredAsmt^*$ is an optimal recommendation set, $\revtwo(\Tilde{\offeredAsmt}) \leq \revtwo(\offeredAsmt^*)$. Hence, we conclude $\revtwo(\Tilde{\offeredAsmt}) = \revtwo(\offeredAsmt^*)$.
We can use the above argument repeatedly to construct a sequence of recommendation sets, each having revenue equal to $\revtwo(\offeredAsmt^*)$ until a revenue-ordered recommendation set is obtained. $\blacksquare$
\endproof


\begin{assumption} \label{assume:small_vij} Model parameters satisfy: $ \max_{ i,j \in \prodUniverse,  i \neq j } \Vij \leq \epsilon \min_{k \in \prodUniverse \cup {\phi}} V_{\{k\}}$.
\end{assumption}

\begin{theorem} \label{thm:small_vij}
    Under Assumption \ref{assume:small_vij}, the revenue-ordered heuristic satisfies:  $\revtwo(\offeredAsmt_{revord}^*) \geq \frac{2 - \epsilon |\offeredAsmt^*_{MNL}|  }{2 + 4\epsilon|\offeredAsmt^*|}   \revtwo(\offeredAsmt^*)$, where $\offeredAsmt^*_{revord} \in \argmax_{ \offeredAsmt \in \{ A_1, A_2 \cdots A_\numberOfItems \} } \revtwo(\offeredAsmt) $, and $\offeredAsmt^*_{MNL}$ and $\offeredAsmt^*$ are the optimal solutions of the unconstrained problem under the MNL and the \rcm{} models respectively.
\end{theorem}

\noindent Prior to giving proof of Theorem \ref{thm:small_vij}, we define some notation and lemmas that will be useful. We define $R_{MNL}(\offeredAsmt) = \frac{\sum_{i \in \offeredAsmt} V_{\{i\}} r_i}{v_0 + \sum_{i \in \offeredAsmt} V_{\{i\}}},  \ \offeredAsmt^*_{MNL} \in  \argmax_{\offeredAsmt \in \feasibleAsmtSet} R_{MNL}(\offeredAsmt),  \ \offeredAsmt^*_{revord} \in \argmax_{ \offeredAsmt \in \{ A_1, A_2 \cdots A_\numberOfItems \} } \revtwo(\offeredAsmt)  \text{ and }  \offeredAsmt^*  \in  \argmax_{\offeredAsmt \in \feasibleAsmtSet} R(\offeredAsmt) .$
When $\feasibleAsmtSet = 2^\prodUniverse$, we take  $\offeredAsmt^*_{MNL}$ to be a revenue-ordered set.

\noindent \begin{lemma}
    \label{lem:MNL_ub}
    Under Assumption \ref{assume:small_vij}, $   R_{MNL}(\offeredAsmt) \leq \frac {2}{2 -  \epsilon  |C|}\revtwo(\offeredAsmt).$
\end{lemma}
\proof{Proof.}
\ifisVerbose{
$\begin{aligned}[t]
R_{MNL}(\offeredAsmt) - \revtwo(\offeredAsmt)  \ &= \frac{\sum_{i \in \offeredAsmt} \Vi r_i}{v_0 + \sum_{i \in \offeredAsmt} \Vi} - \frac{\sum_{i \in \offeredAsmt} \Vi r_i + \sum_{i, j \in \offeredAsmt, j >i } \Vij (r_i + r_j)}{v_0 +\sum_{i \in \offeredAsmt} \Vi +  \sum_{i, j \in \offeredAsmt, j >i } \Vij} \\
&= \frac{  \left( \sum_{i, j \in \offeredAsmt, j >i } \Vij \right)  \sum_{i \in \offeredAsmt} \Vi r_i  - \left( v_0 + \sum_{i \in \offeredAsmt} \Vi \right) \sum_{i, j \in \offeredAsmt, j >i } \Vij (r_i + r_j)  }{ \left( v_0 + \sum_{i \in \offeredAsmt} \Vi \right) \left(v_0 +\sum_{i \in \offeredAsmt} \Vi +  \sum_{i, j \in \offeredAsmt, j >i } \Vij \right) } \\
&\leq \frac{  \left( \sum_{i, j \in \offeredAsmt, j >i } \Vij \right)  \sum_{i \in \offeredAsmt} \Vi r_i    }{ \left( v_0 + \sum_{i \in \offeredAsmt} \Vi \right) \left(v_0 +\sum_{i \in \offeredAsmt} \Vi +  \sum_{i, j \in \offeredAsmt, j >i } \Vij \right) } \\
&\leq \frac{  \left( \sum_{i, j \in \offeredAsmt, j >i } \Vij \right)  \sum_{i \in \offeredAsmt} \Vi r_i    }{ \left( v_0 + \sum_{i \in \offeredAsmt} \Vi \right)^2} \\
&\leq \frac{  \left( \sum_{i, j \in \offeredAsmt, j >i } \epsilon \underbar{V} \right)  \sum_{i \in \offeredAsmt} \Vi r_i    }{ \left( v_0 + \sum_{i \in \offeredAsmt} \Vi \right)^2} \\
&= \frac{ |C| \left( |\offeredAsmt| - 1 \right) \epsilon \underbar{V} \sum_{i \in \offeredAsmt} \Vi r_i}{2\left( v_0 + \sum_{i \in \offeredAsmt} \Vi \right)^2} \\
&\leq \frac{ |\offeredAsmt| \left( |\offeredAsmt| - 1 \right) \epsilon }{2 \left(1+|\offeredAsmt| \right)} R_{MNL}(\offeredAsmt)\\
& \leq \frac{ \epsilon |\offeredAsmt|   }{2}R_{MNL}(\offeredAsmt) \end{aligned} $
where $ \underbar{V} = \min_{k \in \prodUniverse \cup {\phi}} V_{\{k\}} $.
} \else {
$\begin{aligned}[t]
R_{MNL}(\offeredAsmt) - \revtwo(\offeredAsmt)
&\leq \frac{  \left( \sum_{i, j \in \offeredAsmt, j >i } \Vij \right)  \sum_{i \in \offeredAsmt} \Vi r_i    }{ \left( v_0 + \sum_{i \in \offeredAsmt} \Vi \right) \left(v_0 +\sum_{i \in \offeredAsmt} \Vi +  \sum_{i, j \in \offeredAsmt, j >i } \Vij \right) } \\
& \leq \frac{ |\offeredAsmt| \left( |\offeredAsmt| - 1 \right) \epsilon \left( \min_{k \in \prodUniverse \cup {\phi}} V_{\{k\}} \right) \sum_{i \in \offeredAsmt} \Vi r_i}{2\left( v_0 + \sum_{i \in \offeredAsmt} \Vi \right)^2}  \hspace{5mm}  \leq \;\;\frac{ \epsilon |\offeredAsmt|   }{2}R_{MNL}(\offeredAsmt). \quad \blacksquare
\end{aligned} $ } \fi
\endproof

\begin{lemma}
    \label{lem:MNL_lb}
    Under Assumption \ref{assume:small_vij}, $ \revtwo(\offeredAsmt)  \leq \left( 1+  2 \epsilon  |C| \right) R_{MNL}(\offeredAsmt).$
\end{lemma}
\proof{Proof.}
\ifisVerbose{
$\begin{aligned}[t]
\revtwo(\offeredAsmt) - R_{MNL}(\offeredAsmt) &=\frac{ \left( v_0 + \sum_{i \in \offeredAsmt} \Vi \right) \sum_{i, j \in \offeredAsmt, j >i } \Vij (r_i + r_j) - \left( \sum_{i, j \in \offeredAsmt, j >i } \Vij \right)  \sum_{i \in \offeredAsmt} \Vi r_i    }{ \left( v_0 + \sum_{i \in \offeredAsmt} \Vi \right) \left(v_0 +\sum_{i \in \offeredAsmt} \Vi +  \sum_{i, j \in \offeredAsmt, j >i } \Vij \right) } \\ 
&\leq  \frac{ \left( v_0 + \sum_{i \in \offeredAsmt} \Vi \right) \sum_{i, j \in \offeredAsmt, j >i } \Vij (r_i + r_j)   }{ \left( v_0 + \sum_{i \in \offeredAsmt} \Vi \right) \left(v_0 +\sum_{i \in \offeredAsmt} \Vi +  \sum_{i, j \in \offeredAsmt, j >i } \Vij \right) } \\ 
&\leq \frac{  \sum_{i, j \in \offeredAsmt, j >i } 2 \epsilon \Vi r_i  }{ v_0 +\sum_{i \in \offeredAsmt} \Vi +  \sum_{i, j \in \offeredAsmt, j >i } \Vij  } \\
& \leq \frac{  2 \epsilon |\offeredAsmt | \sum_{i\in \offeredAsmt }  \Vi r_i  }{ v_0 +\sum_{i \in \offeredAsmt} \Vi  } \\
&= 2 \epsilon |\offeredAsmt | R_{MNL}(\offeredAsmt).
\end{aligned} $ 
} \else{ 
$\begin{aligned}[t]
\revtwo(\offeredAsmt) - R_{MNL}(\offeredAsmt) 
&\leq  \frac{ \left( v_0 + \sum_{i \in \offeredAsmt} \Vi \right) \sum_{i, j \in \offeredAsmt, j >i } \Vij (r_i + r_j)   }{ \left( v_0 + \sum_{i \in \offeredAsmt} \Vi \right) \left(v_0 +\sum_{i \in \offeredAsmt} \Vi +  \sum_{i, j \in \offeredAsmt, j >i } \Vij \right) } \\ 
& \leq \frac{  2 \epsilon |\offeredAsmt | \sum_{i\in \offeredAsmt }  \Vi r_i  }{ v_0 +\sum_{i \in \offeredAsmt} \Vi  } \hspace{5mm}  = \;\;2 \epsilon |\offeredAsmt | R_{MNL}(\offeredAsmt). \quad \blacksquare
\end{aligned} $ 
} \fi
\endproof

\begin{lemma}
    \label{lem:MNL_optbound}
    Under Assumption \ref{assume:small_vij},   $\revtwo(\offeredAsmt_{MNL}^*) \geq \frac{2 - \epsilon |\offeredAsmt^*_{MNL}|  }{2 + 4\epsilon|\offeredAsmt^*|}  \revtwo(\offeredAsmt^*). $
\end{lemma}

\proof{Proof.}
\ifisVerbose{
$\begin{aligned}[t]
 \revtwo(\offeredAsmt^*) - \revtwo(\offeredAsmt_{MNL}^*)  &= \left( \revtwo(\offeredAsmt^*) - R_{MNL}(\offeredAsmt^*) \right) + \left( R_{MNL}(\offeredAsmt^*) - R_{MNL}(\offeredAsmt^*_{MNL})\right) + \left(  R_{MNL}(\offeredAsmt^*_{MNL}) - \revtwo(\offeredAsmt_{MNL}^*) \right) \\
&\leq \left( \revtwo(\offeredAsmt^*) - R_{MNL}(\offeredAsmt^*) \right) + \left(  R_{MNL}(\offeredAsmt^*_{MNL}) - \revtwo(\offeredAsmt_{MNL}^*) \right) \\
&\leq 2\epsilon |\offeredAsmt^*| R_{MNL}(\offeredAsmt^*) + \epsilon \frac{|\offeredAsmt^*_{MNL}|}{2} R_{MNL}(\offeredAsmt^*_{MNL}) \hspace{30mm} \text{(Using Lemma \ref{lem:MNL_ub} and \ref{lem:MNL_lb})} \\
&\leq \frac{4|\offeredAsmt^*| + |\offeredAsmt^*_{MNL}|}{2} \epsilon R_{MNL}(\offeredAsmt^*_{MNL}) \\
&\leq \frac{4|\offeredAsmt^*| + |\offeredAsmt^*_{MNL}|}{2} \frac{2}{2- \epsilon|\offeredAsmt^*_{MNL}|}  \epsilon  \revtwo(\offeredAsmt^*_{MNL}).
\end{aligned} $
Thus, we have
\begin{align*}
\revtwo(\offeredAsmt_{MNL}^*) \geq \frac{2 - \epsilon |\offeredAsmt^*_{MNL}|  }{2 + 4\epsilon|\offeredAsmt^*|}  \revtwo(\offeredAsmt^*).
\end{align*}
} \else{ 
$\begin{aligned}[t]
\revtwo(\offeredAsmt^*) - \revtwo(\offeredAsmt_{MNL}^*)  &= \left( \revtwo(\offeredAsmt^*) - R_{MNL}(\offeredAsmt^*) \right) + \left( R_{MNL}(\offeredAsmt^*) - R_{MNL}(\offeredAsmt^*_{MNL})\right) + \left(  R_{MNL}(\offeredAsmt^*_{MNL}) \right.\\
&\quad\quad  \left. - \revtwo(\offeredAsmt_{MNL}^*) \right)
\quad \leq\quad 2\epsilon |\offeredAsmt^*| R_{MNL}(\offeredAsmt^*) + \epsilon \frac{|\offeredAsmt^*_{MNL}|}{2} R_{MNL}(\offeredAsmt^*_{MNL}) \hspace{30mm} \\
&\leq \frac{4|\offeredAsmt^*| + |\offeredAsmt^*_{MNL}|}{2} \epsilon R_{MNL}(\offeredAsmt^*_{MNL})   \hspace{5mm} \leq \;\; \frac{4|\offeredAsmt^*| + |\offeredAsmt^*_{MNL}|}{2} \frac{2}{2- \epsilon|\offeredAsmt^*_{MNL}|}  \epsilon  \revtwo(\offeredAsmt^*_{MNL}).
\end{aligned} $ 
The inequalities are obtained using Lemmas \ref{lem:MNL_ub} and \ref{lem:MNL_lb}. $\blacksquare$
} \fi

\endproof

\proof{Proof of Theorem \ref{thm:small_vij} .}
  As $\offeredAsmt^*_{MNL}$ is a revenue-ordered recommendation set, using Lemma \ref{lem:MNL_optbound},   $\revtwo(C^*_{revord}) \geq \revtwo(\offeredAsmt_{MNL}^*) \geq \frac{2 - \epsilon |\offeredAsmt^*_{MNL}|  }{2 + 4\epsilon|\offeredAsmt^*|}  \revtwo(\offeredAsmt^*)  \geq \frac{2 - \epsilon \numberOfItems  }{2 + 4\epsilon \numberOfItems}  \revtwo(\offeredAsmt^*).$ $\blacksquare$
\endproof

\begin{remark}Guarantees similar to the above can be obtained for \binsearcheff{} if we can obtain an approximation guarantee for the \textsc{Compare-Step} under the same assumptions. We omit this analysis here due to space constraints.\end{remark}

%% file: 94_more_experiments.tex
\section{Additional Experimental Details} \label{sec:add_exp}

\noindent\textbf{Additional Data Descriptive Statistics.} A summary of the fractions of product bundles purchased across datasets (including transactions of all sizes) is presented in Table~\ref{tab:product-percentages-full}.

\begin{table}
\centering
\resizebox{\textwidth}{!}{%
\begin{tabular}{|c|c|c|c|c|c|c|c|c|}
\hline
\textbf{Purchase bundle size} &
  \textbf{Bakery} &
  \textbf{Instacart} &
  \textbf{Kosarak} &
  \textbf{LastFM Genres} &
  \textbf{Walmart Items} &
  \textbf{Yoochoose Items} &
  \textbf{Ta Feng} &
  \textbf{UCI Online Retail} \\ \hline
1                & 4.8\%  & 4.9\% & 15.4\% & 37.8\% & 21.3\%  & 0.0\%  & 14.9\% & 22.6\% \\ \hline
2                & 18.1\% & 5.8\% & 20.0\% & 15.6\% & 17.8\%   & 2.0\%  & 12.7\% & 6.2\%  \\ \hline
3                & 32.9\% & 6.4\% & 17.6\% & 9.0\%  & 10.8\%  & 7.8\%  & 11.2\% & 4.2\%  \\ \hline
4                & 22.7\% & 6.9\% & 12.1\% & 6.1\%  & 8.1\%    & 13.7\% & 9.3\%  & 3.1\%  \\ \hline
5                & 11.5\% & 7.1\% & 7.4\%  & 4.5\%  & 6.1\%    & 22.4\% & 8.0\%  & 3.0\%  \\ \hline
6                & 5.1\%  & 7.1\% & 4.6\%  & 3.5\%  & 4.8\%    & 0.4\%  & 6.8\%  & 2.6\%  \\ \hline
7                & 2.9\%  & 6.8\% & 3.0\%  & 2.8\%  & 3.9\%    & 1.3\%  & 5.6\%  & 2.5\%  \\ \hline
8                & 2.0\%  & 6.3\% & 2.2\%  & 2.4\%  & 3.2\%    & 2.0\%  & 4.7\%  & 2.4\%  \\ \hline
9                &        & 5.7\% & 1.8\%  & 2.0\%  & 2.7\%    & 3.7\%  & 3.9\%  & 2.4\%  \\ \hline
10               &        & 5.1\% & 1.5\%  & 1.7\%  & 2.3\%    & 4.4\%  & 3.3\%  & 2.1\%  \\ \hline
11               &        & 4.6\% & 1.2\%  & 1.4\%  & 2.0\%    & 11.6\% & 2.8\%  & 2.2\%  \\ \hline
12               &        & 4.1\% & 1.0\%  & 1.3\%  & 1.8\%    & 0.4\%  & 2.4\%  & 1.9\%  \\ \hline
13               &        & 3.6\% & 0.9\%  & 1.1\%  & 1.6\%    & 0.7\%  & 2.0\%  & 2.0\%  \\ \hline
14               &        & 3.2\% & 0.8\%  & 1.0\%  & 1.3\%    & 1.1\%  & 1.7\%  & 2.0\%  \\ \hline
15               &        & 2.9\% & 0.7\%  & 0.8\%  & 1.2\%    & 1.8\%  & 1.4\%  & 2.1\%  \\ \hline
16               &        & 2.5\% & 0.6\%  & 0.8\%  & 1.1\%    & 2.4\%  & 1.2\%  & 2.2\%  \\ \hline
17               &        & 2.2\% & 0.5\%  & 0.7\%  & 0.9\%    & 7.1\%  & 1.1\%  & 1.8\%  \\ \hline
18               &        & 2.0\% & 0.5\%  & 0.6\%  & 0.8\%    & 0.3\%  & 0.9\%  & 1.7\%  \\ \hline
19               &        & 1.7\% & 0.4\%  & 0.5\%  & 0.7\%    & 0.4\%  & 0.8\%  & 1.9\%  \\ \hline
20               &        & 1.5\% & 0.4\%  & 0.5\%  & 0.7\%    & 0.6\%  & 0.7\%  & 1.7\%  \\ \hline
\textgreater{}20 &        & 9.5\% & 7.2\%  & 6.0\%  & 6.8\%  & 15.8\% & 4.7\%  & 29.3\% \\ \hline
\end{tabular}%
}
\caption{Summary of transaction sizes across the $8$ real-world datasets.}
\label{tab:product-percentages-full}
\end{table}

\begin{table}
\tiny
\centering
\resizebox{.99\textwidth}{!}{
\begin{tabular}{ |c|c|c|c|c|  }
\hline 
\textbf{Heuristic} & LAGUNA2009HCE & BURER2002 & DUARTE2005 &
    FESTA2002VNS\\ \hline
\textbf{Number of Times Ranked First} &   22.3\% & 20.2\% & 15.9\%  & 14.8 \% \\ \hline
\end{tabular}
}
\caption{Performance of QUBO heuristics: fraction of times the top heursitics gave the best solution. \label{fig:prod5000rcm}}
\end{table}

\noindent\textbf{Selecting QUBO heuristics for \noisybinsearcheff.} 
We consider $\sim$20 heuristics discussed in ~\cite{dunning2018works} (available at \url{https://github.com/MQLib/MQLib}), and solve several QUBO instances at the \textsc{Compare-Step}s from optimizing over multiple synthetically generated \rcm{} models. Our results show that none of the heuristics consistently outperform others in terms of solution quality. Thus, we measure the performance of each heuristic in terms of the proportion of times it gave the best solution among all heuristics. Table \ref{fig:prod5000rcm} summarizes the performance of the top four heuristics ranked in this manner. Thus, for our experiments, we run these four heuristics in parallel at each comparison step of \noisybinsearch, and select the best solution to resolve the comparison.

\noindent\textbf{Qualitative Results on the UCI Online Retail Dataset from Estimated \bmvl-2.}
To complement the detailed quantitative analysis using Instacart dataset in Section~\ref{sec:beta}, we summarize some qualitative insights using another dataset here. In particular, we take a brief look at the estimated parameters of \bmvl-2 for the UCI dataset, which has (only) product names available. 
Based on our exploratory analysis, product pairs that have high $\beta_{ij}$ values in the estimated \rcm \ model include: (a) necklaces and earrings, (b) necklaces and bracelets, (c) placemats and coasters, and (d) wall art for gents and wall art for ladies, among others. These pairs of products are clearly complementary. The estimated \rcm \ model thus, is able to learn the complementary relation between these pairs of products, and our optimization schemes would be able to use this information effectively for recommendation set optimization.

In addition to the product pairs which are clearly complementary, we also see other pairs of products having high $\beta_{ij}$s. For instance, necklaces in two different colors, key rings with two different letters etched, numbered tiles with consecutive numbers, and candle plates and candle holders, which don't seem complementary at a first glance but still end up having high $\beta_{ij}$ values. Some possible reasons for simultaneous purchase of two such similar items could be: a) if a customer has strong affinity for a product, they might want to have multiple versions of it (such as in different colors), or b) customers might want to purchase two similar items with the intention of choosing one among the two later on and eventually returning one of the products. In either case, information about these less obvious product pairs is also valuable for the decision maker.

\noindent\textbf{Empirical Fit of \bmvl-2 on Additional Datasets.} In Table~\ref{tab:estimdatasetresults}, we report the likelihood fit of multiple choice models on six datasets, showing that it is extremely competitive with competing models. In the table, the MMNL model was estimated with $5$ components, and the \ucm~model was estimated with varying proportions of correction sets. Noe that the estimation technique for MMNL failed to converge for some datasets due to the number of observations.

\begin{table}[H]
	\tiny
	\centering
	\begin{tabular}{ |c||c|c|c||c|c|c||c|c|c| }
		\hline
		\textbf{Dataset} & \multicolumn{3}{c||}{\textbf{Bakery Dataset}} & \multicolumn{3}{c||}{\textbf{Kosarak Dataset}} & \multicolumn{3}{c||}{\textbf{Walmart Items Dataset}}\\
		\hline
		\textbf{Model}& \textbf{No. Params} & \textbf{Train LL} & \textbf{Test LL} & \textbf{No. Params} & \textbf{Train LL} & \textbf{Test LL} & \textbf{No. Params} & \textbf{Train LL} & \textbf{Test LL}\\
		\hline
		MNL    & 50 & -91140 &   -22736& 2621    & -1317416 &   -331857 & 1075 & -160299 & -40526 \\
		\hline
		MMNL (5)   & 255 & -91362 &  -22785 & 13110  & -1337913& -345109 & 5380 & -158568& -40039 \\ 
		\hline
		\ucm~(0\%)& 50  & -90977   & -22691 &   2621  & -1404888 & -353659 &  1075  & -168250   & -42494 \\
		\hline
		\ucm~(1\%)& 62 & -77674 &  -19289 & 2812 & -1389932 &  -350060 & 1098 & -142794 &  -35985\\
		\hline
		\ucm~(5\%)& 110 & -77596 & -19303 & 3580 & -1386661 &  -349553 & 1194 & -138402 &  -35017\\
		\hline
		\ucm~(20\%)&   292  & -77451 & -19373 &   6460  & -1379610 & -349239 &    1551  & -135825 & -34871\\
		\hline
		\ucm~(50\%)& 655 & -77214 & -19431 & 12220 & -1358675 & -350582& 2265 & -131954   & -35018\\
		\hline
		\ucm~(100\%)& 1261 & -76791 & -19571 & 21819 & -1326125 & -436080 & 3456 & -125531 & -48916\\
		\hline
		\rcm & 1261 & -76791 & \textbf{-19281}& 21733 & -1325751 & \textbf{-294304} & 3384 & -125410 & \textbf{-26518}\\
		\hline
		\hline
		\textbf{Dataset} & \multicolumn{3}{c||}{\textbf{LastFM Genres Dataset}} & \multicolumn{3}{c||}{\textbf{\textbf{Yoochoose Items Dataset}}}  & \multicolumn{3}{c||}{\textbf{Instacart Dataset}} \\
		\hline
        \textbf{Model}& \textbf{No. Params} & \textbf{Train LL} & \textbf{Test LL} & \textbf{No. Params} & \textbf{Train LL} & \textbf{Test LL} & \textbf{No. Params} & \textbf{Train LL} & \textbf{Test LL}\\
		\hline
		MNL    & 443    & -1981969 &   -495887& 52677    & -256610 &  -64795 &  5981    & -2668742 &   -670192 \\
		\hline
		MMNL (5)   & 2220 & - &  -& 263390 & -& -& 29910 & -& - \\
		\hline
		\ucm~(0\%)& 443  & -2138089   & -534898 &   52679  & -303703   &-76847.8 &  5981  & -2749875   &-690323  \\
		\hline
		\ucm~(1\%)&  529 & -2128021&  -532535& 52985 & -291047&  -74024 & 6767 & -2722321 &  -684258\\
		\hline
		\ucm~(5\%)& 873 & -2112185 & -528698 & 54209 & -287511 & -73371 & 9914 & -2690143 &  -679500 \\
		\hline
		\ucm~(20\%)&   2166 & -2096532 & -525428 &   58801  & -286012 & -73212 &  21714  & -2624928 & -676805\\
		\hline
		\ucm~(50\%)& 4750 & -2086600 & -524088 & 67985 & -285872   & -73462 &  45313 & -2510259  & -681712 \\
		\hline
		\ucm~(100\%)&9058 & -2076535 & -509677 & 83292 & -283825 & -73320 &  84646 & -2346723 & -1170856\\
		\hline
		\rcm~& 8871 & -2009631 & \textbf{-493188}& 34776 & -178693 & \textbf{-38903} & 84545 & -2346233 & \textbf{-393099}\\
		\hline
	\end{tabular}
	\caption{Log-likelihood (LL) fits for different models on multiple real datasets.}
	\label{tab:estimdatasetresults}
\end{table}

\begin{figure}
	\begin{subfigure}[b]{.3\linewidth}
		\centering
	 \includegraphics[width=\linewidth]{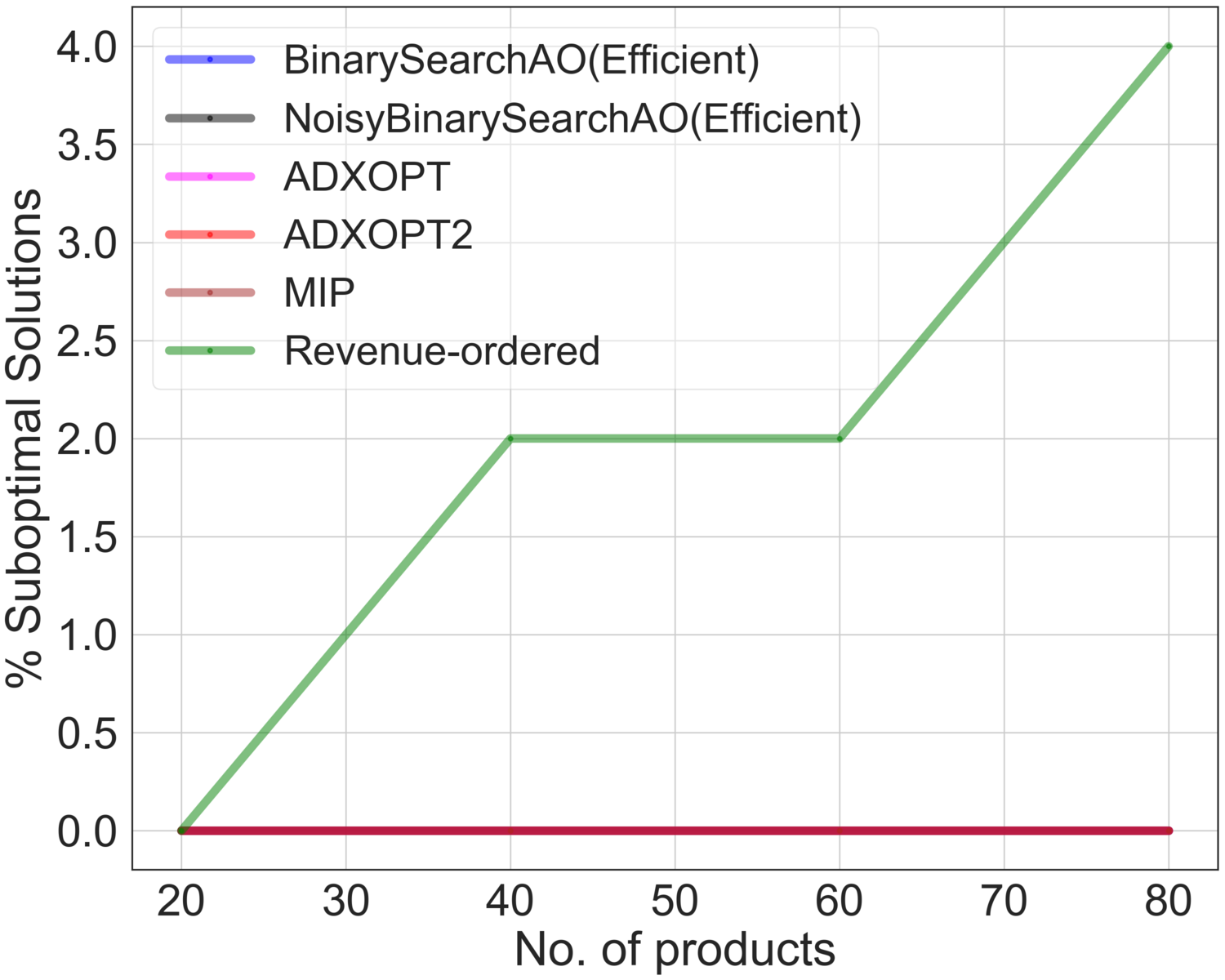}
		\caption{\label{fig10a}}
	\end{subfigure}\hfill
	\begin{subfigure}[b]{.3\linewidth}
		\centering
	\includegraphics[width=\linewidth]{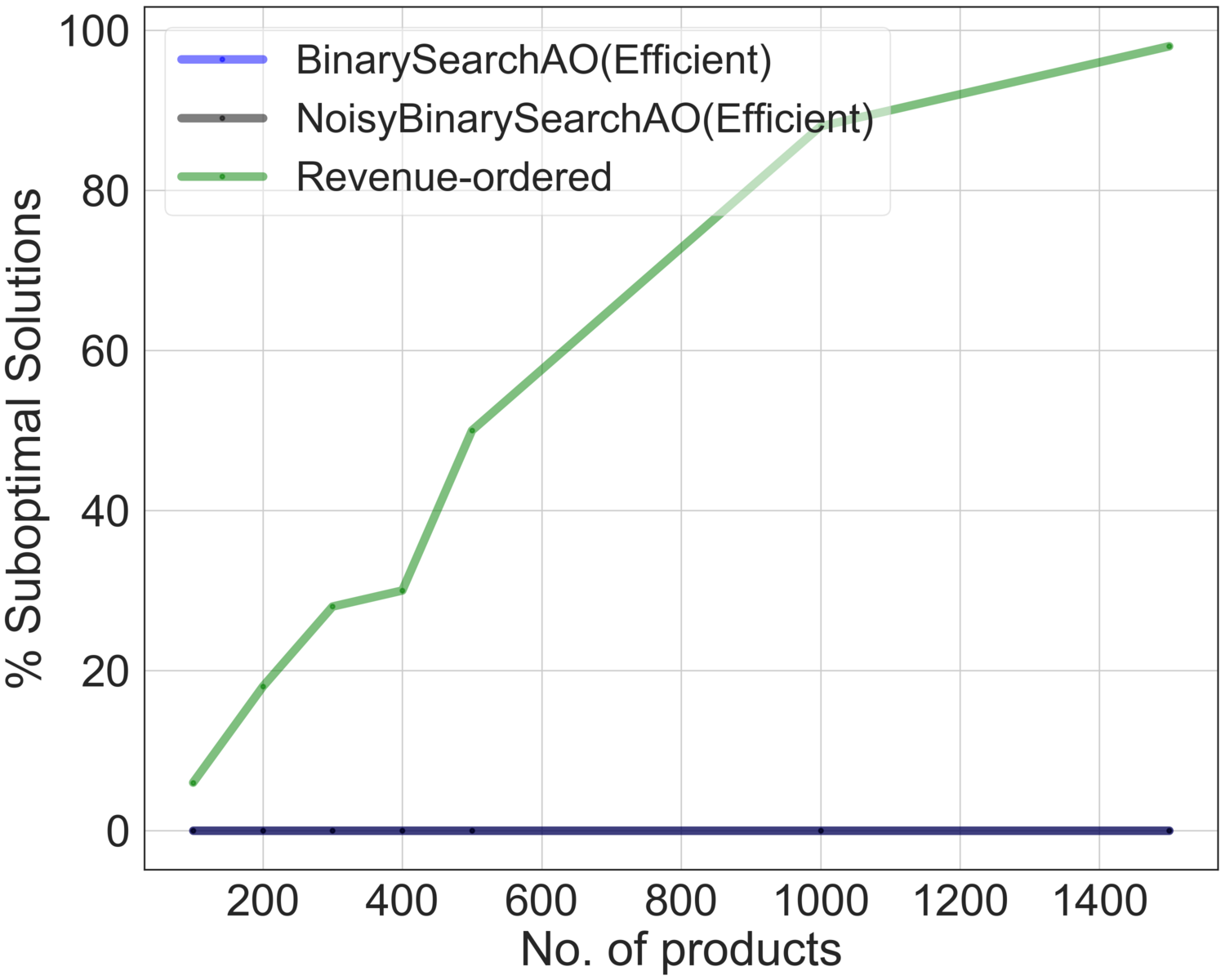}      
		\caption{\label{fig10b}}
	\end{subfigure}\hfill
	\begin{subfigure}[b]{.3\linewidth}
		\centering
		 \includegraphics[width=\linewidth]{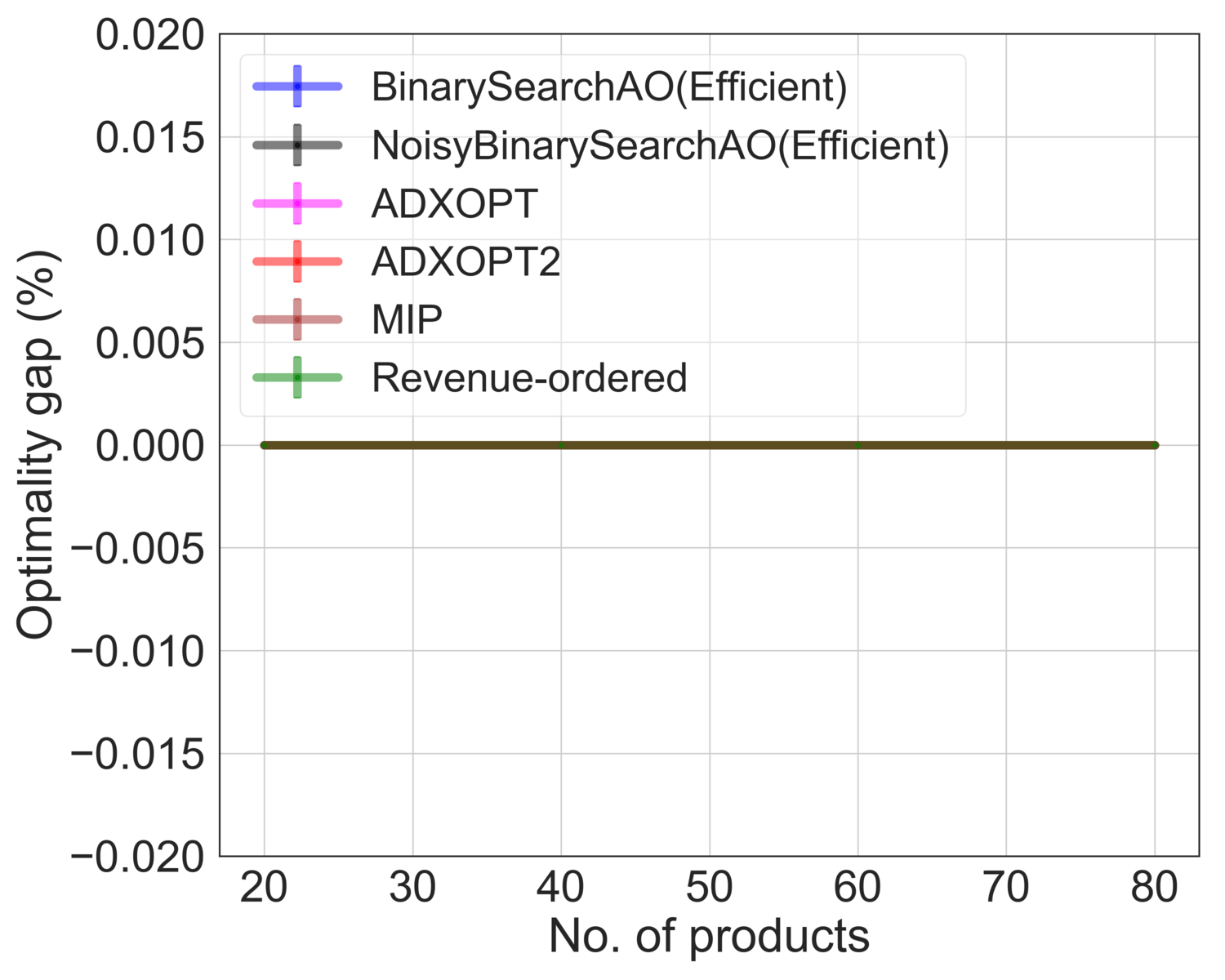}
		\caption{\label{fig10c}}
	\end{subfigure}\vfill
	\begin{subfigure}[b]{.3\linewidth}
		\centering
		\includegraphics[width=\linewidth]{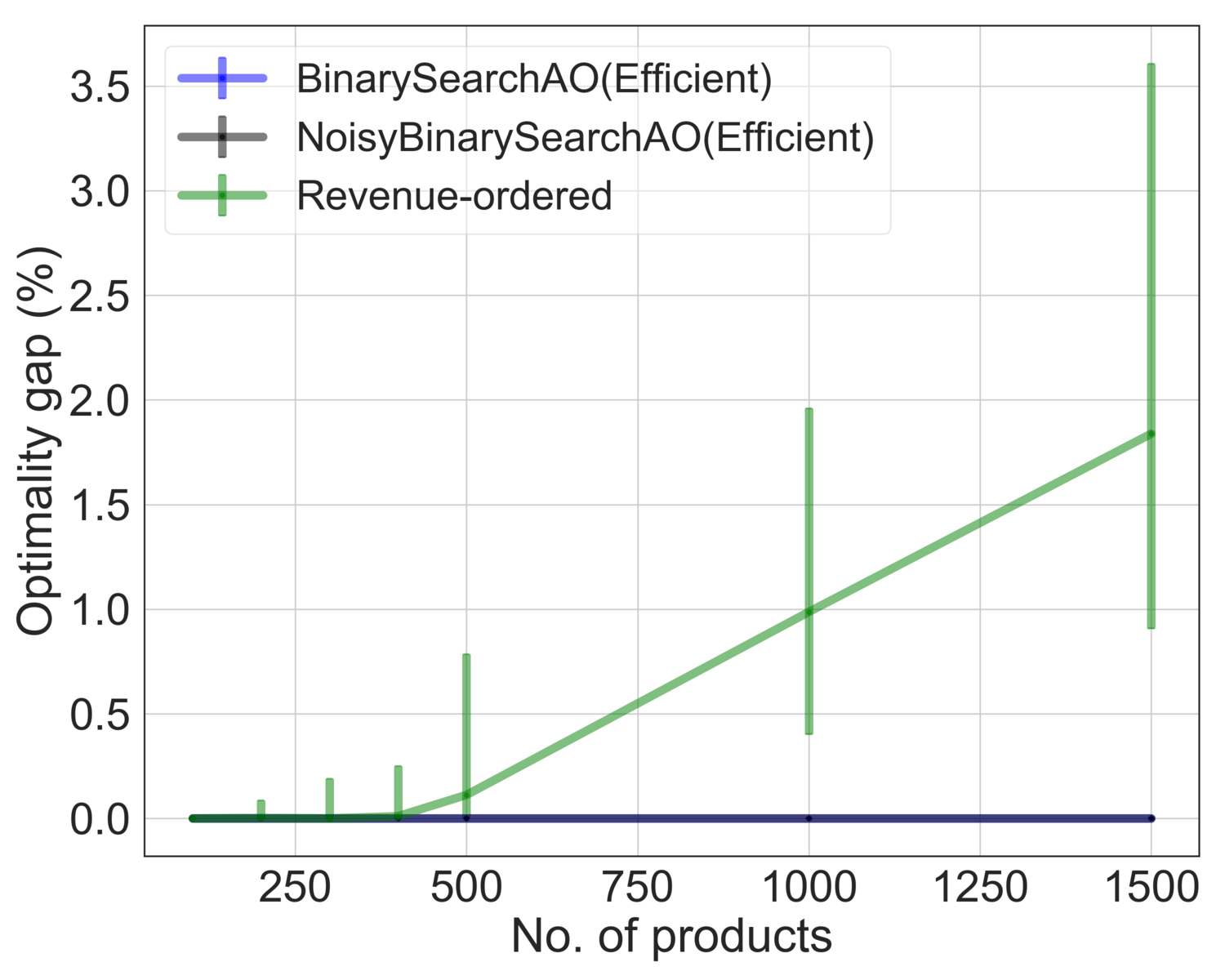}  
		\caption{\label{fig10d}}
	\end{subfigure}\hfill
    \begin{subfigure}[b]{.3\linewidth}
    	\centering
    	\includegraphics[width=\linewidth]{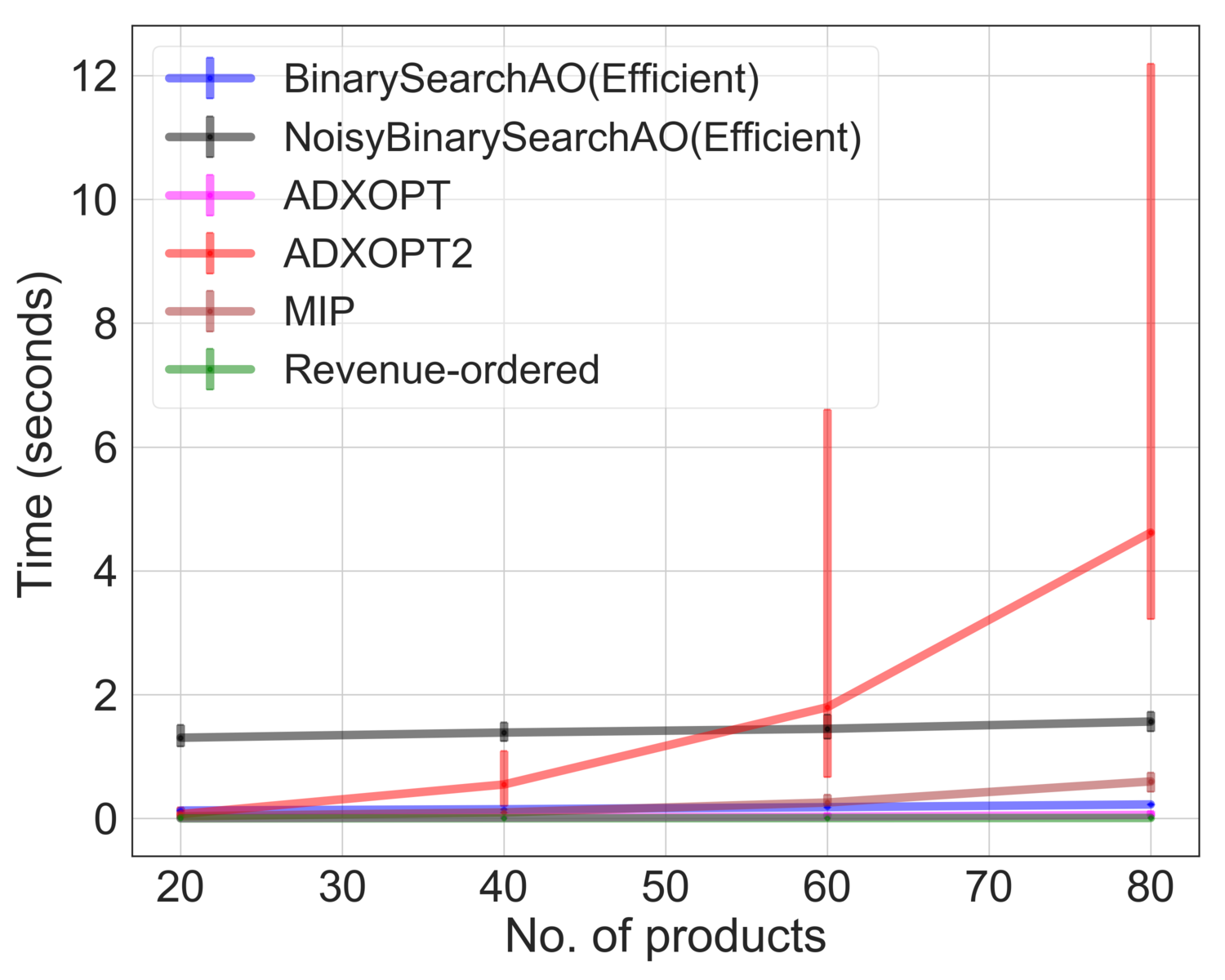}
    	\caption{\label{fig10e}}
    \end{subfigure}\hfill
    \begin{subfigure}[b]{.3\linewidth}
    	\centering
    	\includegraphics[width=\linewidth]{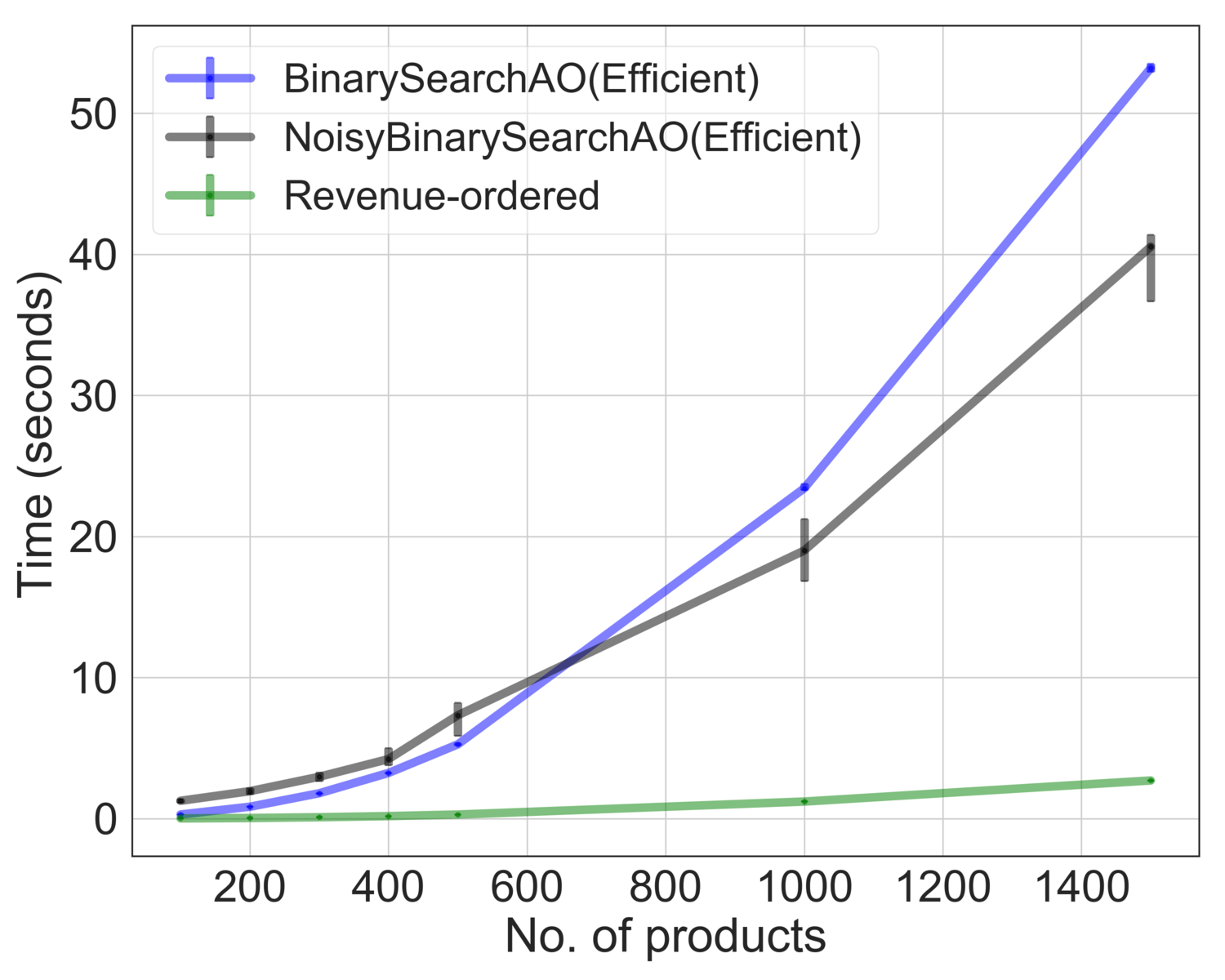}  
    	\caption{\label{fig10f}}
    \end{subfigure}\vfill
    
	\caption{Optimality and run-time plots using the UCI dataset in the unconstrained setting for \bmvl-2.\label{fig:e1ucidatasetssuboptimalityanalysis}}
\end{figure}

\begin{figure}
	\begin{subfigure}[b]{.3\linewidth}
		\centering
		\includegraphics[width=\linewidth]{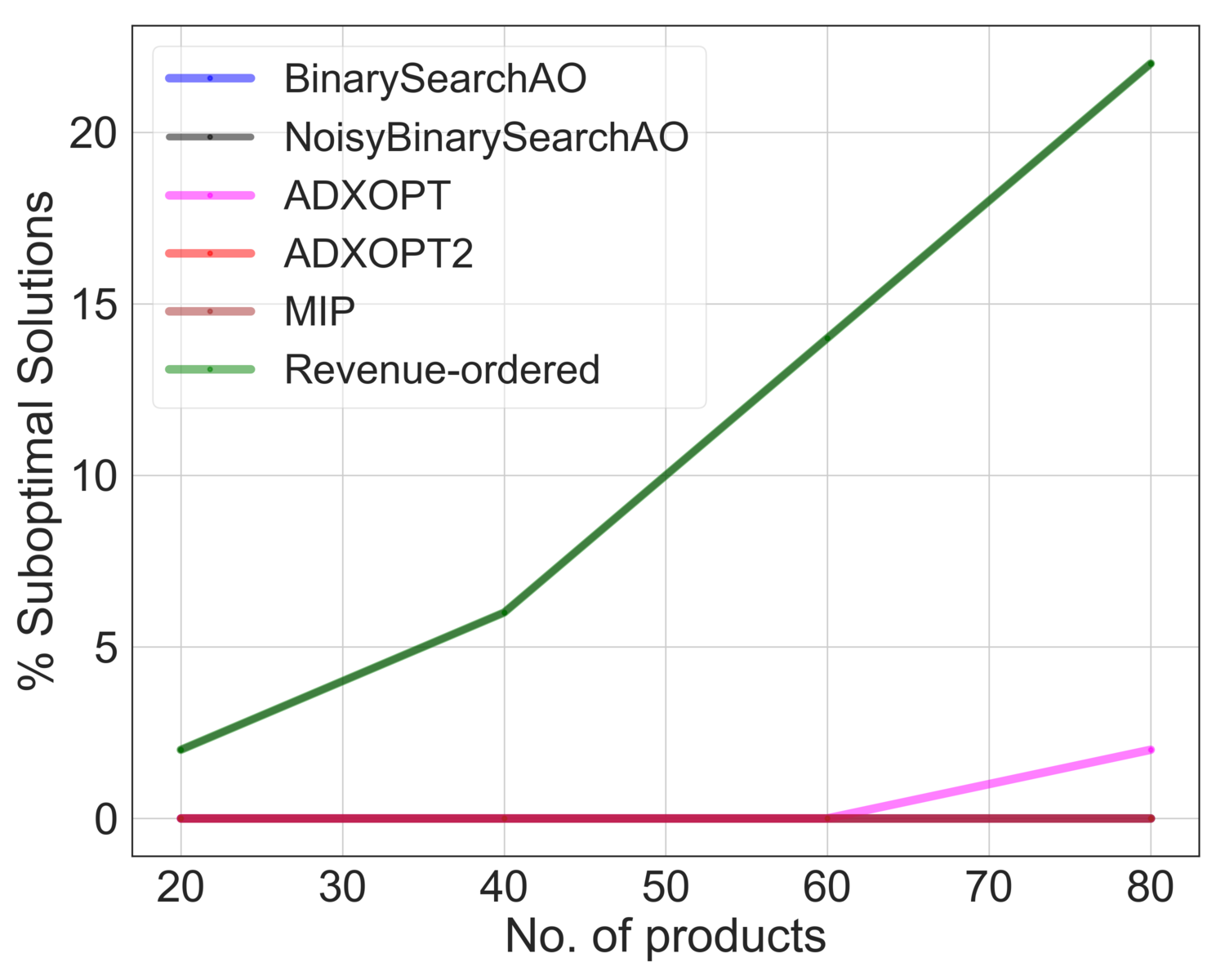}
		\caption{\label{fig11a}}
	\end{subfigure}\hfill
	\begin{subfigure}[b]{.3\linewidth}
		\centering
		\includegraphics[width=\linewidth]{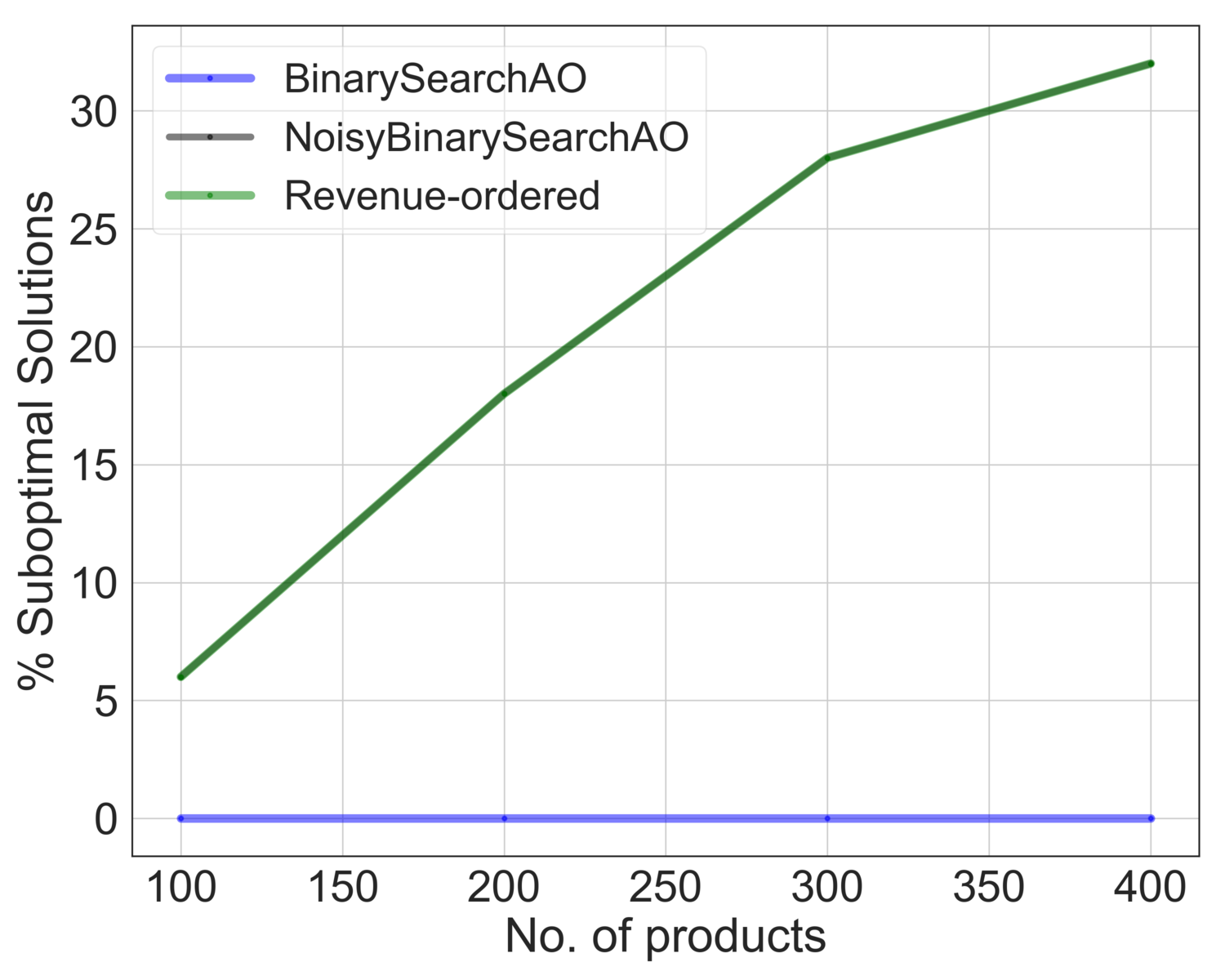}      
		\caption{\label{fig11b}}
	\end{subfigure}\hfill
	\begin{subfigure}[b]{.3\linewidth}
		\centering
		\includegraphics[width=\linewidth]{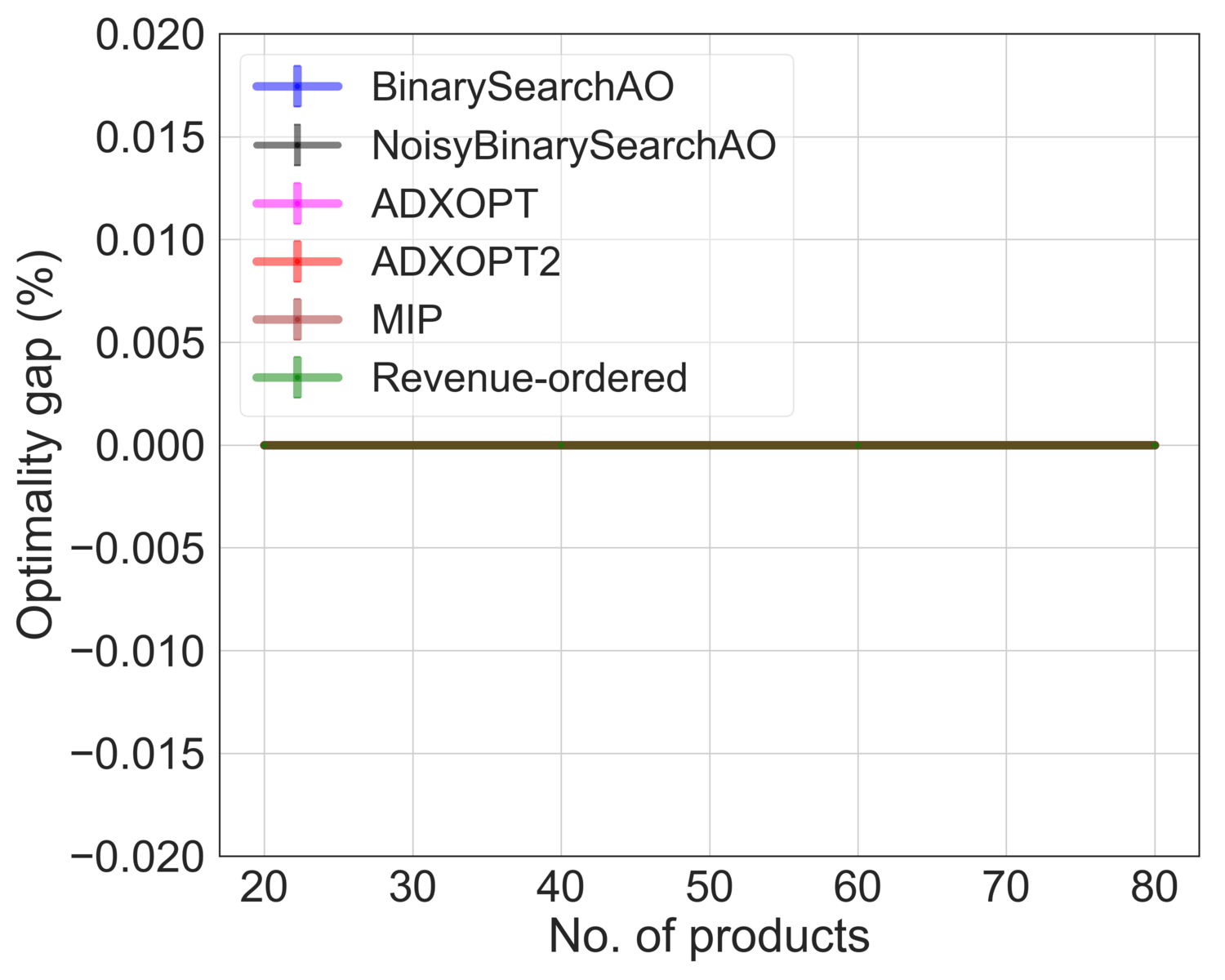}
		\caption{\label{fig11c}}
	\end{subfigure}\vfill
	\begin{subfigure}[b]{.3\linewidth}
		\centering
		\includegraphics[width=\linewidth]{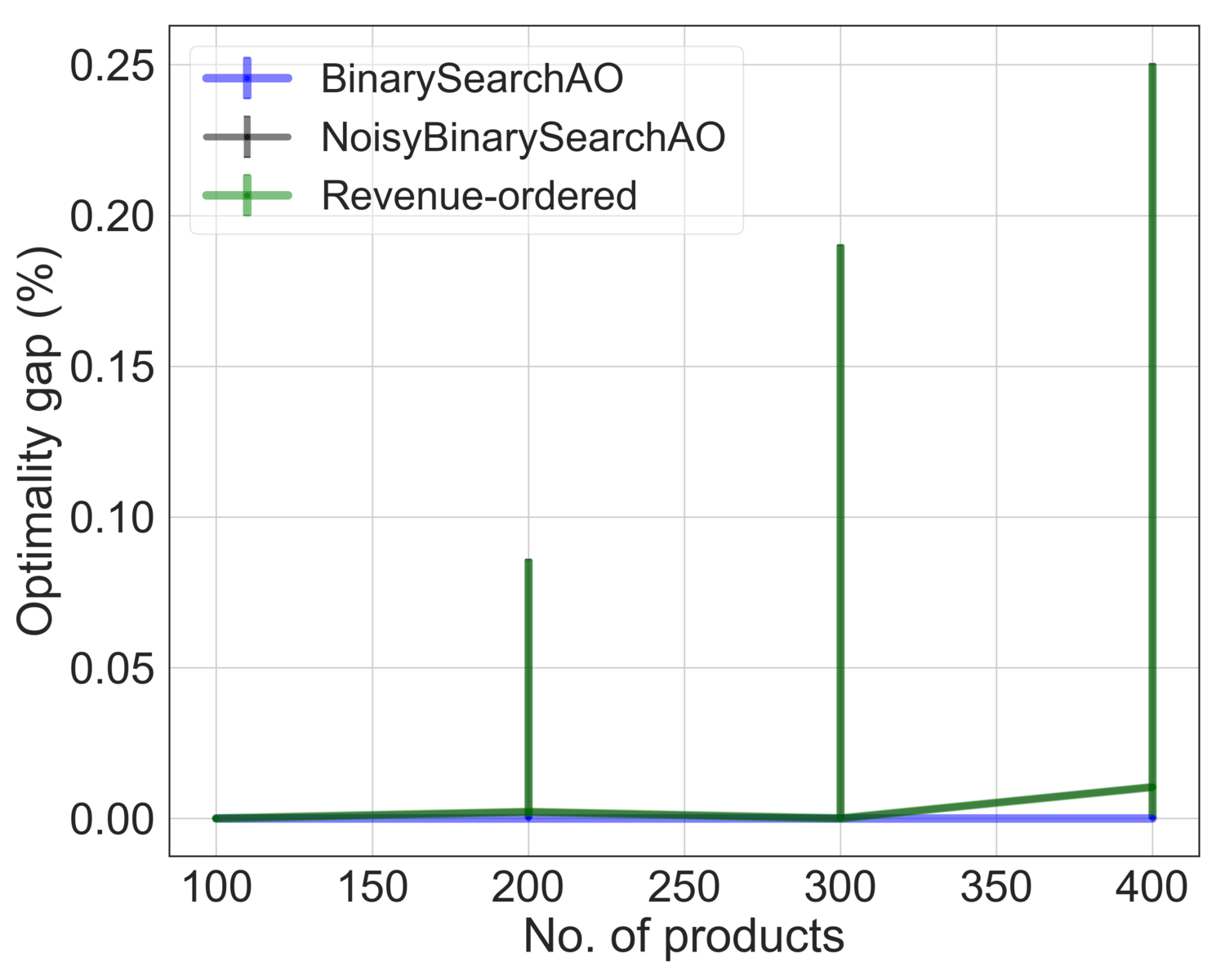}  
		\caption{\label{fig11d}}
	\end{subfigure}\hfill
	\begin{subfigure}[b]{.3\linewidth}
		\centering
		\includegraphics[width=\linewidth]{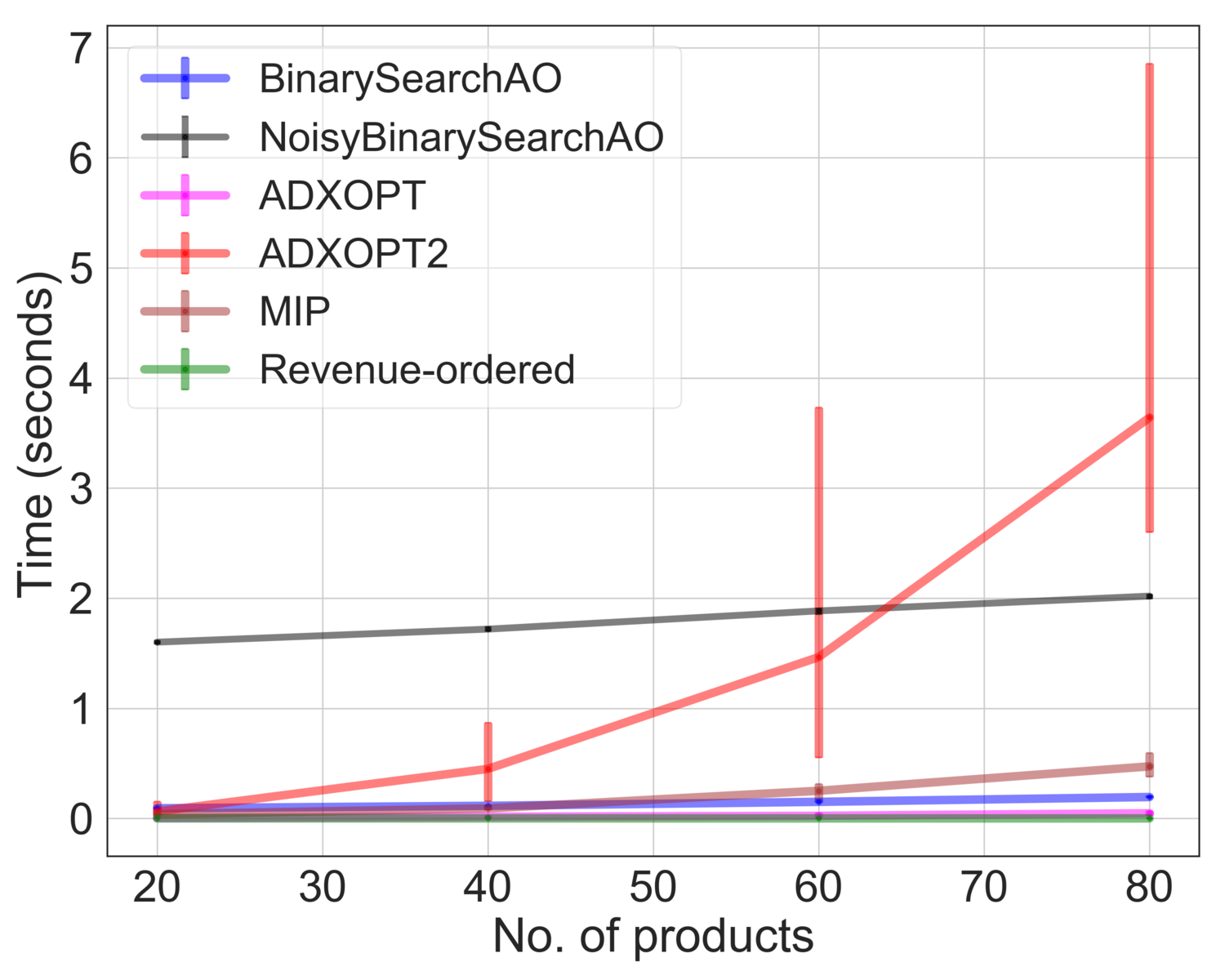}
		\caption{\label{fig11e}}
	\end{subfigure}\hfill
	\begin{subfigure}[b]{.3\linewidth}
		\centering
		\includegraphics[width=\linewidth]{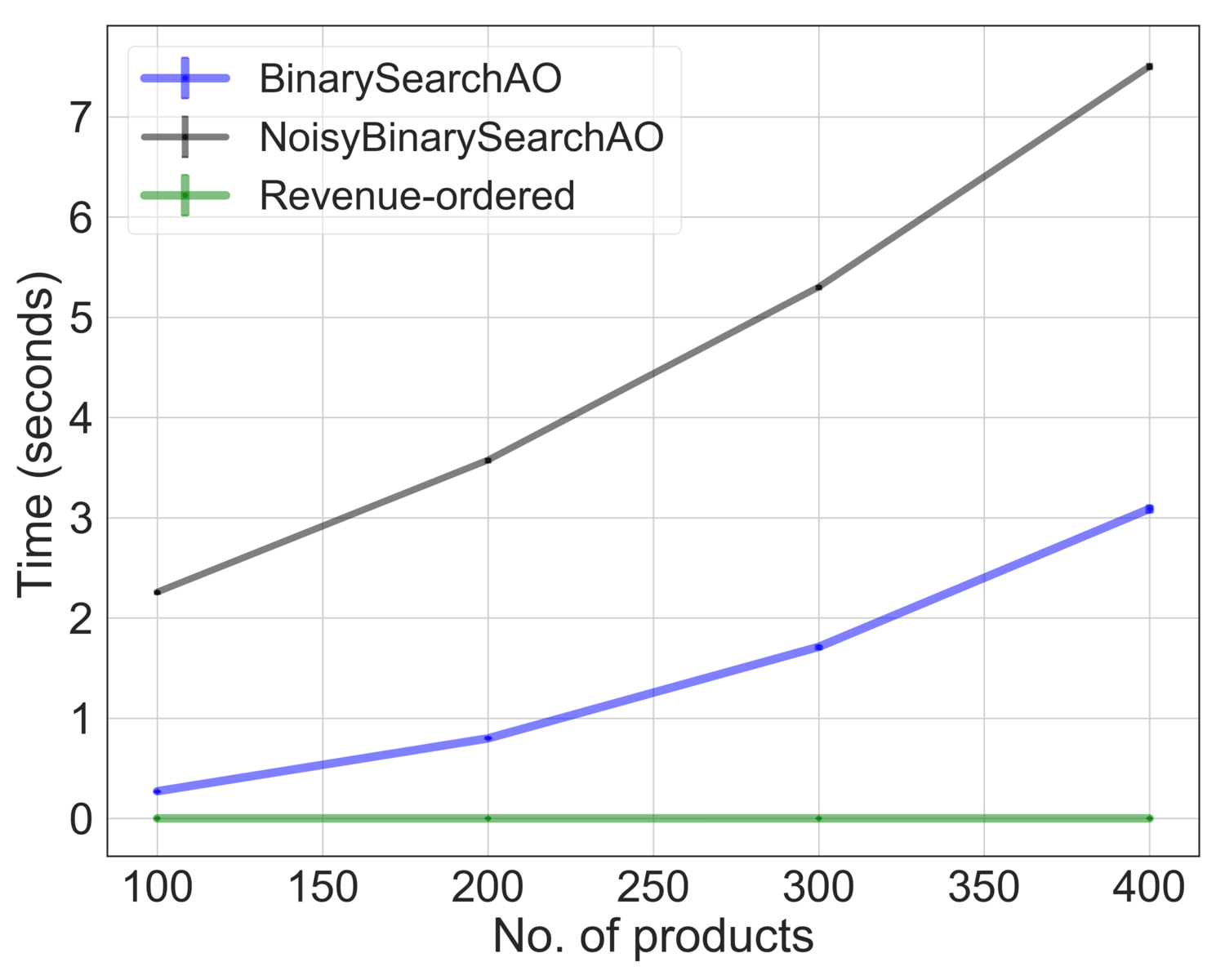}  
		\caption{\label{fig11f}}
	\end{subfigure}\vfill
	
	\caption{Optimality and run-time plots using the UCI dataset in the constrained setting for \bmvl-2.\label{fig:e1ucidatasetssuboptimalityanalysisconstrained}}

\end{figure}

\noindent\textbf{Run-times for Computing \rcm{} based Recommendation Sets using UCI.}
We report similar results for UCI, analogous to Section~\ref{sec:run-time-experiments} in Figures~\ref{fig:e1ucidatasetssuboptimalityanalysis} and~\ref{fig:e1ucidatasetssuboptimalityanalysisconstrained}. Similar results for synthetic datasets are omitted due to space constraints.

\begin{figure}
	\begin{subfigure}[b]{.3\linewidth}
		\centering
		\includegraphics[width=\linewidth]{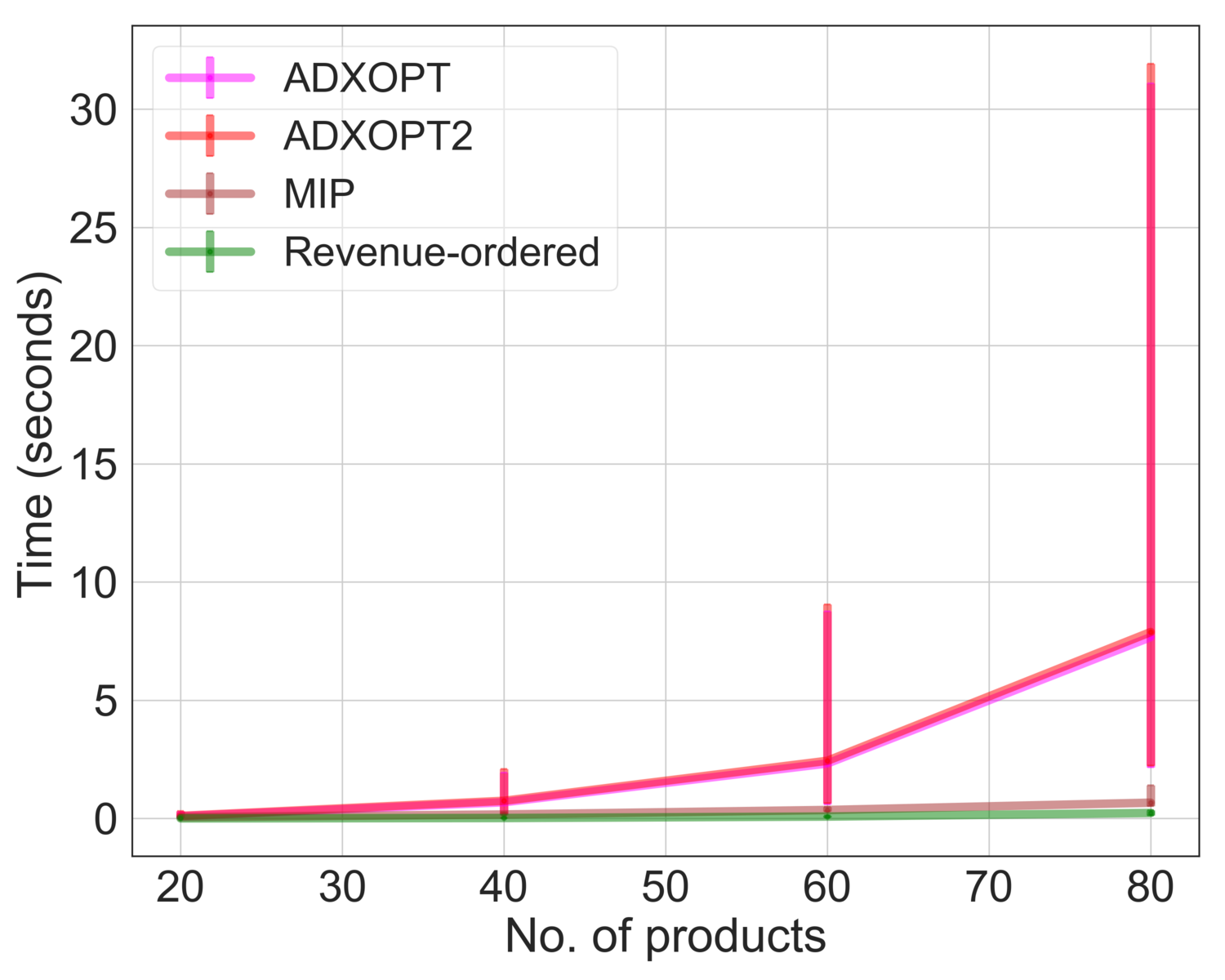}
		\caption{\label{fig12a}}
	\end{subfigure}\hfill
	\begin{subfigure}[b]{.3\linewidth}
		\centering
		\includegraphics[width=\linewidth]{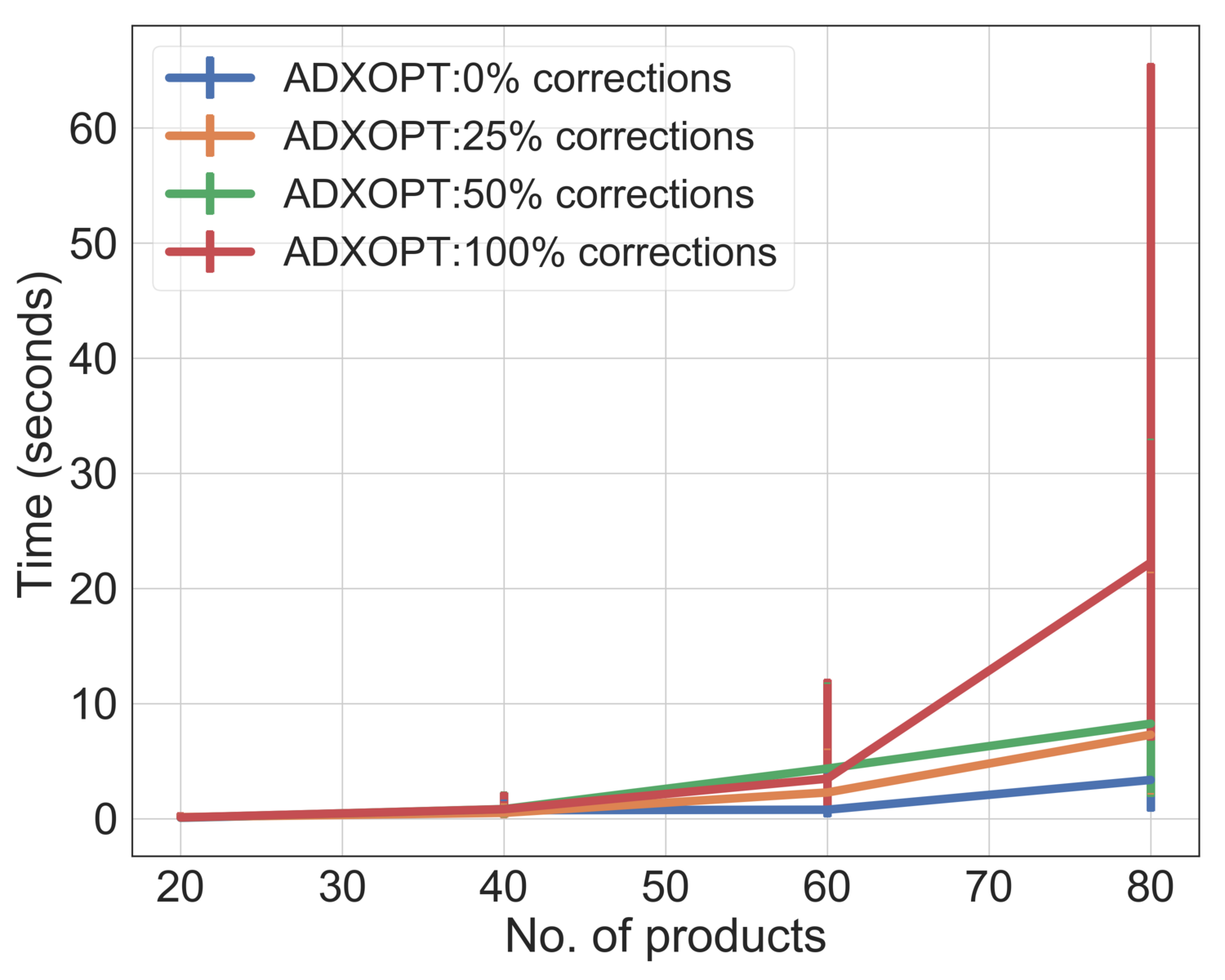}      
		\caption{\label{fig12b}}
	\end{subfigure}\hfill
	\begin{subfigure}[b]{.3\linewidth}
		\centering
		\includegraphics[width=\linewidth]{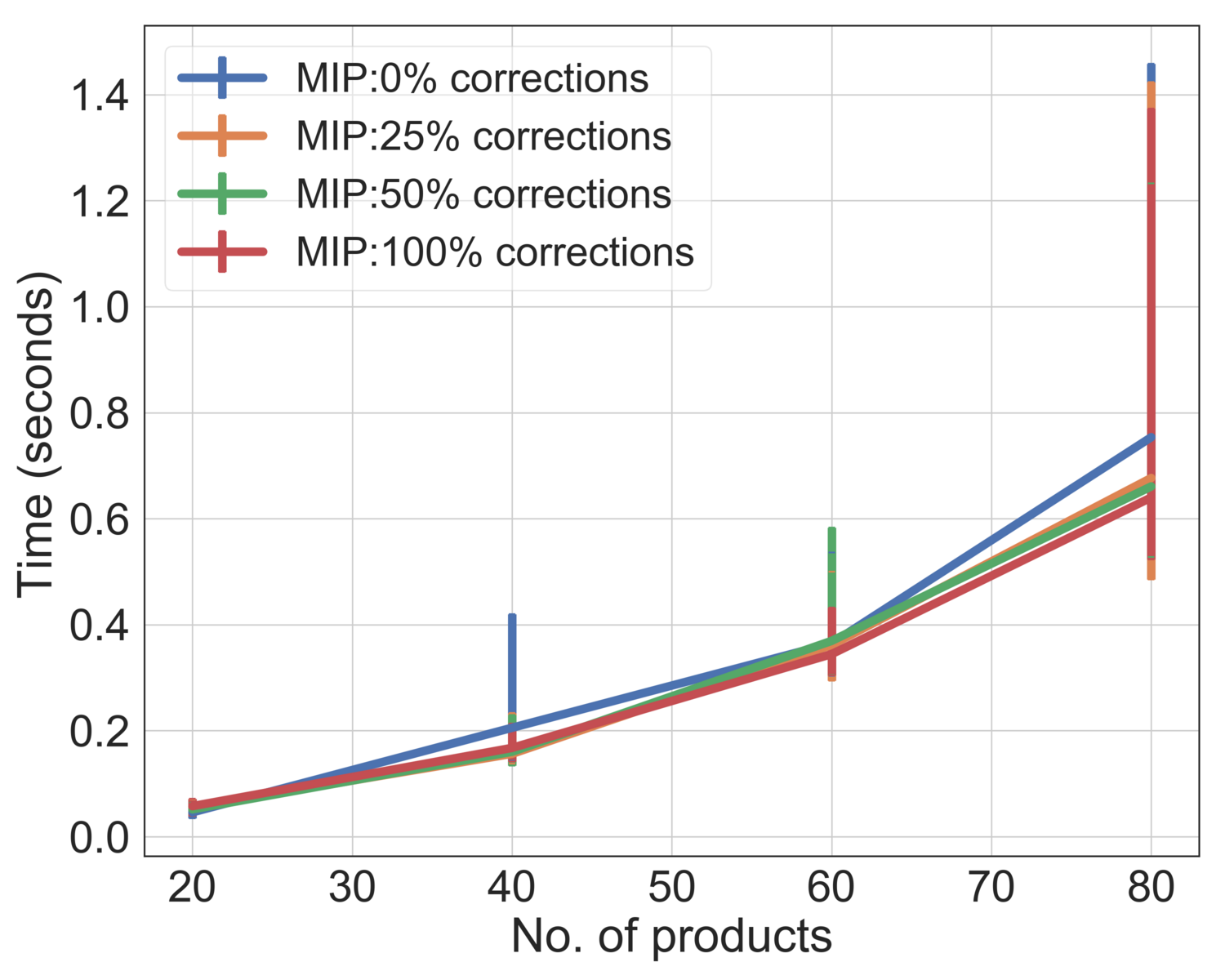}
		\caption{\label{fig12c}}
	\end{subfigure}\hfill

	\caption{Under the \ucm~model with synthetic data: 
		(\subref{fig12a}) Run-times of algorithms,
		(\subref{fig12b}) \& (\subref{fig12c}) Sensitivity of run-times to the number of correction sets for \adxopt~and MIP. 
		\label{fig:e2mmctimeanalysis}}

\end{figure}

\noindent\textbf{Computing \ucm{} based Recommendation Sets.} We report run-times using a novel mixed integer program (MIP) under the \ucm~ model in the small products regime (Figure~\ref{fig12a}), as neither constrained or unconstrained optimization is addressed in~\cite{benson2018discrete}. We observe that \adxopt~and \adxopttwo~have similar ranges of run-times, thus they significantly overlap with each other. In Figures~\ref{fig12b} \& \ref{fig12c}, we observe that for a fixed number of products, the time taken by \adxopt~ increases as the number of correction sets (H-sets) increase, while it remains unaffected for the MIP. Overall, revenue-ordered heuristic is likely the only approach that can scale to much larger instances for this model.

%% file: multi_purchase_msom_all.bbl
\begin{thebibliography}{30}
\providecommand{\natexlab}[1]{#1}
\providecommand{\url}[1]{\texttt{#1}}
\providecommand{\urlprefix}{URL }

\bibitem[{Benson et~al.(2018)Benson, Kumar, \protect\BIBand{}
  Tomkins}]{benson2018discrete}
Benson AR, Kumar R, Tomkins A (2018) A discrete choice model for subset
  selection. \emph{Eleventh ACM International Conference on Web Search and Data
  Mining}, 37--45 (ACM).

\bibitem[{Berbeglia et~al.(2022)Berbeglia, Garassino, \protect\BIBand{}
  Vulcano}]{berbeglia2022comparative}
Berbeglia G, Garassino A, Vulcano G (2022) A comparative empirical study of
  discrete choice models in retail operations. \emph{Management Science}
  68(6):4005--4023.

\bibitem[{Besag(1974)}]{besag1974spatial}
Besag J (1974) Spatial interaction and the statistical analysis of lattice
  systems. \emph{Journal of the Royal Statistical Society: Series B
  (Methodological)} 36(2):192--225.

\bibitem[{Blanchet et~al.(2016)Blanchet, Gallego, \protect\BIBand{}
  Goyal}]{blanchet2016markov}
Blanchet J, Gallego G, Goyal V (2016) A {M}arkov chain approximation to choice
  modeling. \emph{Operations Research} 64(4):886--905.

\bibitem[{Bonmin()}]{bonmin}
Bonmin (2023) Bonmin solver. \url{https://www.coin-or.org/Bonmin/}, accessed:
  2023-03-07.

\bibitem[{Burnashev \protect\BIBand{} Zigangirov(1974)}]{burnashev1974interval}
Burnashev MV, Zigangirov K (1974) An interval estimation problem for controlled
  observations. \emph{Problemy Peredachi Informatsii} 10(3):51--61.

\bibitem[{Cox(1972)}]{cox1972analysis}
Cox DR (1972) The analysis of multivariate binary data. \emph{Applied
  Statistics} 113--120.

\bibitem[{Davis et~al.(2013)Davis, Gallego, \protect\BIBand{}
  Topaloglu}]{davis2013assortment}
Davis J, Gallego G, Topaloglu H (2013) Assortment planning under the
  multinomial logit model with totally unimodular constraint structures.
  Technical report, School of ORIE, Cornell University.

\bibitem[{Dunning et~al.(2018)Dunning, Gupta, \protect\BIBand{}
  Silberholz}]{dunning2018works}
Dunning I, Gupta S, Silberholz J (2018) What works best when? a systematic
  evaluation of heuristics for {Max-Cut and QUBO}. \emph{INFORMS Journal on
  Computing} 30(3):608--624.

\bibitem[{Feldman et~al.(2022)Feldman, Zhang, Liu, \protect\BIBand{}
  Zhang}]{alibaba2019}
Feldman J, Zhang DJ, Liu X, Zhang N (2022) Customer choice models vs. machine
  learning: Finding optimal product displays on {Alibaba}. \emph{Operations
  Research} 70(1):309--328.

\bibitem[{Glover \protect\BIBand{} Kochenberger(2018)}]{glover2018tutorial}
Glover F, Kochenberger G (2018) A tutorial on formulating {QUBO} models.
  \emph{ArXiv 1811.11538} .

\bibitem[{Hruschka et~al.(1999)Hruschka, Lukanowicz, \protect\BIBand{}
  Buchta}]{hruschka1999cross}
Hruschka H, Lukanowicz M, Buchta C (1999) Cross-category sales promotion
  effects. \emph{Journal of Retailing and Consumer Services} 6(2):99--105.

\bibitem[{Jagabathula(2014)}]{jagabathula2014assortment}
Jagabathula S (2014) Assortment optimization under general choice. \emph{SSRN
  2512831} .

\bibitem[{K{\"o}k et~al.(2008)K{\"o}k, Fisher, \protect\BIBand{}
  Vaidyanathan}]{kok2008assortment}
K{\"o}k AG, Fisher ML, Vaidyanathan R (2008) Assortment planning: Review of
  literature and industry practice. \emph{Retail Supply Chain Management},
  99--153 (Springer).

\bibitem[{Kopalle et~al.(1999)Kopalle, Krishna, \protect\BIBand{}
  Assuncao}]{kopalle1999role}
Kopalle PK, Krishna A, Assuncao JL (1999) The role of market expansion on
  equilibrium bundling strategies. \emph{Managerial and Decision Economics}
  20(7):365--377.

\bibitem[{Manchanda et~al.(1999)Manchanda, Ansari, \protect\BIBand{}
  Gupta}]{manchanda1999shopping}
Manchanda P, Ansari A, Gupta S (1999) The shopping basket: A model for
  multicategory purchase incidence decisions. \emph{Marketing Science}
  18(2):95--114.

\bibitem[{McCardle et~al.(2007)McCardle, Rajaram, \protect\BIBand{}
  Tang}]{mccardle2007bundling}
McCardle KF, Rajaram K, Tang CS (2007) Bundling retail products: Models and
  analysis. \emph{European Journal of Operational Research} 177(2):1197--1217.

\bibitem[{M{\'e}ndez-D{\'\i}az et~al.(2014)M{\'e}ndez-D{\'\i}az, Miranda-Bront,
  Vulcano, \protect\BIBand{} Zabala}]{mendez2014branch}
M{\'e}ndez-D{\'\i}az I, Miranda-Bront JJ, Vulcano G, Zabala P (2014) A
  branch-and-cut algorithm for the latent-class logit assortment problem.
  \emph{Discrete Applied Mathematics} 164:246--263.

\bibitem[{Palmer(2016)}]{speed}
Palmer O (2016) How does page load time impact engagement?
  \url{https://www.optimizely.com/insights/blog/how-does-page-load-time-impact-engagement/},
  accessed: 2023-03-07.

\bibitem[{Pardalos \protect\BIBand{} Jha(1992)}]{pardalos1992complexity}
Pardalos PM, Jha S (1992) Complexity of uniqueness and local search in
  quadratic 0--1 programming. \emph{Operations Research Letters}
  11(2):119--123.

\bibitem[{Plackett(1975)}]{plackett1975analysis}
Plackett RL (1975) The analysis of permutations. \emph{Journal of the Royal
  Statistical Society: Series C (Applied Statistics)} 24(2):193--202.

\bibitem[{Rusmevichientong et~al.(2010{\natexlab{a}})Rusmevichientong, Shen,
  \protect\BIBand{} Shmoys}]{rusmevichientong2010dynamic}
Rusmevichientong P, Shen ZJM, Shmoys DB (2010{\natexlab{a}}) Dynamic assortment
  optimization with a multinomial logit choice model and capacity constraint.
  \emph{Operations Research} 58(6):1666--1680.

\bibitem[{Rusmevichientong et~al.(2010{\natexlab{b}})Rusmevichientong, Shmoys,
  \protect\BIBand{} Topaloglu}]{rusmevichientong2010assortment}
Rusmevichientong P, Shmoys D, Topaloglu H (2010{\natexlab{b}}) Assortment
  optimization with mixtures of logits. Technical report, School of ORIE,
  Cornell University.

\bibitem[{Russell \protect\BIBand{} Petersen(2000)}]{russell2000analysis}
Russell GJ, Petersen A (2000) Analysis of cross category dependence in market
  basket selection. \emph{Journal of Retailing} 76(3):367--392.

\bibitem[{Seetharaman et~al.(2005)Seetharaman, Chib, Ainslie, Boatwright, Chan,
  Gupta, Mehta, Rao, \protect\BIBand{} Strijnev}]{seetharaman2005models}
Seetharaman P, Chib S, Ainslie A, Boatwright P, Chan T, Gupta S, Mehta N, Rao
  V, Strijnev A (2005) Models of multi-category choice behavior.
  \emph{Marketing Letters} 16(3-4):239--254.

\bibitem[{Singh et~al.(2005)Singh, Hansen, \protect\BIBand{}
  Gupta}]{singh2005modeling}
Singh VP, Hansen KT, Gupta S (2005) Modeling preferences for common attributes
  in multicategory brand choice. \emph{Journal of Marketing Research}
  42(2):195--209.

\bibitem[{Train(2009)}]{train2009discrete}
Train KE (2009) \emph{Discrete choice methods with simulation} (Cambridge
  University Press).

\bibitem[{Tulabandhula et~al.(2021)Tulabandhula, Sinha, \protect\BIBand{}
  Karra}]{sinha2017optimizing}
Tulabandhula T, Sinha D, Karra S (2021) Optimizing revenue while showing
  relevant assortments at scale. \emph{European Journal of Operations
  Research.} 300:561--570, \urlprefix\url{https://arxiv.org/abs/2003.04736}.

\bibitem[{US~International Trade~Administration(2023)}]{marketsize}
US~International Trade~Administration (2023) {eCommerce} sales \& size
  forecast. \url{https://www.trade.gov/ecommerce-sales-size-forecast},
  accessed: 2023-03-07.

\bibitem[{Yang \protect\BIBand{} Sudharshan(2019)}]{yang2019examining}
Yang Z, Sudharshan D (2019) Examining multi-category cross purchases models
  with increasing dataset scale-{A}n artificial neural network approach.
  \emph{Expert Systems with Applications} 120:310--318.

\end{thebibliography}
